\documentclass[twocolumn]{aastex631}

\usepackage{CJKutf8}

\usepackage[caption=false]{subfig}
\usepackage{tabularx}
\usepackage{capt-of}
\usepackage{hhline}
\usepackage{amsmath}	
\usepackage{amssymb}	
\usepackage{textcomp}
\usepackage{upgreek}
\usepackage{listings}
\newcommand{\Msun}{\mathrm{M}_{\odot}}

\newcommand{\cmark}{\ding{51}} 
\newcommand{\xmark}{\ding{55}} 
\usepackage{pifont} 

\begin{document}

\title{How do Massive Primordial Black Holes Impact the Formation of the First Stars and Galaxies?}

\author[0000-0003-1541-177X]{Saiyang Zhang\begin{CJK*}{UTF8}{bsmi}(張賽暘）\end{CJK*}}

\affiliation{Department of Physics, University of Texas at Austin, Austin, TX 78712, USA}

\affiliation{Weinberg Institute for Theoretical Physics, Texas Center for Cosmology and Astroparticle Physics, \\ University of Texas at Austin, Austin, TX 78712, USA}


\author[0000-0002-4966-7450]{Boyuan Liu\begin{CJK*}{UTF8}{bsmi}(劉博遠)\end{CJK*}}
\affiliation{Department of Astronomy, University of Texas at Austin, Austin, TX 78712, USA}
\affiliation{Institute of Astronomy, University of Cambridge, Madingley Road, Cambridge, CB3 0HA, UK}
\affiliation{Universit\"at Heidelberg, Zentrum fur Astronomie, Institut f\"ur Theoretische Astrophysik, D-69120 Heidelberg, Germany}

\author[0000-0003-0212-2979]{Volker Bromm}
\affiliation{Weinberg Institute for Theoretical Physics, Texas Center for Cosmology and Astroparticle Physics, \\ University of Texas at Austin, Austin, TX 78712, USA}
\affiliation{Department of Astronomy, University of Texas at Austin, Austin, TX 78712, USA}

\author[0000-0002-6038-5016]{Junehyoung Jeon}
\affiliation{Department of Astronomy, University of Texas at Austin, Austin, TX 78712, USA}

\author[0000-0002-9604-343X]{Michael Boylan-Kolchin}
\affiliation{Weinberg Institute for Theoretical Physics, Texas Center for Cosmology and Astroparticle Physics, \\ University of Texas at Austin, Austin, TX 78712, USA}
\affiliation{Department of Astronomy, University of Texas at Austin, Austin, TX 78712, USA}

\author[0000-0002-1528-1920]{Florian K\"uhnel}
\affiliation{Max-Planck-Institut f{\"u}r Physik, Boltzmannstr.~8, 85748 Garching, Germany}
\affiliation{Arnold Sommerfeld Center, Ludwig-Maximilians-Universit{\"a}t, Theresienstr.~37, 80333 M{\"u}nchen, Germany}



\begin{abstract}
We investigate the impact of massive primordial black holes (PBHs; $m_{\rm BH}\sim 10^6~\Msun$) on the star formation and first galaxy assembly process using high-resolution hydrodynamical simulations from $z = 1100$ to $z \sim 9$. We find that PBH accretion is self-regulated by feedback, suppressing mass growth unless feedback is weak. PBHs accelerate structure formation by seeding dark matter halos and gravitationally attracting gas, but strong feedback can delay cooling and suppress star formation. In addition, the presence of baryon–dark matter streaming creates an offset between the PBH location and the peaks induced in gas density, promoting earlier and more efficient star formation compared to standard $\Lambda$CDM. By $z \sim 10$, PBH-seeded galaxies form dense star clusters, with PBH-to-stellar mass ratios comparable to observed high-$z$ AGN like UHZ-1. Our results support PBHs as viable SMBH seeds but do not exclude alternative scenarios. We emphasize that PBH-seeding provides a natural explanation for some of the newly-discovered overmassive SMBHs at high redshift, in particular those with extreme ratios of BH-to-dynamical (virial) mass that challenge standard formation channels. Future studies with ultra-deep JWST surveys, the Roman Space Telescope, and radio surveys with facilities such as SKA and HERA will be critical in distinguishing PBH-driven SMBH growth from other pathways. 

\end{abstract}




\section{Introduction} \label{sec:intro}

The formation of the first, so-called Population~III (Pop~III), stars and the origin of supermassive black holes (SMBHs) in the early Universe remain among the most compelling mysteries in cosmology \citep[for general reviews, see e.g.,][]{Bromm2013, Smith2019:BHreview, Woods2019, Inayoshi:2020, Klessen:2023FirstStars}. In the standard $\Lambda$CDM cosmology,  
Pop~III star formation is predicted to start at redshifts $z \sim 20-30$, marking the end of the cosmic dark ages \citep[e.g.,][]{Barkana2001:FirstStar,Abel2002, Bromm2002ApJ...564...23B, Yoshida2008:ProtoStar}. 
However, the formation of the first stars and black holes depends sensitively on the physical nature of dark matter (DM). The corresponding differences in the power spectrum of density fluctuations on small scales and non-gravitational interactions between DM and baryons can significantly impact the environments where the first sources of light emerge \citep[e.g.,][]{Dayal2017,Hirano2018,Sullivan2018,Liu2019,Liu2019bdms,Liu2020,Kulkarni2022,Qin2024}. Their observational signatures can thus act as effective probes of the underlying dark matter physics.

The unprecedented sensitivity of the James Webb Space Telescope (JWST) has allowed us to explore the early evolutionary stages of the Universe \citep{Gardner:2006JWST,Robertson2022}. Recent JWST observations have discovered a surprising population of UV-bright galaxies with stellar masses in excess of $10^9\,\Msun$ at redshifts $z\gtrsim 10$ \citep[e.g.,][]{Maisies:2022,GHz2,GLASSz13,Castellano2023ApJ...948L..14C, Labbe2023Natur.616..266L, Adams2024, Donnan2024}{}{}, many of them confirmed with follow-up spectroscopy \citep[e.g.,][]{Arrabal2023ApJ...951L..22A, Carniani2024}. 

The emerging JWST first-galaxy phenomenology is providing a powerful stress test for standard $\Lambda$CDM cosmology \citep[e.g.,][]{Inayoshi2022ApJ...938L..10I, Boylan2023}. To alleviate the tension and account for this rapid galaxy assembly and the highly efficient star formation within a short cosmic time ($\sim 300 ~\rm Myr$), multiple solutions have been proposed, including starburst regimes that are relatively free from supernova and other stellar feedback \citep[see e.g.,][]{Dekel:2023FFB,Shen2023, boylan2024} and boosted light-to-mass ratios from a top-heavy initial mass function or peculiar stellar evolution \citep[e.g.,][]{Trinca2024, Wang2024ApJ...963...74W,Liu2024che}. Additionally, modifications to the underlying cosmology have also been suggested, such as accelerated structure formation by invoking a blue tilt in the primordial power spectrum \citep[e.g.,][]{Padmanbha2023,Hirano2024ApJ...963....2H} or early dark energy \citep[e.g.,][]{Shen2024}. In extreme cases, such scenarios could trigger first star formation as early as $z \gtrsim 100$ \citep{Hirano2015ApJ...814...18H, Ito2024PASJ...76..850I}.
 
Furthermore, JWST has revealed an abundant population of SMBHs at $z\gtrsim 8$  \citep[e.g.,][]{Goulding2023ApJ...955L..24G, Larson_2023_BH, Bogdan:2023UHZ1,Greene2024,Kovacs2024ApJ...965L..21K, Maiolino2024A&A,Natarajan:2023UHZ1,GHZ9Napolitano2024arXiv241018763N}. Among them is the extremely distant SMBH discovered in the well-constrained GN-z11 system \citep{Tacchella_2023,Maiolino2024}. 
At slightly lower redshifts ($z=4-9$) lie a previously unknown class of objects, ultra-compact and dust-obscured sources often referred to as ``little red dots'' (LRDs) that may represent an early evolutionary phase of black hole growth. LRDs therefore may provide unique insight into the formation and evolution of early cosmic structures \citep[e.g.,][]{Labbe2023Natur.616..266L, KokorvLRD2024ApJ...968...38K, KocevskiLRD2024arXiv240403576K, Leung2024:LRDarXiv,Taylor2024,LabbeLRD2025ApJ...978...92L}. 
Intriguingly, most of the newly-discovered high-$z$ SMBHs exhibit ratios of their inferred mass to the host stellar mass exceeding the local relation \citep[e.g.,][]{PacucciLRD2023ApJ...957L...3P,InayoshiLRD2024ApJ...973L..49I}. This suggests that the conditions under which these early structures formed were significantly different from those of galaxies in the present-day Universe.  

The existence of massive SMBHs at high redshifts raises questions of the seeding mechanism and co-evolution with their host galaxies~\citep[e.g.,][]{ Regan2024,Silk2024ApJ...961L..39S}. Light-seed models, typically linked to the remnants of Pop~III stars~\citep[for reviews, see][]{Smith2019:BHreview, Inayoshi:2020}{}{}, struggle to account for the existence of these SMBHs at such early times due to the limited growth period and the challenges associated with maintaining accretion rates at, or in excess of, the Eddington limit~\citep[e.g.,][]{Milosavljevic2009,Jeon2023, Bogdan:2023UHZ1}{}{}. Consequently, alternative scenarios have been proposed to explain the rapid emergence of these early SMBHs. 

One such scenario is the formation of direct-collapse black holes (DCBHs), which could form from the (nearly) free-fall collapse of massive primordial gas clouds in metal-poor environments with little to no fragmentation~\citep{Loeb1994ApJ...432...52L,BrommDCBH2003ApJ...596...34B, Begelman2006:DCBH, Lodato:2006DCBH}. These DCBHs can have initial masses of around $10^4-10^6\,\Msun$, providing a head start for the growth into SMBHs by $z \sim 7-10$~\citep{Jeon2023, Jeon2025ApJ...979..127J}. However, the formation of DCBHs is thought to be facilitated by special conditions, such as the delayed onset of star formation due to a strong flux of soft-UV, Lyman-Werner (LW) radiation, suppressing molecular hydrogen cooling \citep[e.g.,][]{AgarwalDCBH2014MNRAS.443..648A,Habouzit2016,OBrennan2025}.

Another promising scenario invokes massive primordial black holes (PBHs) as seeds for early SMBHs~\citep[for analytical work see, e.g.,][]{DuchtingPBH2004PhRvD..70f4015D, Bernal:2017nec, DeLuca2023PhRvL.130q1401D, Dayal2024A&A...690A.182D}. PBHs as a well-motivated dark matter candidate~\citep[reviewed in][]{Carr2020ARNPS..70..355C}, theorized to form from the collapse of overdensities shortly after the Big Bang \citep{Zeldovich1967SvA....10..602Z, hawking1971gravitationally,Carr1975ApJ...201....1C, Belotsky2019, Escriva2022}, present a plausible mechanism for addressing some of the challenges posed by recent JWST observations \citep[for a review of various PBH formation scenarios, see][]{ESCRIVA2024261}. Unlike stellar remnants or DCBHs, PBHs do not require a pre-existing star-forming environment, making them an attractive candidate for explaining the presence of SMBHs at very high redshifts \citep{Yuan2024SCPMA..6709512Y, Huang2024arXiv241005891H, Ziparo2025JCAP...04..040Z}.

Even if PBHs were to constitute only a small fraction of the dark matter, they could significantly influence cosmic history by generating a rich phenomenology~\citep[for a summary of the constraints, see][]{Carr2021}. Moreover, there is now an increasing set of observations which might be allocated to PBHs, such as various microlensing events \citep[e.g.,][]{2019PhRvD..99h3503N, 2017ApJ...836L..18M, 2020A&A...636A..20W, 2022MNRAS.512.5706H, 2024MNRAS.527.2393H}, correlations in the cosmic x-ray and infrared background fluctuations \citep{2005Natur.438...45K}, absence of certain ultra-faint dwarf galaxies \citep[e.g.,][]{2018PDU....22..137C}, several LIGO/Virgo/KAGRA subsolar black hole merger candidates \citep{2021arXiv210511449P, 2023MNRAS.526.6234L, 2023PDU....4201285M}, or select supernov{\ae} (Calcium-rich gap-transients) which do not trace the stellar population, but could well be explained by PBH-triggered white dwarf explosions \citep{2024PhRvL.132o1401S}. These and further hints for PBHs are discussed in \citet{2024PhR..1054....1C}.

On large scales, PBHs contribute to the growth of density perturbations through the Poisson effect, which accelerates early structure formation \citep{Meszaros1975A&A....38....5M, Afshordi2003ApJ...594L..71A, Kashlinsky2021PhRvL.126a1101K, Cappelluti2022ApJ}. On smaller scales, individual PBHs can act as seeds for dark matter halos (i.e., the seed effect), facilitating the formation and growth of early massive galaxies and SMBHs \citep[e.g.,][]{Mack2007ApJ, Ricotti2007ApJI, Ricotti2008ApJII, Carr2018MNRAS.478.3756C, Inman2019PhRvD.100h3528I, Boyuan2022ApJ,Boyuan2023arXiv231204085L,Su2023,Colazo2024, Zhang:2024PBH, Matteri:2025klg}. By accreting surrounding gas and radiating away the resulting heat energy, PBHs can also influence the thermodynamic properties of their environment \citep[e.g.,][]{Ali-Haimoud2017PhRvD, Deluca2020JCAP...06..044D, Lu2021, Takhistov2022, Ziparo2022, Zhang2024MNRAS.528..180Z, Casanueva-Villarreal2024}, potentially altering the processes of first star formation \citep{Boyuan2022MNRAS.514.2376L,Boyuan2023arXiv231204085L}. 

However, much of the previous PBH research relied heavily on analytical models or dark-matter-only simulations, which do not capture the complex interactions between PBHs, surrounding gas, and the larger-scale environment. Black hole feedback ---- where radiation and winds from accreting black holes influence their surroundings ---- can alter the thermodynamic state of the interstellar medium \citep[e.g.,][]{Lu2021, Takhistov2022, Casanueva-Villarreal2024}. This feedback, if sufficiently strong, may even suppress the collapse of gas into stars\footnote{One caveat is that, since PBHs are also able to accelerate structure formation, more minihalos would form earlier. When considering both effects together, there may be no significant change to the overall Pop~III star formation rate \citep{Boyuan2022MNRAS.514.2376L}.}, thereby delaying star formation, as previously explored in simulations with stellar mass ($10-100\,\Msun$) PBHs \citep{Boyuan2022MNRAS.514.2376L}. Moreover, analytical models are inherently limited in their ability to simulate the three-dimensional complexity of accretion disks around black holes and the subsequent feedback processes. Such limitations restrict our understanding of the non-linear, multi-scale interactions that are crucial for accurately modeling the evolution of early black hole seeds and the first galaxies, thus emphasizing the need for hydrodynamical simulations.

In this paper, we simulate the changes to the formation history of first generation stars and to the assembly of the first galaxies, taking into account the enhancement of initial density perturbations by an isolated massive PBH ($\sim 10^6\,\Msun$). In contrast to \cite{Boyuan2022MNRAS.514.2376L}, we simulate the BH accretion and feedback starting from the baryon-photon decoupling epoch at $z$=1100. This PBH mass scale is motivated by a cosmic thermal history that involves a change in the equation of state during $e^+e^-$ annihilation, when the temperature of the Universe is about $\sim 1\rm \ MeV$ \citep{Carr2021PDU....3100755C}. We use a two-step approach that begins with a particle dark matter (PDM)-only simulation to set the initial conditions for a subsequent hydrodynamical simulation. In the hydrodynamical simulations, we specifically model the effect of BH-accretion feedback on the primordial chemistry, as well as the thermal properties of gas clouds within the overdense structure seeded by the PBH. We analyze the resulting black hole growth rate for different feedback efficiencies, the timing of first star formation, and the assembly history of a select first galaxy and compare to an analytical model, a standard $\Lambda$CDM simulation, and recent JWST observations.


In Section~\ref{sec:Method}, we describe the detailed numerical setup and initial conditions for our simulations, the black hole accretion and feedback models, and our recipe to insert sink particles for gas undergoing runaway collapse. When presenting our simulation results in Section~\ref{sec:Results}, we focus on the behavior of primordial gas, the impact of black hole feedback, and the PBH-galaxy coevolution. After addressing the limitations of our work and potential caveats in Section~\ref{sec:caveat}, we summarize our findings and discuss avenues for future research in Section~\ref{sec:Summary}.

\section{Methodology} \label{sec:Method}

Our cosmological hydrodynamic simulations are carried out using the \textsc{gizmo} code \citep{Hopkins2015MNRAS.450...53H}, which utilizes the Lagrangian meshless finite-mass (MFM) hydro solver (with a kernel size of $N_{\mathrm{ngb}}=32$ nearest neighbors), together with the parallelization scheme and Tree+PM gravity solver from \textsc{gadget-3} \citep{springel2005cosmological}. The hydro and gravity solvers are combined with a non-equilibrium primordial chemistry and cooling network\footnote{In addition, LiH is in principle an efficient coolant, but it is not abundant enough to affect the evolution of primordial gas \citep{LiuBromm2018}.} for 12 species ($\rm H$, $\rm H^{+}$, $\rm H^{-}$, $\rm H_{2}$, $\rm H_{2}^{+}$, $\rm He$, $\rm He^{+}$, $\rm He^{2+}$, $\rm D$, $\rm D^{+}$, $\rm HD$, $\rm e^{-}$), following \citet{Bromm2002ApJ...564...23B} and \citet{Johnson2006MNRAS.366..247J}. In all simulations, we adopt \textit{Planck18} cosmological parameters  \citep{Plank2020A&A...641A...6P}: $\Omega_{\rm m} = 0.3111$, $\Omega_{\rm b}=0.04897$, $h = 0.6776$, $\sigma_8 = 0.8102$, $n_{\rm s} = 0.9665$.

In this work, we focus on star formation and galaxy assembly around a single, isolated PBH at high redshift. To resolve the initial nonlinear structure around this PBH, we first run a pathfinder PDM-only simulation to establish the cosmological initial conditions for subsequent simulations including the effect of baryons, as described in detail in Section~\ref{subsec:PathFinder}. At such high redshifts, extending to $z\gtrsim 100$, the presence of CMB photons will become important in suppressing the formation of $\mathrm{H}_2$ molecules. In addition, we consider the effect of LW photons from the accretion feedback, also destroying $\mathrm{H}_2$ molecules and thus changing the gas cooling process at $z\lesssim 100$. Therefore, we discuss the impact of CMB and LW photons on the primordial chemistry network in Section~\ref{subsec:Chem}. We introduce our modeling of BH accretion and feedback in Section~\ref{subsec:BHParam}, and discuss our treatment of Jeans-unstable gas, leading to the insertion of sink particles that represent stars, in Section~\ref{subsec:Jeans}. For the convenience of the reader, we summarize the key parameters and settings for each simulation run in Table~\ref{Table:SimParam}.

\begin{table*}[ht!]
    \centering
    \caption{Summary of key simulation parameters and results. $L$ is the length of the simulation box in comoving kpc. $z_{\rm ini}$ is the initial redshift where the simulation starts, and  $z_{*}$ the redshift when stars begin to form around the central PBH. $N_{\rm eff}$ is the total number of particles within the simulation box. $\epsilon_r$ is the thermal feedback coupling efficiency. $\Delta m_{\rm BH}$ is the total mass accreted by the central black hole by the end of the simulation (note that some runs terminate at $z > 9$), and $m_{\star}$ is the total stellar mass formed within the PBH-hosting halo, or within the PDM halo in the case of the \texttt{CDM} run, by $z = 9$.  
    STR is a flag indicating whether the relative streaming of PDM and gas particles is included (\cmark) or not (\xmark), and the value represents the amplitude with respect to the root-mean-square streaming velocity $\sigma_{\rm b\chi}$. BH\_LW is another flag to control whether we include (\cmark) the local LW feedback from BH accretion or not (\xmark).  }
    \begin{tabular}{cccccccccc}
    \hline
        Run & $L$\,[ckpc] &$z_{\rm ini}$& $z_{*}$  & $N_{\rm eff}$  & $\epsilon_{\rm r}$& $\Delta m_{\rm BH} [\Msun]$& $m_{\star} [\Msun]$   & STR & BH\_LW\\
    \hline
        \texttt{PDMonly}  & 250 & 3400 & - & $ 256^3$  & - & - & -  & - & -  \\
        \texttt{CDM} & 250 & 1100 & 14.9 & $2\times 256^3$   &- & -& $1.61\times 10^6 $ & \xmark & -  \\
        \texttt{CDM\_str} & 250 & 1100 & 13.7 & $2\times 256^3$   &- & -& - \footnote{Due to the limitation of the computation resource and storage space, we stop the \texttt{CDM\_str} and \texttt{CDM\_sstr} simulations once the first star formation takes place in each case.} & 0.8 \cmark & -  \\
        \texttt{CDM\_sstr} & 250 & 1100 & 11.2 & $2\times 256^3$   &- & -& - & 1.6 \cmark & -  \\
        \hline
        \texttt{PBH\_fd05} & 250 & 1100 & - &  $2\times 256^3$   &0.05 &$7.06\times 10^3 $& -  & \xmark & \xmark \\
        \texttt{PBH\_fd005} & 250 & 1100 & 29.6 &  $2\times 256^3$   &0.005 &$4.62\times 10^4 $& $2.06\times 10^6$  & \xmark & \xmark \\
        \texttt{PBH\_fd0005} & 250 & 1100 & 32.2 &  $2\times 256^3$   &0.0005 &$4.58\times 10^5 $& $5.63 \times 10^6$   & \xmark & \xmark \\
        \texttt{PBH\_str\_fd005} & 250 & 1100 & 70.6 &  $2\times 256^3$   &0.005& $7.32\times 10^4$& $1.95\times 10^6$  & 0.8 \cmark  & \xmark\\
        \texttt{PBH\_sstr\_fd005} & 250 & 1100 & 113.1 &  $2\times 256^3$   &0.005& $3.91 \times 10^4$& $4.56\times 10^6$  & 1.6 \cmark  & \xmark \\
        \hline
        \texttt{PBH\_LW\_fd05} & 250 & 1100 & - &  $2\times 256^3$  &  0.05& $1.32\times 10^4$& -  & \xmark & \cmark  \\
        \texttt{PBH\_LW\_fd005} & 250 & 1100 & - &  $2\times 256^3$   &0.005 &$7.84\times 10^4 $ & - & \xmark & \cmark \\
        \texttt{PBH\_nfeed} & 250 & 1100 & - &  $2\times 256^3$ & 0 & $1.10\times 10^7$& -  & \xmark & \xmark   \\
 \hline 

    \hline
    \end{tabular}
    \label{Table:SimParam}
\end{table*}

\subsection{Pathfinder Simulation and Initial Conditions} \label{subsec:PathFinder}

We start with a PDM-only pathfinder simulation, denoted as \texttt{PDMonly}, using the initial conditions generated by the \textsc{MUSIC} code \citep{hahn2011multi} at $z_{\rm eq}=3400$, which marks the beginning of the matter-dominated epoch and the growth of structure, placing a single PBH at the center of the simulation box. We use $m_{\rm BH}$ to represent the PBH mass, initially set to $10^6\,\Msun$ throughout all simulation runs that contain the PBH\footnote{In this work, we effectively assume that the mass fraction of PBHs in the PDM component is less than $6\times10^{-4}$, representing an upper limit set by the ratio of the PBH mass to the total mass enclosed within the simulation volume (calculated for the average cosmological matter density). This value is consistent with existing constraints from large-scale structure formation~\citep{Carr2018MNRAS.478.3756C} and dynamical friction effects~\citep{Carr1999ApJ...516..195C}. It also agrees with simulated results from previous work, exploring varying PBH density fractions \citep[e.g.,][]{Zhang:2024PBH,Casanueva2025}.}. We assign a box side length of $L \sim 250\ \rm ckpc$ throughout, with $N_{\rm PDM} =256^3$ setting the resolution of PDM particles. To add perturbations and calculate the velocities of PDM particles surrounding the PBH, we adopt the numerical recipe from previous work \citep{Ali-Haimoud2017PhRvD,Inman2019, Boyuan2023arXiv231204085L, Zhang:2024PBH}, employing the Zel'dovich approximation~\citep{Zeldovich1970A&A.....5...84Z}. 


We first run this PDM-only simulation to capture the perturbations induced around the single PBH all the way to the baryon-photon decoupling redshift at $z = 1100$. Due to the effect of Silk damping at $z \gtrsim 1100$, we assume that the baryons are initially not perturbed at the scales represented within our box. At $z = 1100$, 
we include a uniform baryonic matter field with the same spatial resolution as for PDM, resulting in a total of $N=2\times 256^3$ particles. In this new initial condition set-up, the mass of the PDM (gas) particles becomes $\sim100$ ($18$)~$\Msun$. 
In this way, we approximately represent the initial PDM perturbations around a PBH, and subsequently let gas particles collapse onto the PBH-seeded halo.

In running these simulations, we set the softening length of PDM and gas to $\epsilon_{\rm PDM} = \epsilon_{\rm gas} \sim 0.01\, L/N_{\rm PDM}^{1/3} \simeq 0.01~ h^{-1}\rm kpc$. In addition, we assign initial abundances for the 12 chemical species to values predicted for the IGM at $z=1100$, as obtained in \citet[see their figs.~1 and 2]{Galli2013ARA&A..51..163G}. For some of the simulations, we include a universal velocity offset between the gas and PDM particles to account for the effect of PDM-gas streaming motions. Previous work studied the effect of streaming on structure formation under large-scale perturbations (i.e., the Poisson effect) of stellar-mass PBHs ($m_{\rm BH} \sim 10-100\,\Msun$), finding a weaker impact for the PBH regime compared to the \texttt{CDM} case~\citep{Kashlinsky2021PhRvL.126a1101K,Atrio2022, Boyuan2022MNRAS.514.2376L}. However, since the streaming effect has not been previously investigated for the seed effect of massive PBHs, we will consider this velocity offset $v_{\rm b\chi}$ by varying the multipliers of the root-mean-square streaming velocity\footnote{We first take a typical value of $v_{\rm b\chi} = 0.8~\sigma_{\rm b\chi}$, corresponding to the peak of the velocity distribution, following previous studies by \cite{Schauer2019MNRAS.484.3510S}, as the \texttt{PBH\_str\_fd005} run. Then, we separately explore the case for a larger offset of $1.6~ \sigma_{\rm b\chi}$ to assess the difference from the standard case, labeled as \texttt{PBH\_sstr\_fd005}.} $\sigma_{\rm b\chi} = 30 ~\rm km/s$ at the recombination redshift $z = 1100$.

\subsection{Photochemistry of CMB and LW Photons} \label{subsec:Chem}

\begin{table*}[ht!]
\centering

\caption{Summary of chemical reaction rates involving CMB and LW photons implemented in the GIZMO code. In the first block of the table, the couplings of CMB photons to $\rm H^{-}$, $\rm H_{2}$, $\rm H_{2}^{+}$ are expressed with fitting formulae taken from \cite{Galli1998A&A...335..403G} and \cite{CC2011ApJS..193....7C}. Here, $f_{\text{nth}}$ is a numerical factor tuning the reaction rate for H$^-$ and non-thermal photons, with a value of $f_{\text{nth}} = 0.1$, to reproduce the evolution of IGM chemical abundances in \cite{Galli2013ARA&A..51..163G}. On the other hand, the interaction rates of LW photons with $\rm H_2$ and $\rm H^-$ are given in \cite{DB1996ApJ...468..269D,Abel1997NewA....2..181A}, with modification factors $\alpha_{\text{kde}}$ and $\beta_{\text{kdi}}$ to account for the shape of the input spectra \citep{Agarwal2015MNRAS.446..160A}. All rates are in units of $\rm s^{-1}$. The LW intensity $J_{\rm LW}$ is written in units of $J_{\rm 21}=10^{-21}\ \rm erg\ s^{-1}\ cm^{-2}\ sr^{-1}\ Hz^{-1}$.}
\begin{tabular}{cc}
\hline
\textbf{Reaction} & \textbf{Reaction Rate (s$^{-1}$)} \\ \hline
H$^-$ + $\gamma_{\text{cmb}}$ = H + e$^-$ & $k_{30} = 1.1 \times 10^{-1} \cdot T_{\text{cmb}}^{2.13} \cdot \exp{\left(\frac{-8823.0}{T_{\text{cmb}}}\right)} + 8 \times 10^{-8} \cdot T_{\text{cmb}}^{1.3} \cdot \exp{\left(\frac{-2300.0}{T_{\text{cmb}}}\right)} \cdot f_{\text{nth}}$ \\ 
H$_2^+$ + $\gamma_{\text{cmb}}$ = H + H$^+$ & $k_{31} = \left(20 \cdot T_{\text{cmb}}^{1.59} \cdot \exp{\left(\frac{-82000.0}{T_{\text{cmb}}}\right)} + 1.63 \times 10^7 \cdot \exp{\left(\frac{-32400.0}{T_{\text{cmb}}}\right)}\right) \cdot 0.5$ \\ 
H$_2^+$ + $\gamma_{\text{cmb}}$ = 2H$^+$ + e$^-$ & $k_{32} = 90.0 \cdot T_{\text{cmb}}^{1.48} \cdot \exp{\left(\frac{-335000.0}{T_{\text{cmb}}}\right)}$ \\ 
H$_2$ + $\gamma_{\text{cmb}}$ = H$_2^+$ + e$^-$ & $k_{33} = 2.9 \times 10^2 \cdot T_{\text{cmb}}^{1.56} \cdot \exp{\left(\frac{-178500.0}{T_{\text{cmb}}}\right)}$\\ \hline 

$\text{H}^- + \gamma_{\text{LW}} \rightarrow \text{H} + e^-$ & $k_{24} = 1.1 \times 10^{-10} \times (J_{\text{LW}}/J_{21}) \times \alpha_{\text{kde}} \times f_{\text{shield}}$ \\ 

 $\text{H}_2 + \gamma_{\text{LW}} \rightarrow \text{H} + \text{H}$  & $k_{27} = 1.38 \times 10^{-12}\times (J_{\text{LW}}/J_{21}) \times \beta_{\text{kdi}} \times f_{\text{shield}} $ \footnote{These factors are taken to be $\alpha_{\text{kde}} = 1.71$ and $\beta_{\text{kdi}} = 0.97$, obtained by integrating the shape of the SMBH accretion spectrum with the reaction cross section, following \cite{Sugimura2014MNRAS.445..544S, Takhistov2022}.} \\
\hline    

\hline
\end{tabular}

\label{tab:reaction}
\end{table*}

 Previous semi-analytical work \citep{Ito2024PASJ...76..850I} has considered the effect of CMB photons on the process of first star formation within a halo that virialized at an extremely high redshift of $z \gtrsim 100$. Their presence suppresses the abundance of $\rm H_{2}$ molecules and other chemical species \citep[see the summary in][]{Galli2013ARA&A..51..163G} enough that the atomic cooling channel will initially dominate. 
 Since our simulations are initialized at similarly high redshifts, the impact of CMB photons on shaping the primordial chemical abundances and gas cooling channels has to be taken into account. Specifically, in the \textsc{GIZMO} code, rates for the reactions of select chemical species, primarily $\rm H^{-}$, $\rm H_{2}$, $\rm H_{2}^{+}$, with CMB photons have been implemented using the fitting formulae in \cite{Galli1998A&A...335..403G} and \cite{CC2011ApJS..193....7C}. 

Furthermore, with PBHs being placed into the primordial gas, their accretion feedback will change the properties of the primordial gas by injecting additional radiation fields into their surroundings \citep[as discussed in e.g.][]{Boyuan2022MNRAS.514.2376L, Ziparo2022,Boyuan2023arXiv231204085L, Zhang2024MNRAS.528..180Z}. Of particular importance is the presence of a LW background, as it will interact with $\rm H_{2}$ through photo-dissociation and with $\rm H^-$ through photo-detachment, further suppressing their abundances.
As we will discuss in Section~\ref{sec:Results}, the inclusion of this radiation background will affect the gas cooling efficiency and the timing of the initial collapse, especially at $z\lesssim 100$. Here, we consider local LW radiation from BH accretion (see Section~\ref{subsubsec:bhacret}), where the reaction rate will be proportional to the product of the LW radiation intensity, $J_{\text{LW}}$, with the local gas shielding factor $f_{\text{shield}}$. The former term is related to the magnitude of the accretion rate $\dot{m}_{\rm PBH}$, whereas the local shielding factor $f_{\text{shield}}$ depends on the $\rm H_{2}$ and $\rm H^-$ relative abundances, as well as the local gas density and temperature, with the fitting formulae given in \cite{DB1996ApJ...468..269D,Wolcott2011MNRAS.418..838W}. In Table~\ref{tab:reaction}, we summarize the relevant chemical reaction processes with CMB and LW photons, as implemented in our simulation.

\subsection{Black Hole Physics}
\label{subsec:BHParam}

Due to the numerical limitations of the TreePM gravity solver \citep[][]{springel2005cosmological}, we do not treat the PBH as a point object. We instead apply a (physical) softening length to the BH particle, but using a value much smaller than the average gas/PDM softening length, specifically $\epsilon_{\rm PBH,phy} \sim 0.0004~h^{-1}\rm kpc$. Additionally, the force softening kernel for gas particles around the BH has a fixed size of $\epsilon_{\rm g,PBH} = 2.8 \epsilon_{\rm PBH,phy} \sim 0.0016~\rm kpc$. In our simulations, the mass of the BH is much larger than the average mass of the gas particles, $m_{\rm BH}/m_{\rm gas} \gtrsim 5 \times 10^4 \gg 1$, and also of PDM particles, with $m_{\rm BH}/m_{\rm PDM} \gtrsim 10^4 \gg 1$. Therefore, the dynamical friction at small scales is automatically accounted for in our simulations.

We also model the effect of accretion and its feedback on gas surrounding the central PBH, as it could play an important role at high redshifts. In Sections~\ref{subsubsec:bhacret} and \ref{subsubsec:acretfdbk}, we describe the subgrid model for BH accretion \citep{springel2005cosmological, tremmel2015off, tremmel2017romulus}, and the implementation of the resulting thermal and LW feedback \citep{Takhistov2022,Boyuan2022MNRAS.514.2376L}.


\subsubsection{Black Hole Accretion}\label{subsubsec:bhacret}

In the early Universe where the density is almost uniform, we use a Bondi-Hoyle formalism to calculate the spherical accretion rate of the BH, $\dot{m}_{\rm acc}$, given as 
\begin{align}
	\dot{m}_{\mathrm{acc}} & = \frac{4\uppi (G m_{\mathrm{BH}})^{2} \rho_{\mathrm{gas}}}{\tilde{v}^{3}} = \frac{4\uppi (G m_{\mathrm{BH}})^{2} \rho_{\mathrm{gas}}}{(c_{s}^{2} + v_{\mathrm{gas}}^{2})^{3/2}} \notag \\
& \simeq 0.0072~\Msun \mathrm{yr}^{-1} \left( \frac{10~\rm km/s}{\tilde{v}} \right)^{3}\notag\\
&\times \left( \frac{n_{\rm H}}{1 \ \rm cm^{-3}} \right) \left(\frac{m_{\mathrm{BH}}}{10^6~\Msun}\right)^{2}, \label{eq:bondi}
\end{align}
where $\rho_{\rm gas} \simeq \mu m_{\rm H} n_{\rm H}$ is the mean density of the gas within the BH accretion kernel, and $\mu = 1.22$ is the mean molecular weight of the primordial gas. $c_{s}$ is the sound speed of the surrounding gas calculated from the mass-weighted average temperature, and $v_{\mathrm{gas}}$ is the velocity dispersion of the gas around the BH, also sampled from the gas particles within the BH accretion kernel. Here, the BH accretion kernel is defined as the region in which gas properties determine the BH accretion rate. We set the BH accretion kernel size to the Bondi radius $r_{\rm Bondi} \sim 2Gm_{\rm BH}/c_s^2$ (using the $m_{\rm BH}$ and $c_s$ values from the last timestep).

Once we calculate the accretion rate, the BH mass is updated after each time step $\delta t$ by $\delta m_{\mathrm{BH}} = \dot{m}_{\mathrm{acc}} \delta t$. However, the surrounding gas mass does not decrease smoothly. We adopt the algorithm from \cite{springel2005cosmological}, where the BH particle stochastically swallows surrounding gas particles to ensure consistency with the average accretion rate, as calculated, and to enforce mass conservation. In addition, following \cite{springel2005cosmological}, a drag force is applied when the BH starts to accrete gas to guarantee momentum conservation.

\subsubsection{Accretion Feedback}\label{subsubsec:acretfdbk}

In our simulation, we consider the effects of both LW radiation, which suppresses the formation of $\mathrm{H}_2$ molecules, and thermal feedback in terms of photo-ionization heating. In contrast to the previous work of \cite{Boyuan2022MNRAS.514.2376L}, where BH feedback is only enabled at $z\lesssim 100$, we turn on the BH feedback once the BH starts to accrete gas at $z \lesssim 1100$ to study the impact of feedback on the surrounding primordial gas.

\begin{figure}
\centering
    \includegraphics[width=1\columnwidth]{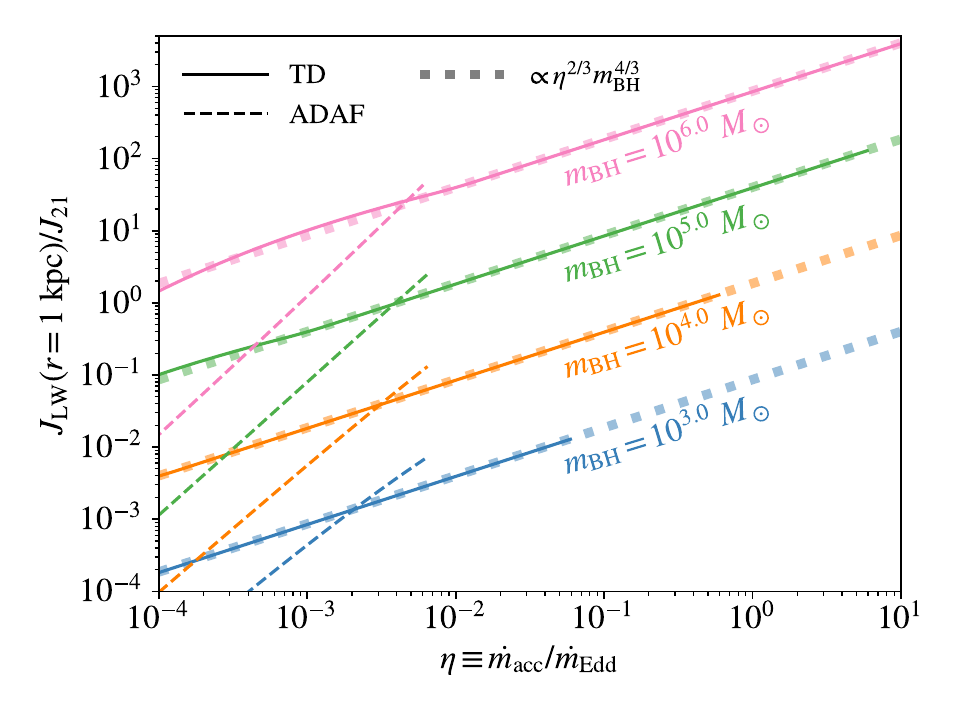}
    \caption{Normalized Lyman–Werner (LW) intensity, $J_{\mathrm{LW}} / J_{21}$, as a function of the Eddington ratio, $\dot{m}_{\mathrm{acc}} / \dot{m}_{\mathrm{Edd}}$, for black holes of varying masses ($m_{\mathrm{BH}} = 10^{3}, 10^{4}, 10^{5}, 10^{6}\,\Msun$). The $J_{\mathrm{LW}}$ values are computed at a distance of $r = 1~\mathrm{kpc}$ from the black hole. The results based on the semi-analytical spectra models for TD and ADAF accretion profiles in \cite{Takhistov2022} are shown by the solid and dashed lines, respectively. Here, the ADAF regime is only expected to occur at low accretion rates ($\eta\lesssim 0.007$).  The TD regime satisfies the scaling relation $J_{\mathrm{LW}} \propto \eta^{2/3} m_{\mathrm{BH}}^{4/3}$ (Eq.~\ref{jlw}; thick dotted lines), which for simplicity is adopted in our simulations as an optimistic estimate of LW radiation from PBHs. 
}
    \label{fig:jlw}
\end{figure}

Considering a combination of radiatively efficient thin-disk (TD) and radiatively inefficient, but geometrically thick, advection-dominated accretion flow (ADAF) models, 
we calculate the spectra of BH accretion flows according to \citet{Takhistov2022}. For simplicity, we use fitting formulae only for the TD model as an optimistic estimate of the LW feedback intensity, and illustrate the difference to the ADAF regime in Fig.~\ref{fig:jlw}. In future improved work, a broken power-law model should be applied to take into account the ADAF regime. We can write the specific intensity (in units of $J_{\rm 21}=10^{-21}\ \rm erg\ s^{-1}\ cm^{-2}\ sr^{-1}\ Hz^{-1}$) of LW radiation ($h\nu\sim 11.2 -13.6$~eV) originating from a BH of mass $m_{\rm BH}$ and accretion rate $\dot{m}_{\rm acc}$ at a distance $r$ as: 
\begin{align}
    \frac{J_{\rm LW}}{J_{21}} \simeq 8.8 \times 10^{2} \eta^{2/3} \left( \frac{m_{\rm BH}}{10^6\ \rm M_\odot} \right)^{4/3} \left( \frac{r}{\rm kpc} \right)^{-2} , \label{jlw}
\end{align}
where $\eta \equiv \dot{m}_{\rm acc}/\dot{m}_{\rm Edd}$ is the Eddington ratio, with the Eddington rate given by
\begin{align}
    \dot{m}_{\mathrm{Edd}} = 0.047\ \Msun\ \rm yr^{-1}\ \left( \frac{m_{\mathrm{BH}}}{10^{6}\ \Msun} \right) \left( \frac{\epsilon_{0}}{0.057} \right)^{-1} . \label{eq:mdot_edd}
\end{align}
Here we adopt $\epsilon_0 \simeq 0.057$ as the radiative efficiency for non-rotating BHs in the thin-disk model.

In the accretion process, we also implement thermal feedback by injecting energy within the Bondi radius surrounding the BH, following the prescription in \cite{springel2005cosmological}. After each time step $\delta t$, the total amount of energy injected is $\delta E = \epsilon_{r} L_{\mathrm{BH}} \delta t$, where $L_{\mathrm{BH}} = \epsilon_{\rm EM} \dot{m}_{\mathrm{acc}} c^{2}$, and $\epsilon_{r}$ is the thermal coupling coefficient. Here, $\epsilon_{\rm EM}$ is the radiation efficiency, calculated using the method in \cite{Negri2017MNRAS.467.3475N} to capture the transition from ADAF to thin-disk model, as follows:
\begin{align}
\epsilon_{\mathrm{EM}}=\frac{\epsilon_{0}A\eta}{1+A\eta}\ ,\label{epsilonEM}
\end{align}
where we take $A =100$.
Due to lack of understanding of the feedback process at ultra-high redshifts, we treat $\epsilon_{r}$ as a free parameter and study how the gas cloud evolves under different choices for this coefficient. As a limiting case, we set this parameter to $0$, thus exploring the collapse of gas onto the PBH-seeded halo without this additional heating. The effect of varying this parameter on first star formation is discussed in Section~\ref{subsec:popIII}.

%

\subsection{Creation of Sink Particles} \label{subsec:Jeans}

Since we have a PBH located at the center of the halo accreting dense gas, we aim to evaluate the feasibility of this central gas clump leading to star formation. Below, we describe the algorithm used to identify collapsing gas clouds near the PBH. Once the densest gas cloud reaches $n_{\rm H} \sim 10^4~\rm cm^{-3}$, we begin to assess the temperature $T$ and average number density $n_{\rm H}$ for each individual gas particle to calculate a local Jeans mass $M_{\rm J}$, expressed as:

\begin{align}
M_{\rm J} \simeq & 10 \left( \frac{1}{4 \pi \rho_{\rm gas}  } \right)^{1/2}\left( \frac{5 k_B T}{3G \mu m_{\rm H}} \right)^{3/2} \,\notag \\ &\simeq 2\times 10^4~\Msun \left( \frac{T}{10^3 \ \rm K} \right)^{3/2} \left( \frac{10^4 \ \rm cm^{-3}}{n_{\rm H}} \right)^{1/2}. \label{eq:Jeans}
\end{align}

\noindent This value $M_{\rm J}$ is then compared to $N_{\rm J} \, m_{\rm gas}$, where $N_{\rm J} = 64$ represents the number of particles required to resolve the collapsing gas cloud.

To determine whether the gas particle collapses or is accreted by the black hole, we calculate the local free-fall time of the gas:

\begin{equation}
t_{\rm ff} = \sqrt{\frac{3 \pi}{32 \, G \rho_{\rm gas} }} \simeq 0.47~{\rm Myr} \left( \frac{10^4 \ \rm cm^{-3}}{n_{\rm H}} \right)^{1/2},
\end{equation}
using the local gas density $\rho_{\rm gas} $. We record this timescale once the Jeans instability criterion is met, i.e., $N_{\rm J} \, m_{\rm gas} \gtrsim M_{\rm J}$, and compare it to the survival time $t_{\rm survive}$ of the collapsing particle. In our simulations, $t_{\rm survive}$ is evaluated by accumulating the time between each time-step during which the gas survives as a collapsing particle, resulting in $\delta t_{\rm survive} = \delta t$. If the gas particle survives the accretion or reheating by black hole thermal feedback (i.e., if $t_{\rm survive} \gtrsim t_{\rm ff}$), we mark it as part of the collapsing cloud, indicating that it would not be accreted by the central PBH. In the process, we convert this particle into a sink stellar particle\footnote{Here we make the simplifying assumption that the star formation or mass conversion efficiency of this collapsing gas cloud into a proto-stellar system is 1. However, a more realistic value is about $\sim 0.5$ \citep{Liu2024MNRAS.534..290L} for low metallicity gas clouds. This change might result in a slightly lower total stellar mass, but without affecting the overall picture.}, but do not assign any stellar population or turn on the stellar feedback, thus assuming that PBH feedback dominates. On the other hand, if the gas particle fails to meet the collapse criterion in subsequent time steps, while still exhibiting $t_{\rm survive} \lesssim t_{\rm ff}$, we reset the timer to $t_{\rm survive} = 0$.

 We further define $N_{\rm col}$ as the cumulative number of collapsing (stellar) particles. 
 With this framework, we can simulate the coevolution of the PBH and stars within a first galaxy all the way until $z\sim 9$. 
  In addition, a simulation snapshot is recorded whenever the total number of collapsing particles doubles, starting from $N_{\rm col} = 1$,  so that the evolution of collapsing gas clouds around the PBH is well-captured in time.

\section{Results and Discussion} \label{sec:Results}

With the methodology described in Section~\ref{sec:Method}, we now analyze the select PBH simulation runs, focusing on the conditions that enable the formation of first-generation stars and the assembly of the first galaxies. This section will explore key results and implications of our findings, structured into three subsections: PBH-growth through accretion and feedback in Section~\ref{subsec:Accret}, the triggering of Population III (Pop III) star formation in Section~\ref{subsec:popIII}, and the co-evolution of PBHs with the stellar systems, also considering observational perspectives, in Section~\ref{subsec:coevo}.

\subsection{PBH Accretion and Feedback} \label{subsec:Accret}

\begin{figure*}[ht!]
\centering
    \includegraphics[width=\linewidth]{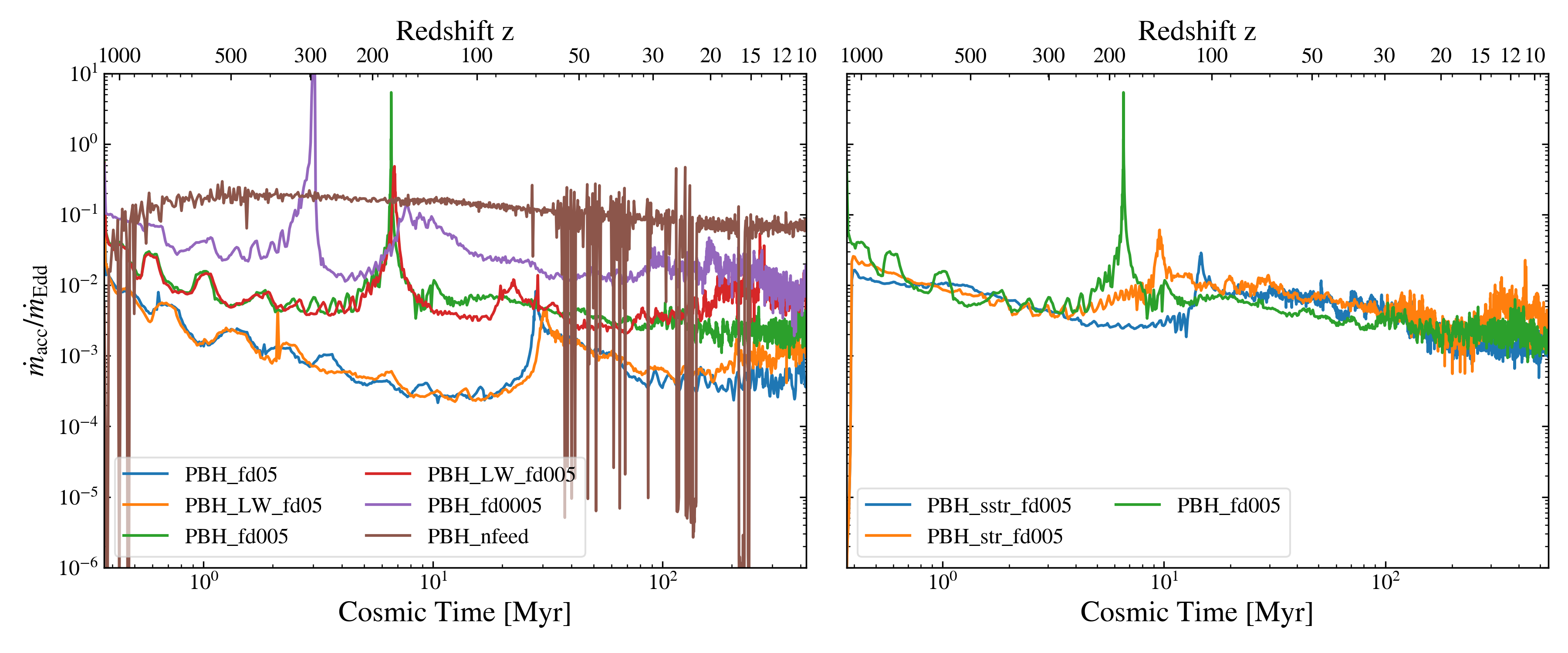}
    \caption{PBH accretion history as a function of cosmic time and redshift, spanning from $z \sim 1100$ to $z \sim 9$. The accretion efficiency is represented by the dimensionless ratio between the physical accretion rate, $\dot{m}_{\mathrm{acc}}$, and the Eddington limit, $\dot{m}_{\mathrm{Edd}}$, calculated using Equ.~(\ref{eq:mdot_edd}). In the \textbf{left panel}, we show results from several simulation runs with varying feedback efficiency factors: $\epsilon_r \sim 0.05$ (\texttt{PBH\_fd05} and \texttt{PBH\_LW\_fd05}), $0.005$ (\texttt{PBH\_fd005} and \texttt{PBH\_LW\_fd005}), $0.0005$ (\texttt{PBH\_fd0005}), and $0$ (\texttt{PBH\_nfeed}). The \texttt{PBH\_LW\_fd05} and \texttt{PBH\_LW\_fd005} simulations include the effect of Lyman–Werner (LW) radiation from PBHs, assuming the same accretion feedback efficiency as the corresponding non-LW runs. In the \textbf{right panel}, we include simulations that assume the same feedback efficiency factor of $\epsilon_r \sim 0.005$ for all runs, but with different initial relative streaming velocities of baryons with respect to dark matter. Specifically, we show cases with $v_{\mathrm{b\chi}} = 0$ (\texttt{PBH\_fd005}), $0.8~\sigma_{\mathrm{b\chi}}$ (\texttt{PBH\_str\_fd005}), and $1.6~\sigma_{\mathrm{b\chi}}$ (\texttt{PBH\_sstr\_fd005}).
}
    \label{fig:AccretHist}
\end{figure*}

In Fig.~\ref{fig:AccretHist}, we show the evolution of the PBH accretion rate over cosmic history for select models\footnote{Due to computational limitations, the \texttt{PBH\_LW\_fd05}, \texttt{PBH\_LW\_fd005}, \texttt{PBH\_fd05}, and \texttt{PBH\_nfeed} simulations terminate at redshifts of $z = 10.29$, $z = 11.48$, $z = 10.56$, and $z = 10.82$, respectively. In these cases, star formation occurs first within minihalos not seeded by a central PBH, and no collapsing particle is identified within PBH-seeded halos.}, extending from $z = 1100$ to $z = 9$.
At $z \gtrsim 100$, when the Universe was still highly isotropic and homogeneous, the gas accretion rate evolves relatively smoothly, as seen in both panels of Fig.~\ref{fig:AccretHist}. However, for $z \lesssim 100$, the accretion rate becomes more stochastic as the Universe becomes clumpier and less isotropic. For massive PBHs (here $10^6\,\Msun$), efficient accretion in dense gas environments produces significant luminosities, which in turn drives strong thermal and radiative feedback, subsequently suppressing the accretion rate. 

This phase of efficient accretion is quickly terminated by shock waves generated from the accretion feedback, which raise the gas temperature near the PBH and inhibit further gas collapse. Notably, when the feedback efficiency is higher, the accretion burst is weaker because stronger shocks are launched more rapidly.  Across all simulation runs, these large variations of the accretion rate are observed, driven by the competition between cold gas inflow from the intergalactic medium and the thermal feedback from BH accretion.

Consequently, the PBH mass growth is negligible ($\Delta m_{\mathrm{BH}} / m_{\mathrm{BH}} \lesssim 0.1$) by $z \sim 10$ in all simulations with feedback efficiency $\epsilon_r \gtrsim 0.005$. Lower feedback efficiencies result in higher accretion rates and more significant black hole growth, as demonstrated in Table~\ref{Table:SimParam} and Fig.~\ref{fig:AccretHist}. In the extreme case of \texttt{PBH\_nfeed}, where thermal feedback is completely neglected, the PBH accretion rate consistently reaches $\sim 10\%$ of the Eddington limit \citep[for a similar case, see e.g.,][]{Regan2019}. In this situation, when the gas near the PBH reaches high densities ($n_{\mathrm{H}} \gtrsim 10^2~\mathrm{cm^{-3}}$), it is quickly accreted by the PBH, lowering the average density within the BH accretion kernel. This self-regulating process results in a sub-Eddington accretion rate and a PBH growth pattern that mirrors the growth of its host halo. By $z = 10$, the PBH mass in the \texttt{PBH\_nfeed} case has grown by a factor of $\sim 10$.

As shown in Fig.~\ref{fig:AccretHist}, the inclusion of LW radiation has a negligible effect on the accretion rate at $z \gtrsim 30$. Afterwards, LW radiation slightly enhances the accretion rate. This effect is indirectly related to differences in star formation across simulations (see Section~\ref{subsec:popIII}): in runs without LW radiation, molecular gas clouds can collapse more readily, forming sink particles that remove gas from the vicinity of the PBH. In contrast, molecular cooling is suppressed in LW-irradiated environments, delaying the formation of sink particles and leaving more gas available for accretion onto the PBH. As a result, PBHs in simulations with LW radiation experience a modest increase in accretion at later times. 

In the case of baryon-DM streaming (see Fig.~\ref{fig:AccretHist}, right panel), stronger relative streaming velocities (at $z \gtrsim 1000$) slightly reduce the initial accretion rate, as the increased effective velocity reduces the Bondi accretion rate (see Equ.~(\ref{eq:bondi})). In addition, streaming motions delay and weaken the first accretion burst by hindering the inflow of dense gas to the PBH and subsequently reducing the Bondi accretion rate. At later stages, the streaming effect on accretion rates becomes more subtle; however, it significantly accelerates the timing of the first star formation, as discussed below.

\subsection{Triggering First Star Formation}\label{subsec:popIII}
\begin{figure*}[ht!]
\centering
    \includegraphics[width= \linewidth]{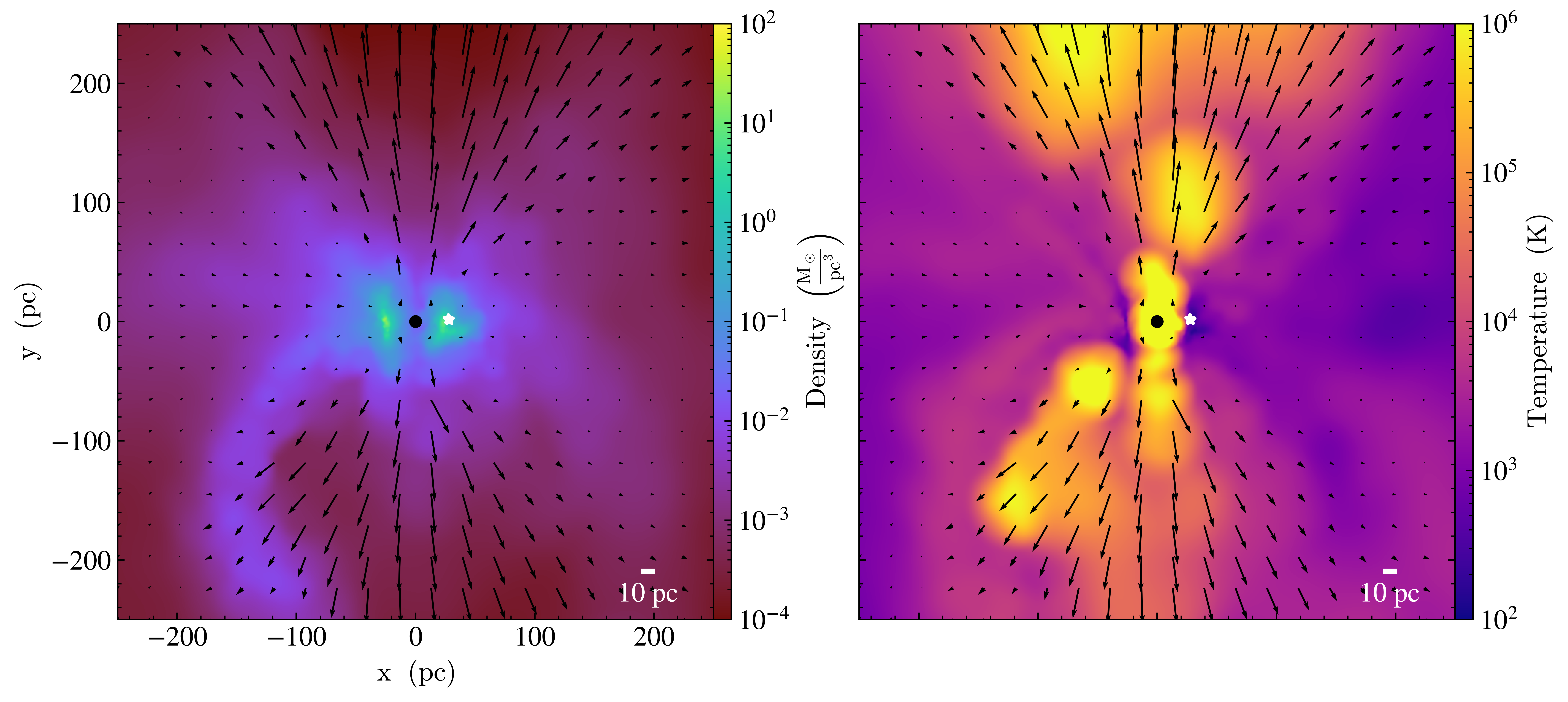}
    \caption{Onset of first star formation. Shown are projections of density (\textbf{left panel}) and temperature (\textbf{right panel}) 
    for the gas surrounding the central PBH, within a physical 500 pc scale. The snapshot is taken at $z \simeq 29.6$, corresponding to the moment when the first collapsing particle in the vicinity of the PBH is identified. We use output from the \texttt{PBH\_fd005} simulation, which assumes a black hole thermal feedback efficiency of $\epsilon_r = 0.5\%$. The black dot marks the position of the PBH, while the white star indicates the location of the collapsing gas cloud, representing a potential site for star formation. Velocity vectors are overlaid on both panels, illustrating the direction and relative magnitude of gas motion. These vectors highlight the inflow of gas toward the PBH, as well as feedback-driven outflows in its vicinity.
 }
    \label{fig:SlicePlot}
\end{figure*}
\begin{figure*}[ht!]
\centering
    \includegraphics[width=  \linewidth]{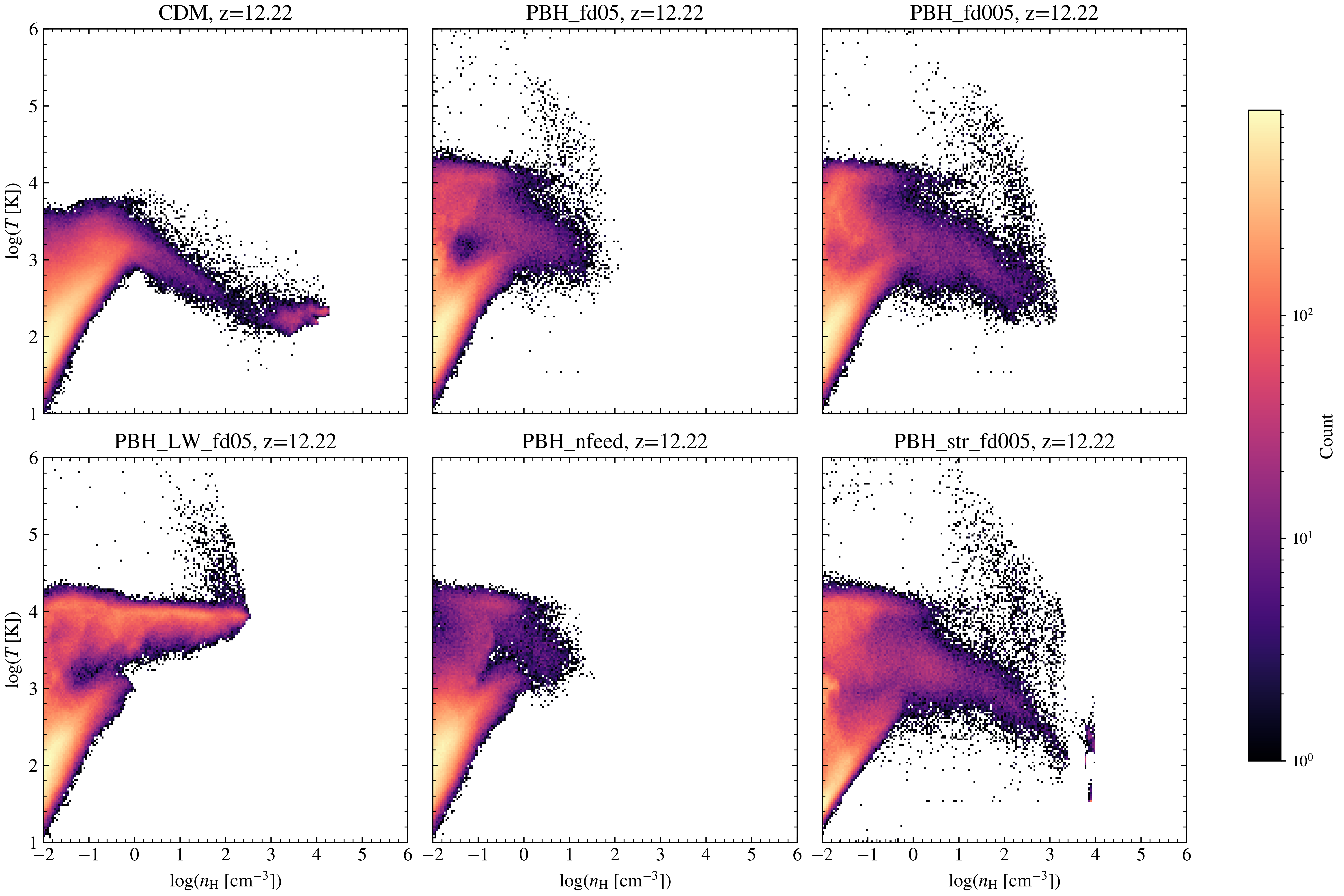}
    \vspace{1ex}
    \caption{Gas properties surrounding the central PBH for several simulation runs at $z \simeq 12$, presented as phase diagrams of temperature ($T$) vs. hydrogen number density ($n_{\mathrm{H}}$). The simulation without a PBH (\texttt{CDM}) is included as a reference to demonstrate the effects of PBH accretion heating on the surrounding gas. The redshift $z \simeq 12$ is chosen such that the initial gas collapse triggered by molecular cooling has already occurred in the \texttt{CDM} run. For the runs with PBHs, different feedback efficiencies are assumed, with $\epsilon_r = 0.05$ (\texttt{PBH\_fd05} and \texttt{PBH\_LW\_fd05}), $\epsilon_r = 0.005$ (\texttt{PBH\_fd005} and \texttt{PBH\_str\_fd005}), and $\epsilon_r = 0$ (\texttt{PBH\_nfeed}). Additional effects are also considered to study their impact on gas properties: the inclusion of Lyman–Werner (LW) radiation in the \texttt{PBH\_LW\_fd05} run, and the effect of baryon–dark matter (DM) streaming in the \texttt{PBH\_str\_fd005} run.
}
    \label{fig:Tn}
\end{figure*}

In the standard model, the formation of first generation stars is triggered by the runaway collapse of dense gas within virialized minihalos at redshifts $z \sim 20-30$, when the gas temperature is cooled by $\rm H_2$ down to $T \sim 100~\mathrm{K}$ and the density approaches $n_{\rm H} \gtrsim 10^3~\mathrm{cm^{-3}}$ \citep[e.g.,][]{Barkana2001:FirstStar, Abel2002, Bromm2002ApJ...564...23B, Yoshida2008:ProtoStar}. In this work, we find that PBHs play a dual role in influencing first star formation: their presence can enhance it by seeding overdense regions, or result in suppression through their impact on the surrounding gas. PBHs act as seeds for small-scale density enhancements, catalyzing the virialization of halos at earlier epochs ($z > 30$). This process accelerates the onset of first star formation by creating localized pockets of high-density gas, similar to the conclusion reached for stellar mass BHs in \citet{Boyuan2022MNRAS.514.2376L} and \citet{Boyuan2023arXiv231204085L}. On the other hand, the heating of the gas from PBH accretion will compete with the gas cooling process. The successful cooling of the gas to trigger the instability needed for first star formation will largely depend on the accretion feedback efficiency factor $\epsilon_{r}$. Here, we find a critical value of $\epsilon_{r,\rm crit}\sim 0.01$, below which star formation could take place in the vicinity of the PBH, as indicated in Table~\ref{Table:SimParam}. Above this value, the gas will be overly heated by accretion feedback, thus preventing collapse, and consequently the formation of first stars.

An example of early star formation seeded by a PBH is encountered at $z\sim 30$ in the \texttt{PBH\_fd005} simulation, as shown in Fig.~\ref{fig:SlicePlot}. In this figure, ionized outflows driven by PBH thermal feedback are evident\footnote{In this work, we did not implement any mechanical feedback or wind particles. We note that the direction of the outflow is stochastic and constantly changing over the course of galaxy evolution.}, with a velocity magnitude of $\gtrsim 400~\mathrm{km~s^{-1}}$ and temperatures in the range of $10^5 - 10^6~\mathrm{K}$. Concurrently, an inflow of cooler gas from the intergalactic medium (IGM) is present in the equatorial plane, with velocities of $\sim 30~\mathrm{km~s^{-1}}$ and temperatures $\lesssim 10^3~\mathrm{K}$. The compression of gas at the intersection of the IGM inflow and the PBH-driven outflow creates a favorable site for first star formation, located approximately $\sim 20~\mathrm{pc}$ from the central black hole. Here, the gas achieves a density of $n_{\rm H} \gtrsim 10^3~\mathrm{cm^{-3}}$ and cools to temperatures as low as $T \sim 100~\mathrm{K}$, where molecular cooling --- primarily through $\mathrm{H_2}$ --- dominates the thermal evolution of the gas, leading to the formation of proto-stars.

Fig.~\ref{fig:Tn} gives further details by showing the phase diagram of the gas within the simulation box at $z \simeq 12$ for several runs. In contrast to the \texttt{CDM} case, gas particles within the runs that include PBH thermal feedback exhibit a phase with elevated temperatures, in the range $1000$ to $10^6 \mathrm{\,K}$, at densities $n_{\rm H} \gtrsim 10~\mathrm{cm}^{-3}$. This phase corresponds to heating of dense gas by the pressure-driven outflow from the BH accretion feedback. In the presence of this additional energy source, in both \texttt{PBH\_fd005} and \texttt{PBH\_str\_fd005} runs, the gas still exhibits similar cooling behavior as that in the \texttt{CDM} run, and star-forming gas particles were identified in those simulation runs at $z\gtrsim 29$, as well.

In contrast, cases with strong Lyman–Werner (LW) feedback (\texttt{PBH\_LW\_fd05} and \texttt{PBH\_LW\_fd005}) or overheating of the gas from efficient PBH accretion feedback (\texttt{PBH\_fd05}) show neither significant gas collapse nor star formation near the PBH. In these scenarios, the destruction of $\mathrm{H_2}$ molecules by LW radiation or the elevated gas temperatures inhibit the conditions necessary for star formation. In a more extreme case where we assume no accretion feedback (\texttt{PBH\_nfeed}), all dense gas once formed will be soon engulfed by the PBH, thus preventing the onset of runaway collapse. Instead, star formation in all these simulations is delayed until $z \sim 10$, occuring in other, PBH-free cold dark matter halos within the simulation box.

Furthermore, we find that the inclusion of DM-baryon relative streaming velocities significantly accelerates first star formation in contrast to the scenario under a standard $\Lambda$CDM cosmology \citep[e.g.,][]{Stacy2011,Schauer2019MNRAS.484.3510S}. Comparing the timing of star formation for the \texttt{PBH\_str\_fd005} ($z\simeq70$) and \texttt{PBH\_sstr\_fd005} ($z\simeq110$) runs, a more prominent enhancement occurs when we impose a larger initial relative streaming velocity $v_{\rm b\chi}$. This enhancement can be attributed to the interaction of the PBH with the gas when traversing through it, leaving localized overdensities in the wake of the PBH trajectory, thus generating possible sites for early star formation that are less vulnerable to BH feedback. We show an illustration of this scenario in the Appendix, where we consider the resulting gas density distribution in the vicinity of the PBH.

\subsection{PBH-Stellar Coevolution}\label{subsec:coevo}

\begin{figure*}[ht!]
\centering
    \includegraphics[width= \linewidth]{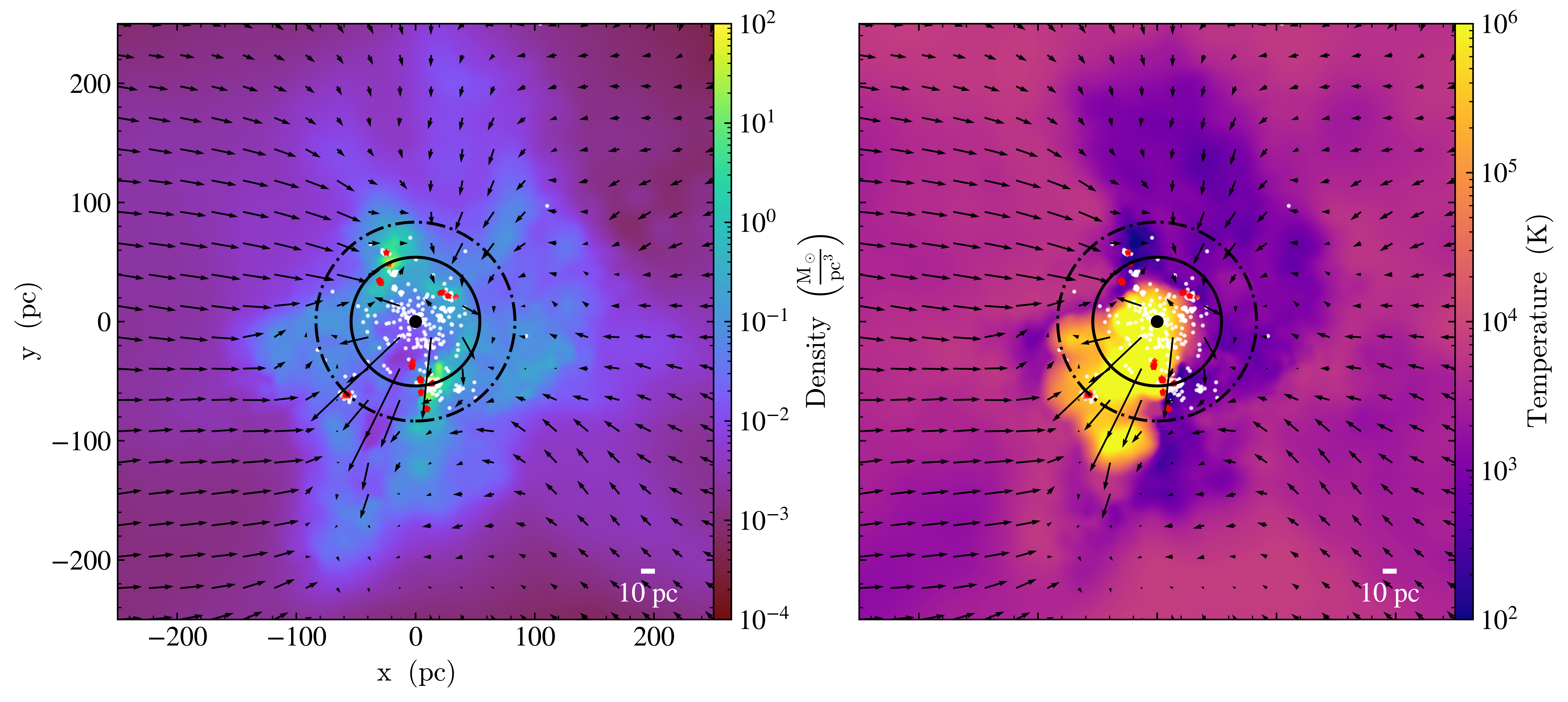}
    \caption{Morphology of a PBH-seeded galaxy. Similar to Fig.~\ref{fig:SlicePlot}, we illustrate the gas properties surrounding the central PBH, showing density (\textbf{left panel}) and temperature (\textbf{right panel}) on a 500~pc scale. The snapshot is taken at $z \simeq 10.5$ from the \texttt{PBH\_fd005} simulation run. The black dot marks the position of the PBH, whereas the white dots represent a sub-sample of the total stars formed (one per 25 star particles). The red symbols indicate the locations of potential star formation sites that are Jeans unstable at the scale of our resolution (with $M_{\rm J}\lesssim 1000\ M_\odot$; see Sec.~\ref{subsec:Jeans}). The black solid and dot-dashed circles denote the radii encompassing 50$\%$ and 80$\%$ of the total stellar mass, respectively.
}
    \label{fig:SlicePlotFinal}
\end{figure*}

As highlighted in the previous section, PBHs can influence the timing and location of the first star formation through their dual roles in catalyzing overdensities and delaying the gas cooling process via radiative and thermal feedback. Building on this foundation, this section explores the subsequent coevolution of PBHs and their host stellar systems, focusing on the assembly history of early galaxies and asking whether there is a connection to some of the more extreme objects now observed by JWST. Our simulation results, as further discussed below, reveal the intricate interplay among physical processes driving this co-evolution, as discussed in Section~\ref{subsubsec:GalAsm} and predict possible observational signatures in the high-redshift Universe (see Section~\ref{subsubsec:obs}).

\begin{figure}[ht!]
\centering
    \includegraphics[width= 1\columnwidth]{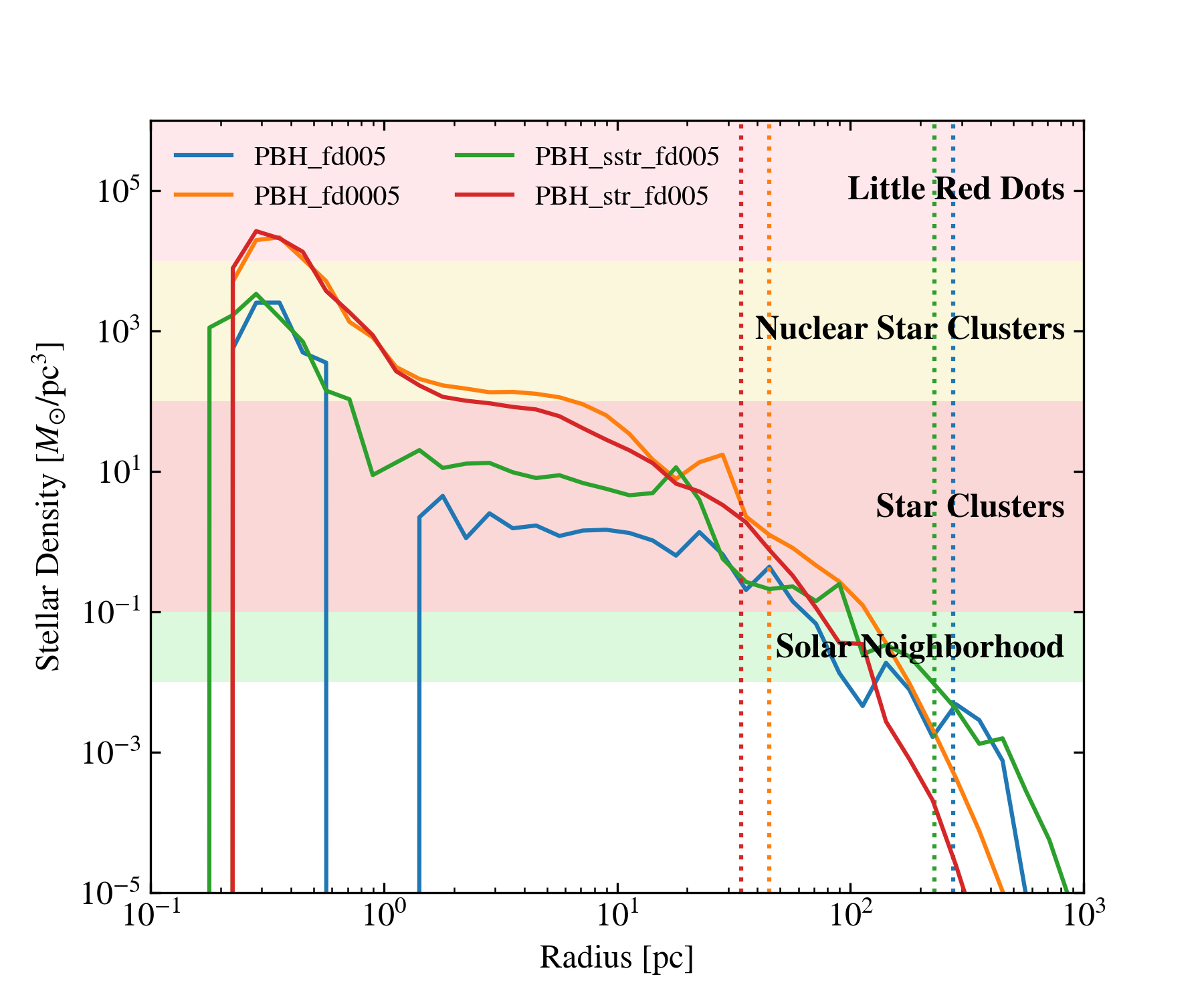}
\caption{Stellar densities in PBH-seeded galaxies at $z \sim 9$. We specifically show radial profiles for the \texttt{PBH\_fd005}, \texttt{PBH\_fd0005}, \texttt{PBH\_sstr\_fd005}, and \texttt{PBH\_str\_fd005} runs, plotting stellar density (in $\Msun\,\mathrm{pc}^{-3}$) vs. distance from the central PBH (in parsecs). The vertical dotted lines indicate the half-mass radius for each simulation run, color-coded to match the respective density profiles. The shaded horizontal bands highlight typical density ranges for various stellar systems, from the solar neighborhood \citep{Xiang:2018azk} to star clusters \citep{Krumholz2019ARA&A..57..227K}, nuclear star clusters \citep{Neumayer2020A&ARv..28....4N}, and Little Red Dots \citep{Guia:2024toq}.}
    \label{fig:stardens}
\end{figure}

\begin{figure}[ht!]
\centering
    \includegraphics[width=1\columnwidth]{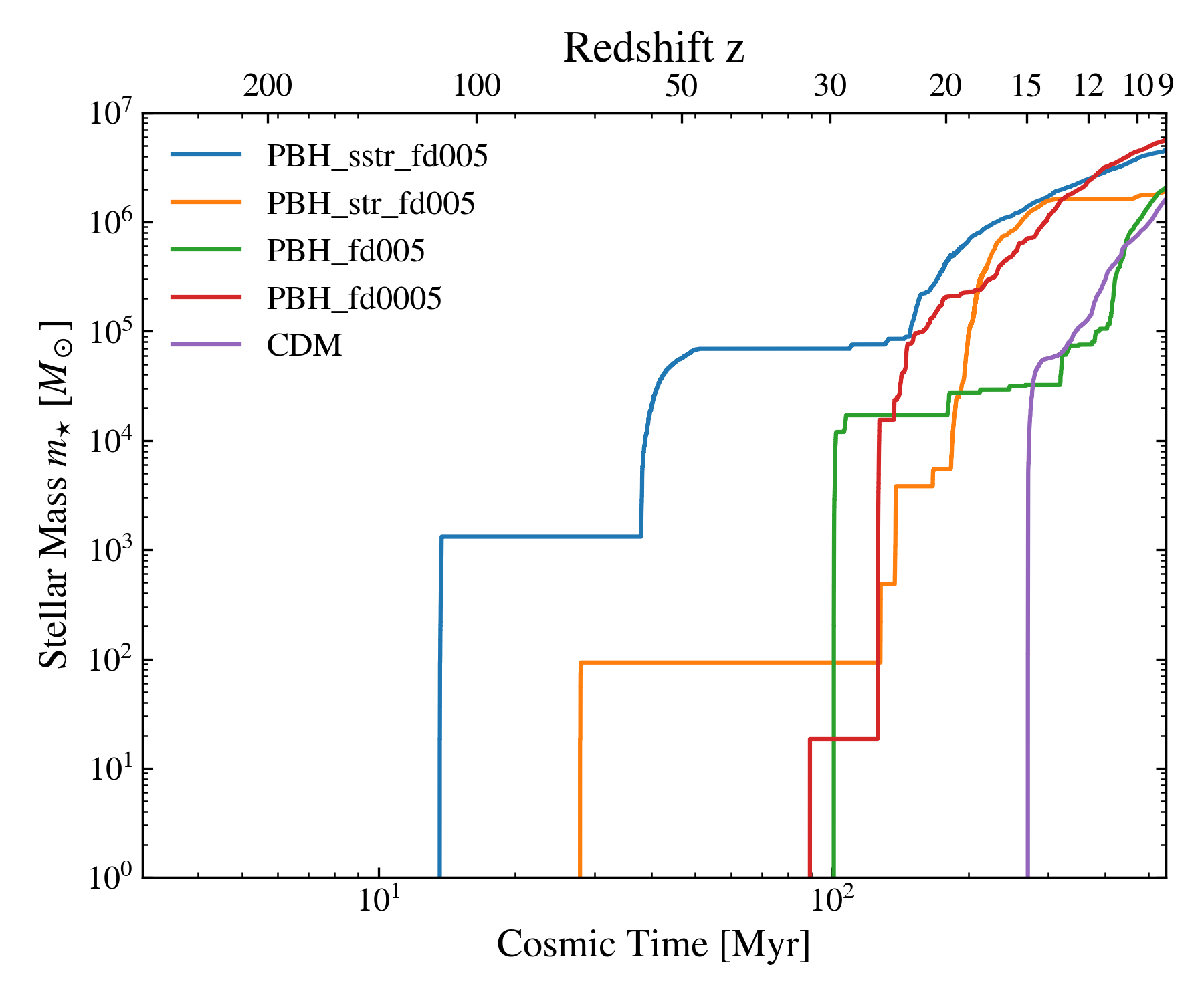}
    \caption{Stellar mass assembly histories for PBH-seeded galaxies. Shown is the total mass contained in stellar sink particles as a function of cosmic time and redshift for different simulation runs. The PBH-seeded runs experience star formation around the central PBH, with feedback efficiencies of $\epsilon_r = 0.005$ (\texttt{PBH\_fd005}, \texttt{PBH\_str\_fd005}, and \texttt{PBH\_sstr\_fd005}) and $\epsilon_r = 0.0005$ (\texttt{PBH\_fd0005}). The \texttt{PBH\_str\_fd005} and \texttt{PBH\_sstr\_fd005} simulations also account for relative baryon–dark matter streaming velocities. As evident in the resulting histories, all runs converge to a similar total stellar mass at $z\sim 9$, but exhibit significant differences at earlier times. In the presence of DM-baryon streaming, PBH-seeded halos trigger star formation much earlier, establishing a `(Pop~III) plateau' at levels of $\sim 10^2-10^3$\,M$_{\odot}$. 
}
    \label{fig:Ncolvsz}
\end{figure}

\begin{figure*}[ht!]
\centering
    \includegraphics[width= \linewidth]{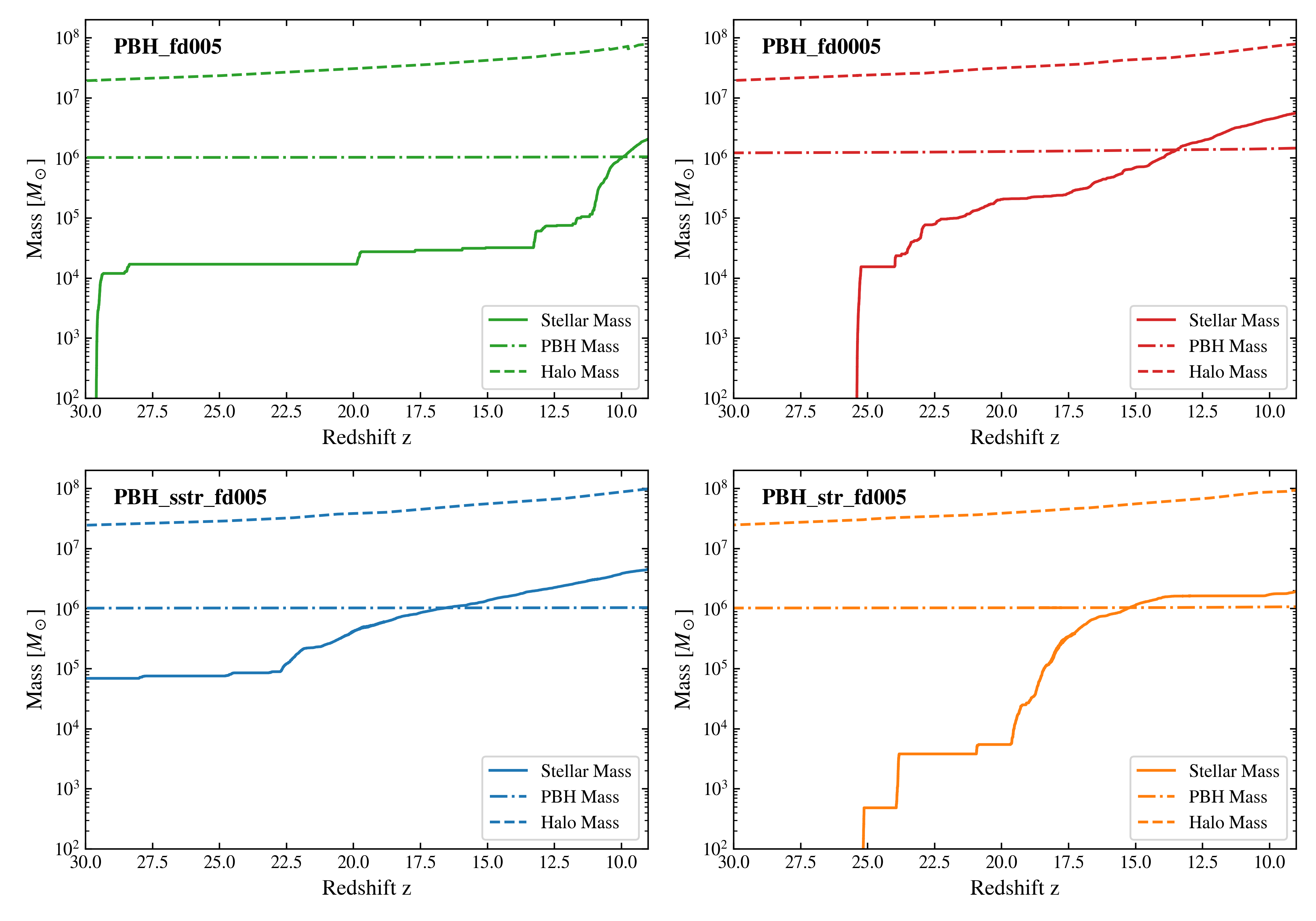}
    \caption{Assembly of the first galaxies seeded by PBHs. Each panel shows the time evolution of different simulation runs (see Table~\ref{Table:SimParam}): \texttt{PBH\_fd005} (\textbf{top left}), \texttt{PBH\_fd0005} (\textbf{top right}), \texttt{PBH\_sstr\_fd005} (\textbf{bottom left}), and \texttt{PBH\_str\_fd005} (\textbf{bottom right}). Within each panel, the co-evolution of PBH mass (dot-dashed lines) and its host dark matter (DM) halo mass (dashed lines) is plotted, alongside the assembly of stellar mass (solid lines), as a function of redshift. As can be seen, the PBH-seeded systems are highly overmassive at early times, with the stellar component only catching up by $z\sim 10$.
}
    \label{fig:MsunMh}
\end{figure*}


\subsubsection{First Galaxy Assembly}\label{subsubsec:GalAsm}

Fig.~\ref{fig:SlicePlotFinal} illustrates the gas properties within a 500\,pc region surrounding the central PBH at a later stage of PBH–stellar coevolution ($z \simeq 10.5$). In this region, the gas density ranges from $\sim 10^{-3}$ to $10^{4}~\mathrm{cm^{-3}}$, while temperatures extend from $\sim 10^2$ to $10^6\mathrm{\,K}$. Velocity vectors reveal inflows with speeds up to $\sim 40~\mathrm{km~s^{-1}}$, alongside localized turbulence near the PBH. Star formation occurs in regions where the gas density exceeds $n_{\mathrm{H}} \sim 10^3~\mathrm{cm^{-3}}$, allowing molecular hydrogen cooling to reduce the temperature to $\sim 100~\mathrm{K}$. 

 In contrast to Fig.~\ref{fig:SlicePlot}, which captures the initial onset of star formation, the newly collapsing sites (marked by red stars in Fig.~\ref{fig:SlicePlotFinal}) reside at the outer part of the galaxy ($\sim 10$–$50~\mathrm{pc}$ from the PBH) and exhibit a pronounced clustering pattern indicative of nascent star clusters \citep[e.g.,][]{Adamo2024}. Furthermore, the gas velocity field shows that the inflowing gas counteracts the outflow from PBH accretion feedback, similar to Fig.~\ref{fig:SlicePlot}, compressing the gas to higher densities that promote efficient cooling and subsequent star formation. On the other hand, the already-formed stars (marked by white dots in Fig.~\ref{fig:SlicePlotFinal}) display a more spherical distribution, with a higher concentration near the central PBH, which implies inside-out star formation. The system remains relatively compact, with a total stellar mass reaching $\sim 1.1 \times 10^6~\Msun$ and a half-mass radius of $\sim 50$ pc at $z \simeq 10.5$.

From the final snapshots of simulations that experience star formation (\texttt{PBH\_fd005}, \texttt{PBH\_fd0005}, \texttt{PBH\_sstr\_fd005}, and \texttt{PBH\_str\_fd005}), we present radial profiles of stellar density at $z \sim 9$ in Fig.~\ref{fig:stardens}. These profiles demonstrate that PBH-seeded systems produce compact and dense stellar structures, with half-mass radii ranging from $\sim 40$ to $300$ pc. The average stellar density within their half-mass radius for these four runs is (in units of $\Msun\,\mathrm{pc}^{-3}$) 0.011 (\texttt{PBH\_fd005}), 7.47 (\texttt{PBH\_fd0005}), 0.045 (\texttt{PBH\_sstr\_fd005}), and 5.90 (\texttt{PBH\_str\_fd005}), respectively. However, within $\sim 1$ pc of the central PBH, the stellar density reaches $\sim 10^4~\Msun\,\mathrm{pc}^{-3}$, forming a dense stellar core with a mass in the range of $10^3 - 10^4~\Msun$. This stellar peak density is comparable to that observed in nuclear star clusters \citep{Neumayer2020A&ARv..28....4N} and ``Little Red Dot'' objects \citep{Guia:2024toq}, suggesting that PBH-seeded star-forming regions could give rise to (a subset of) these high-redshift compact stellar systems.

Fig.~\ref{fig:Ncolvsz} tracks the total mass of collapsing particles as a function of cosmic time and redshift for all simulation runs exhibiting star formation (\texttt{PBH\_fd005}, \texttt{PBH\_fd0005}, \texttt{PBH\_str\_fd005}, and \texttt{PBH\_sstr\_fd005}) and compares them to a fiducial run lacking a PBH (\texttt{CDM}). To further place the assembly of stellar mass into the context of the growing host halo and PBH, we separately plot the mass evolution trajectories for individual PBH runs, as shown in Fig.~\ref{fig:MsunMh}.
Despite variations between different PBH models, all simulations reach stellar masses of at least $\sim 10^6~\Msun$ by $z \sim 9$, differing only by a factor of a few.

Over a period of approximately 400\,Myr from $z \sim 30$ to $10$, continuous gas inflow drives the gradual growth of the galaxy’s stellar component, leading to a transition where the PBH mass becomes subdominant relative to the stellar mass. Despite variations between different PBH models, all simulations reach similar stellar masses of at least $\sim 10^6~\Msun$ by $z=9$: at that point, the stellar masses in the simulations shown in Fig.~\ref{fig:Ncolvsz} are $m_{\star}/10^6~\Msun \simeq 1.6$ (\texttt{CDM}), $2.1$ (\texttt{PBH\_fd005}), $5.6$ (\texttt{PBH\_fd0005}), $1.9 $ (\texttt{PBH\_str\_fd005}), and $4.6$ (\texttt{PBH\_sstr\_fd005}). Notably, all PBH-containing runs exhibit an earlier onset of star formation and yield a higher stellar mass compared to the \texttt{CDM} run. Both the strength of baryon–dark matter streaming and the intensity of accretion feedback are found to affect the mass assembly history. For the \texttt{PBH\_fd005} and \texttt{PBH\_fd0005} runs, the onset of star formation occurs at approximately $z\sim 30$. However, as is evident in Fig.~\ref{fig:MsunMh}, the star formation rate behaves differently at $z \lesssim 25$: the \texttt{PBH\_fd0005} run exhibits an elevated star formation trend, whereas the \texttt{PBH\_fd005} run does not display such a trend until $z \lesssim 12$. Consequently, the total stellar mass in the \texttt{PBH\_fd0005} run is about 2.5 times greater than that in the \texttt{PBH\_fd005} run.
 
 Analysis of the BH accretion history reveals that the accretion rate $\dot{m}_{\rm acc}$ is negatively correlated with the energy feedback efficiency $\epsilon_r$. In simulations with a larger $\epsilon_r$, the increased energy input into the surrounding gas effectively delays cooling, thereby suppressing star formation. On the other hand, streaming motion accelerates star formation, shifting the primary epoch of star formation from $z \sim 12 - 9$ in the \texttt{PBH\_fd005} run to an earlier period of $z \sim 20 - 15$ in the \texttt{PBH\_str\_fd005} run and $z \sim 20 - 10$ in the \texttt{PBH\_sstr\_fd005} run. For the case with moderate streaming (\texttt{PBH\_str\_fd005}), the star formation rate is quenched below $z \sim 15$, allowing the stellar mass in \texttt{PBH\_fd005} (with zero streaming) to eventually catch up by $z \sim 9$. This quenching can be attributed to enhanced BH accretion and feedback at $z \lesssim 15$, which delays gas cooling and subsequently suppresses star formation (see Fig.~\ref{fig:AccretHist}, right panel). Conversely, in the case with particularly strong streaming (\texttt{PBH\_sstr\_fd005}), the accretion rate slightly decreases at $z \lesssim 20$, leading to an extended period of star formation. As a result, by $z \sim 9$, the stellar mass in \texttt{PBH\_sstr\_fd005} surpasses that in the \texttt{PBH\_fd005} run. 


Comparing the growth of stellar, halo, and PBH mass (see Fig.~\ref{fig:MsunMh}) reveals distinct trends. Due to feedback effects, the PBH mass remains nearly constant at $\sim 10^6~\Msun$, as explained in Section~\ref{subsec:Accret}, while the host halo mass steadily increases from $\sim 10^7$ to $\sim 10^8~\Msun$—with the streaming run achieving a slightly higher halo mass. The ratio of stellar mass to halo mass implies a high star formation efficiency by $z\lesssim 10$, $\epsilon_{\star} \sim \mathcal{O}(0.1)$. Moreover, by $z \sim 10$ (approaching the JWST detectability limit), the stellar mass exceeds the PBH mass, although the difference remains within a factor of a few.

\subsubsection{Observational Perspectives}\label{subsubsec:obs}

\begin{figure*}[ht!]
\centering
    \includegraphics[width=1\linewidth]{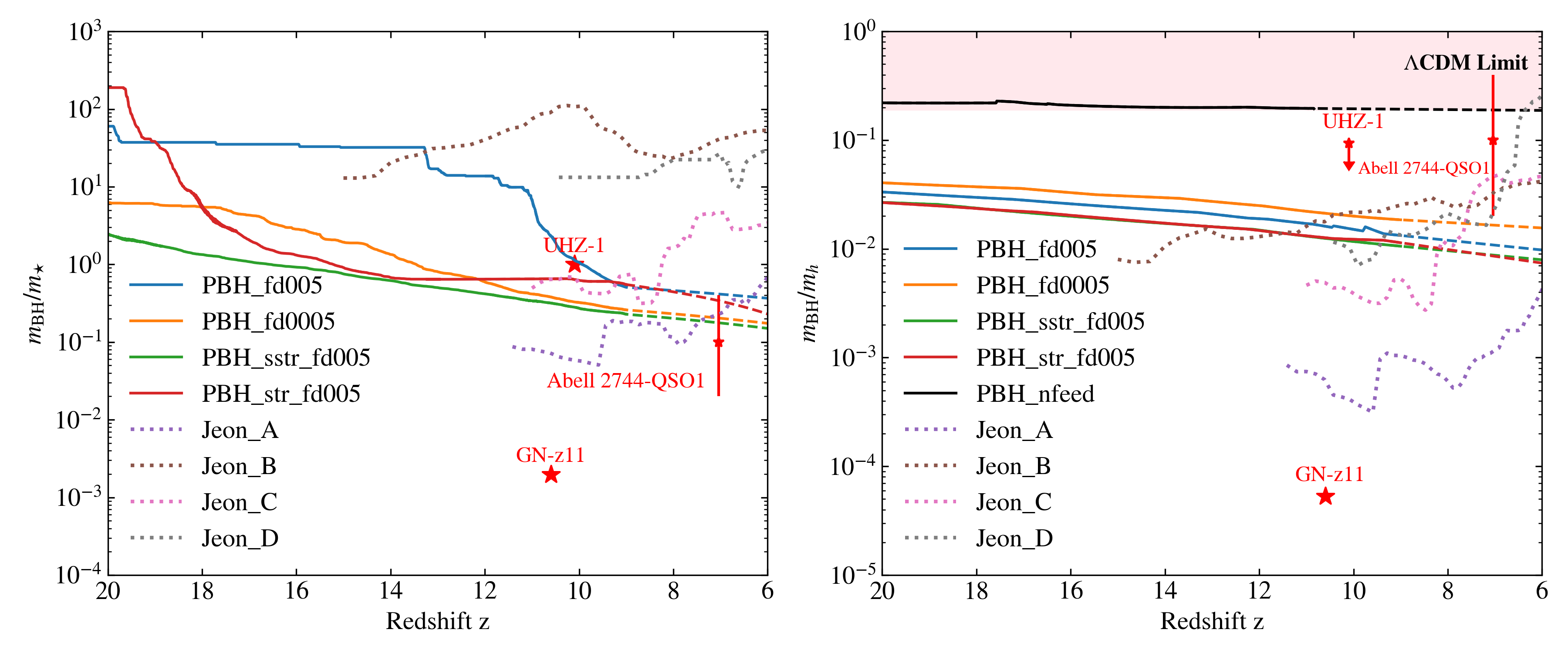}
\caption{Co-evolution of the PBH and its host system. We show the black hole (PBH) to stellar mass ratio ($m_{\mathrm{BH}} / m_{\star}$) (\textbf{left panel}), as well as the PBH to halo mass ratio ($m_{\mathrm{BH}} / m_{h}$) (\textbf{right panel}) for our PBH-seeded simulations, and compare with select AGN sources observed at similar redshifts, including GN-z11 \citep{Maiolino2024Natur.627...59M, Scholtz2024A&A...687A.283S}, UHZ-1 \citep{Bogdan:2023UHZ1}, and Abell 2744-QSO1 \citep{BlackTHUNDER2025arXiv, DEugenio2025arXivBlackTHUNDER}, marked with red stars with error bars. The solid lines represent PBH-seeded simulation runs with feedback efficiencies of $\epsilon_r =0$(\texttt{PBH\_nfeed}), $ 0.005$ (\texttt{PBH\_fd005}, \texttt{PBH\_str\_fd005}, and \texttt{PBH\_sstr\_fd005}) and $\epsilon_r = 0.0005$ (\texttt{PBH\_fd0005}). Simulations \texttt{PBH\_str\_fd005} and \texttt{PBH\_sstr\_fd005} include the effects of relative baryon–dark matter streaming. Here, dashed lines represent the linear extrapolation of those simulation runs. In addition, for comparison, the dotted lines (\texttt{Jeon\_A}, \texttt{Jeon\_B}, \texttt{Jeon\_C}, and \texttt{Jeon\_D}) show the evolution of DCBHs with their host galaxies from several simulation runs in \citet{Jeon2025ApJ...979..127J}. We also indicate (with the red shaded region in the right panel) the range of mass ratios that would violate the $\Lambda$CDM limit, with a lower bound given by the cosmological baryon-to-DM ratio $\Omega_{b} / (\Omega_{b} - \Omega_{m})$. Here, the host halo mass for UHZ-1 is estimated by taking $m_{\rm BH} \simeq  m_{\star}$ according to \cite{Bogdan:2023UHZ1}, resulting in the upper limit shown. The stellar mass of Abell 2744-QSO1 is taken as the maximum value allowed by the host halo (dynamical) mass. This comparison highlights the potential of PBH-seeded systems to explain a subset of the newly-discovered high-redshift AGNs, in particular the most overmassive ones.}
    \label{fig:mbhmstar}
\end{figure*}

Based on our simulation results, we propose several high-redshift observational signatures that could distinguish PBH-seeded galaxies from standard $\Lambda$CDM ones. In our simulations, the overall system exhibits a total stellar mass of $\sim 10^6\,\rm \Msun$ and is morphologically compact (with more than 80\% of stellar mass within $r \sim 100$\,pc as shown in Fig.~\ref{fig:SlicePlotFinal}). Such systems would appear as spatially unresolved objects to JWST, unless gravitational lensing by foreground galaxies enhances their angular size. 

To compare with already observed AGNs such as GN-z11 \citep{Maiolino2024Natur.627...59M} and UHZ-1 \citep{Bogdan:2023UHZ1}, we plot the evolution of the BH-to-stellar mass ratio\footnote{It is important to note that the exact black hole and stellar masses in our simulations may not directly match those inferred from observations. However, since the mass of a PBH-seeded halo is expected to scale linearly with the PBH mass \citep{Mack2007ApJ}, it is reasonable to assume that the stellar mass within PBH-seeded halos also follows a similar scaling. Under this assumption, the black hole-to-stellar mass ratios predicted by our simulations can be meaningfully compared with observations, even for objects with different absolute mass scales.}, $m_{\mathrm{BH}}/m_{\star}$, for select PBH simulation runs in Fig.~\ref{fig:mbhmstar}. The overall decreasing trend of $m_{\mathrm{BH}}/m_{\star}$ over time reflects the relatively faster growth of stellar mass compared to the PBH accretion rate. In all runs, a rapid evolution of $m_{\mathrm{BH}}/m_{\star}$ is observed, transitioning from $\sim 10^2$ at $z \sim 20$ to $\sim 0.1$ by $z \sim 10$. This ratio is significantly higher than the local empirical $M-\sigma$ relation value of $\lesssim 10^{-3}$ \citep[e.g.,][]{Kormendy2013GalaxySMBH, Reines2015ApJ...813...82R}, and also exceeds the typical values of $m_{\mathrm{BH}}/m_{\star} \sim 0.1-0.01$ found in JWST-observed active galaxies at $z \sim 4-7$ \citep{PacucciLRD2023ApJ...957L...3P}. However, it is comparable to the ratios reported for the LRDs at $z\sim 4-10$ \citep{Leung2024:LRDarXiv}. In our simulations, runs with $\epsilon_r = 0.005$ (\texttt{PBH\_fd005} and \texttt{PBH\_str\_fd005}) yield $m_{\mathrm{BH}}/m_{\star}\sim 1$ at $z\sim 10$, whereas runs with $\epsilon_r = 0.0005$ (\texttt{PBH\_fd0005}) or with strong streaming effects (\texttt{PBH\_sstr\_fd005}) produce slightly lower ratios of $\sim 0.3$. The observational constraints for UHZ-1 ($m_{\mathrm{BH}}/m_{\star} \sim 0.1-1$ and $m_{\mathrm{BH}}/m_{h} \lesssim 0.1$) are consistent with these predictions, while GN-z11 ($m_{\mathrm{BH}}/m_{\star} \sim 0.001$ and $m_{\mathrm{BH}}/m_{h}\sim 5\times 10^{-5}$) does not favor a PBH-induced origin. Overall, the presence of overmassive BHs in extremely compact galaxies at high redshifts supports the hypothesis that PBHs may act as seeds for some of them, including a subset of the LRDs.

To distinguish PBH seeding from heavy seeding via DCBHs, we compare our results with the simulated evolution presented in \cite{Jeon2025ApJ...979..127J}, also in Fig.~\ref{fig:mbhmstar}. In their work on galaxy assembly with DCBHs that grow significantly under an optimistic accretion model that takes all the gas available in the vicinity of the BH, the evolution of the $m_{\mathrm{BH}}/m_{\star}$ ratio exhibits a gradual increase with decreasing redshift, in contrast to the decreasing trend found in our PBH cases where the PBHs hardly grow by Bondi accretion. Although a single detection of a UHZ-1–like object may not suffice to distinguish between these scenarios—given that DCBH models predict $m_{\mathrm{BH}}/m_{\star} \sim 0.1-100$ at $z \sim 10$--$15$—a statistically significant sample of AGNs at $z \sim 10-20$ would be required. In such a sample, the rapid evolution of the $m_{\mathrm{BH}}/m_{\star}$ relation would likely favor a PBH origin for the AGN population. Moreover, at $z\gtrsim 10$, our model distinguishes itself from (but does not exclude) scenarios in which SMBHs originate from Pop~III star remnants, as most models predict $m_{\mathrm{BH}}/m_{\star} \lesssim 0.1$ \citep[see also][]{Jeon2012ApJ...754...34J, Pezzulli2016MNRAS.458.3047P, Cammelli2025pop3SMBH}.

Recently, an intriguing object within the LRD class, named Abell-2744-QSO1, has been identified by JWST at $z \simeq 7.04$ \citep{BlackTHUNDER2025arXiv,DEugenio2025arXivBlackTHUNDER}. Spectroscopic studies indicate that its observed properties align well with predictions from our simulations, particularly in terms of the suppression of star formation due to black hole feedback. The luminosity of Abell-2744-QSO1 is primarily attributed to emission from a gas cloud rather than stellar contributions, and its Balmer break is found to be of non-stellar origin. Additionally, the system exhibits an overmassive black hole relative to both its stellar and halo mass\footnote{We make the simplifying assumption that the dynamical mass is equivalent to the halo mass. The stellar mass estimates are taken as the maximum mass allowed by the dynamical mass.}, with a black hole-to-stellar mass ratio of $m_{\mathrm{BH}}/m_{\star} \gtrsim 0.02 - 0.4$ and a black hole-to-halo mass ratio of $m_{\mathrm{BH}}/m_{h} \gtrsim 0.02 - 0.4$. Both ratios fall within the extrapolated range predicted by our PBH-seeded simulations, as shown in Fig.~\ref{fig:mbhmstar}. If the dynamical mass is composed entirely of stellar content, then the object is consistent with our simulations including accretion feedback. Conversely, if the dynamical mass is not dominated by stars, the object closely resembles the \texttt{PBH\_nfeed} run, in which the central black hole grows at a rate similar to the halo’s growth. In contrast, the combination of a high $m_{\mathrm{BH}}/m_{\star} \gtrsim 0.1$ and black hole-to-halo mass ratio would make it more challenging to reconcile this object with the DCBH formation model. Further observations of similar objects would provide strong empirical support for the existence of heavy PBHs, offering a plausible pathway for the formation of early SMBHs, in particular the most overmassive ones.

\section{Limitations and Caveats} \label{sec:caveat}

While our simulations provide valuable insights into the co-evolution of PBHs and their host stellar systems, several limitations and assumptions in our modeling, although not significantly altering our conclusions, should be addressed in future work. These issues primarily arise from simplified physics, computational constraints, and limited resolution. Below, we briefly discuss key limitations and potential improvements.

PBHs are theorized to have formed during the radiation-dominated epoch, well before the starting point of our simulation at $z_{\rm eq} = 3400$. In this work, we take the dark matter to mostly consist of particles with a small admixture of PBHs, making the simplified assumption of an unperturbed DM density background around the PBH at $z \sim z_{\rm eq}$, which allows the halo structure to evolve. However, we also acknowledge that PDM might be gravitationally captured by PBHs, and a PDM halo "clothing" the PBH may have already begun forming during the radiation-dominated epoch at $z \gtrsim z_{\rm eq}$ with substantial subsequent growth \citep[see e.g.][]{Eroshenko:2016yve, Boucenna:2017ghj, Carr:2020mqm, Boudaud:2021irr, Salati:2025hpd}. These exciting aspects, together with accompanying emission signatures (e.g., from WIMP annihilation), could be better addressed in future work.

Due to limitations in our computational resources and the scale of the simulation, our models stop at $z = 9$, restricting our ability to study the long-term evolution of PBH-stellar systems and their host halos at lower redshifts. In addition, at $z \lesssim 10$, thermal feedback from PBH accretion begins to heat up the entire simulation box, rendering the evolution of star clusters around PBH-seeded halos less physical. This excessive heating slows down gas cooling, reduces the star formation rate, and leads to a more extended stellar density distribution than would be expected in a fully resolved system. Future simulations with improved resolution and larger volumes will be necessary to better capture the long-term effects of PBH accretion feedback on star formation and galaxy evolution, particularly at $z \lesssim 10$. Furthermore, the limited box size prevents us from accounting for large-scale structures that may form around a PBH-seeded halo. If a PBH is embedded in an overdensity peak whose scale far exceeds that of the PBH-seeded halo, we would expect a merger with a nearby galaxy, which could boost the stellar mass, disrupt accretion, and alter the $m_{\mathrm{BH}}/m_{\star}$ and $m_{\mathrm{BH}}/m_{h}$ ratios. Therefore, while the current work is valid for a galaxy seeded by an isolated PBH, additional simulations will be needed to explain the early formation of galaxies with $m_{\star} \gg 10^6\,\Msun$ observed at $z \gtrsim 10$ \citep[e.g.,][]{Maisies:2022,GHz2,GLASSz13,Castellano2023ApJ...948L..14C,Labbe2023Natur.616..266L,Adams2024,Donnan2024}.

The efficiency of thermal feedback from PBH accretion, represented by $\epsilon_r$, remains a critical parameter in our simulations. However, our understanding of black hole accretion feedback at early cosmic times is still limited. For simplicity, we implement thermal feedback as energy injection into the gas within the Bondi radius surrounding the PBH, assuming a constant efficiency factor $\epsilon_r$ as a fixed fraction of the accretion luminosity throughout a given simulation run. While this approach provides a reasonable approximation of the average accretion rate over cosmic time, the detailed effects of accretion feedback on the surrounding gas require further investigation~\citep[see e.g.,][]{Boyuan2023arXiv231204085L}. Future work incorporating dynamic accretion models, also including radiative transfer and angular momentum considerations, will be essential to refine our understanding of PBH accretion and its impact on early structure formation. Moreover, our simulations do not account for stellar feedback, which could significantly alter the environment around the PBH. Processes such as supernova explosions, radiation pressure, and ionizing radiation regulate gas cooling, metal enrichment, and subsequent star formation rates \citep[e.g.,][]{Ceverino2009ApJ...695..292C}, and might also reduce the growth rate of the central SMBH \citep{Jeon2025ApJ...979..127J}.

In addition, our current models do not incorporate the effects of metal enrichment from earlier generations of stars, as this lies beyond our current computational capacity. We note that the spectroscopic study of UHZ-1 finds a metallicity of $Z/Z_{\odot} \gtrsim 0.1$, much higher than the metallicity where Pop~III stars could form \citep{Goulding2023ApJ...955L..24G}. Over time, increasing metallicity can enhance gas cooling rates, in particular for $Z/Z_{\odot} \gtrsim 0.01$ in high-density gas regions ($n_{\rm H} \gtrsim 10^4~\mathrm{cm^{-3}}$), and may subsequently shift the initial mass function of newly formed stars to lower-mass stars \citep[e.g.,][]{Omukai2005ApJ...626..627O}. Future work should incorporate these aspects by implementing a stellar evolution prescription to understand how successive generations of stars modify their local environment near the central PBH.

At lower redshifts ($z \lesssim 15$), the formation of rotationally supported gas disks may become relevant \citep[for a general review, see][]{Inayoshi:2020}, within the standard $\Lambda$CDM framework for BH growth.  Typically, the formation of an accretion disk requires the inflow of dense, cold gas within a bound star cluster \citep{Alexander2014Sci...345.1330A}, a scenario that does not apply in our case because an ionized region forms around the central PBH prior to star formation. Our simulations currently employ the Bondi accretion formalism (see Equ.~(\ref{eq:bondi})) to approximate the accretion rate, thus not accounting for angular momentum—despite the fact that angular momentum may play a key role in regulating gas inflow and the formation of accretion disks\footnote{In one of our simulations, \texttt{PBH\_fd0005}, we identify a feature resembling a stellar disk around the PBH, as shown in Fig.~\ref{fig:SlicePlotwf0005} of the Appendix. In this case, transport of angular momentum is efficient as the PBH accretes gas at a relatively high rate ($\dot{m}_{\rm acc}/\dot{m}_{\rm Edd} \gtrsim 0.01$) due to low feedback efficiency ($\epsilon_r = 0.0005$).}. This simplification renders our treatment of gas dynamics incomplete, particularly for regions where gas reaches high densities near the PBH, possibly initiating a super-Eddington accretion phase \citep[see the relevant numerical prescription in][]{tremmel2017romulus}. Within our current implementation, the PBH accretion rate remains largely sub-Eddington. If such low accretion rates were to extend to $z \lesssim 10$, further SMBH growth would remain limited, thus restricting the applicability of our numerical approach to lower-redshift AGN and quasars. Future simulations should identify the criteria for disk formation and incorporate angular momentum considerations to explore their implications for both accretion and feedback.

\section{Summary and Conclusions}\label{sec:Summary}

In this study, we have investigated the role of primordial black holes (PBHs) in influencing the formation of the first stars, their contribution to galaxy assembly, and their implications for high-redshift observations. Below, we summarize the main findings and their broader implications: 

\begin{itemize}
    \item The accretion history of PBHs is strongly influenced by feedback processes. When feedback efficiency is high, energy injection from accretion heats the surrounding gas, forming ionized bubbles and launching shocks, suppressing further accretion, and quickly reaching a dynamical equilibrium. Under such conditions, the growth of the PBH is minimal throughout. In contrast, extremely weak feedback efficiency allows PBHs to accrete near the Eddington limit, leading to substantial mass growth over time.
    
    \item PBHs exhibit a dual influence on the timing and location of star formation. On the one hand, PBHs accelerate structure formation by seeding dark matter halos and gravitationally attracting gas to form dense regions. On the other hand, radiative and thermal feedback from accretion can delay gas cooling, suppressing star formation in cases of strong feedback, and star formation also needs to compete with PBH accretion for the dense gas supply. In cases with weaker yet non-negligible feedback, gas cooling proceeds efficiently, leading to higher densities and enhanced star formation. However, under extremely weak feedback, the PBH devours all dense gas, thus preventing star formation. These diverse behaviors highlight the complexity of PBH effects, which can both suppress and promote star formation depending on the feedback strength.

    \item The inclusion of relative baryon–dark matter streaming enhances star formation by catalyzing the formation of dense gas behind a PBH when it is traveling through the intergalactic medium. Regions with higher-than-average streaming velocities are expected to form stars earlier and host more massive galaxies compared to regions without streaming, somewhat counter-intuitively and in difference from the standard $\Lambda$CDM case \citep[e.g.,][]{Schauer2023}.
   
    \item Under strong Lyman–Werner (LW) feedback, the star formation process is quenched (at least on the scales resolved by our simulation) as the dense gas in the vicinity of the PBH cannot efficiently cool, resulting in near-isothermal evolution dominated by atomic hydrogen cooling.
    
    \item PBHs co-evolve with their host galaxies, catalyzing the formation of dense star clusters around the central black hole. This co-evolution is reflected in the evolution of the PBH-to-stellar mass ratio, which decreases over time as stars form more rapidly relative to PBH mass growth. As shown in Fig.~\ref{fig:mbhmstar}, the simulations yield ratios consistent with observations of select overmassive high-redshift AGN, such as UHZ-1, as well as (a subset of) LRDs that appear around similar redshifts. Considering the even more striking case of the recently discovered Abell-2744-QSO1 LRD that is overmassive with respect to its host dynamical mass, a PBH-seeded scenario may provide a more natural explanation, where other formation pathways seem severely challenged.  

\end{itemize}


To distinguish PBHs from other SMBH formation pathways, further observations of early AGN with frontier facilities, such as the JWST or the Roman Space Telescope (RST), will be crucial to achieve statistically significant samples of the overall SMBH demographics at high redshifts. Furthermore, the dense star clusters in the vicinity of a PBH, as can be inferred from the stellar density profile in Fig.~\ref{fig:stardens}, could imply a high frequency of tidal disruption events, which may be observable with JWST or in future surveys with RST \citep{Inayoshi2024ApJ...966..164I, Wang2025TDE250418144W}.

The additional detection of high-$z$ AGN in radio-wavebands (e.g., with the Atacama Large Millimeter/submillimeter Array, ALMA) would allow a more complete physical characterization, when cross-correlating with the near-infrared (JWST) results \citep[e.g.,][]{Ibar2008MNRAS.386..953I, DeRossi2023,LabbeLRD2025ApJ...978...92L}. Future missions, such as the Square Kilometer Array (SKA) \citep[for constraints on massive PBHs, see][]{Bernal:2017nec}, the Hydrogen Epoch of Reionization Array (HERA) \citep{Koopmans2021ExA....51.1641K}, and lunar-based 21-cm cosmology observatories \citep{Burns2021arXiv210305085B}, offer unique opportunities to probe PBH-driven processes and provide deeper physical insight. For instance, these facilities could detect thermal or radiative signatures originating from PBHs during the cosmic dark ages ($z \sim 30 - 200$), providing indirect evidence of their influence on the early IGM and cosmic radiation background \citep[see also, e.g.,][]{Kashlinsky2016ApJ...823L..25K, Hasinger:2020ptw}. Additionally, such missions could explore the extremely early formation of stars, which might arise due to the blue tilt of the power spectrum introduced by PBHs \citep[see also][]{Hirano2015ApJ...814...18H,Hirano2024ApJ...963....2H}.

Overall, the presence of massive PBHs in the early Universe, even if they were to constitute only a small fraction of the dark matter, could be highly significant, through the seeding of the extreme overmassive systems that are being discovered by the JWST. Similarly, the horizon for galaxies that are sufficiently massive to be detectable with the JWST in future ultra-deep surveys would be significantly extended if such massive PBHs exist, different from standard $\Lambda$CDM cosmology, where galaxies are predicted to disappear at $z \gtrsim 20$. It is exhilarating to contemplate what is out there, in the ultra-deep Universe, and it is clear that the emergence of the first black holes is of particular importance in driving the end of the cosmic dark ages. 


\begin{acknowledgments}
We would like to acknowledge fruitful discussions with Steve Finkelstein and Seiji Fujimoto, our colleagues in UT's Cosmic Frontier Center, as well as with Hui Li, Kohei Inoyashi, Haibo Yu, and James Gurian. The authors acknowledge the Texas Advanced Computing Center (TACC) for providing HPC resources under allocation AST23026. BL gratefully acknowledges the funding of the Royal Society University Research Fellowship and the Deutsche Forschungsgemeinschaft (DFG, German Research Foundation) under Germany's Excellence Strategy EXC 2181/1 - 390900948 (the Heidelberg STRUCTURES Excellence Cluster). MBK acknowledges support from NNSF grants AST-1910346, AST-2108962, and AST-2408247; NASA grant 80NSSC22K0827; HST-GO-16686, HST-AR-17028, HST-AR-17043, JWST-GO-03788, and JWST-AR-06278 from the Space Telescope Science Institute, which is operated by AURA, Inc., under NASA contract NAS5-26555; and from the Samuel T. and Fern Yanagisawa Regents Professorship in Astronomy at UT Austin.

\end{acknowledgments}

%

\vspace{5mm}
\facilities{Lonestar6 and Stampede3 (TACC)}


\software{astropy \citep{2013A&A...558A..33A,2018AJ....156..123A},  
          Colossus \citep{Diemer2018ApJCOLOSSUS}
          }

\begin{figure*}[ht!]
\centering
    \includegraphics[width= \linewidth]{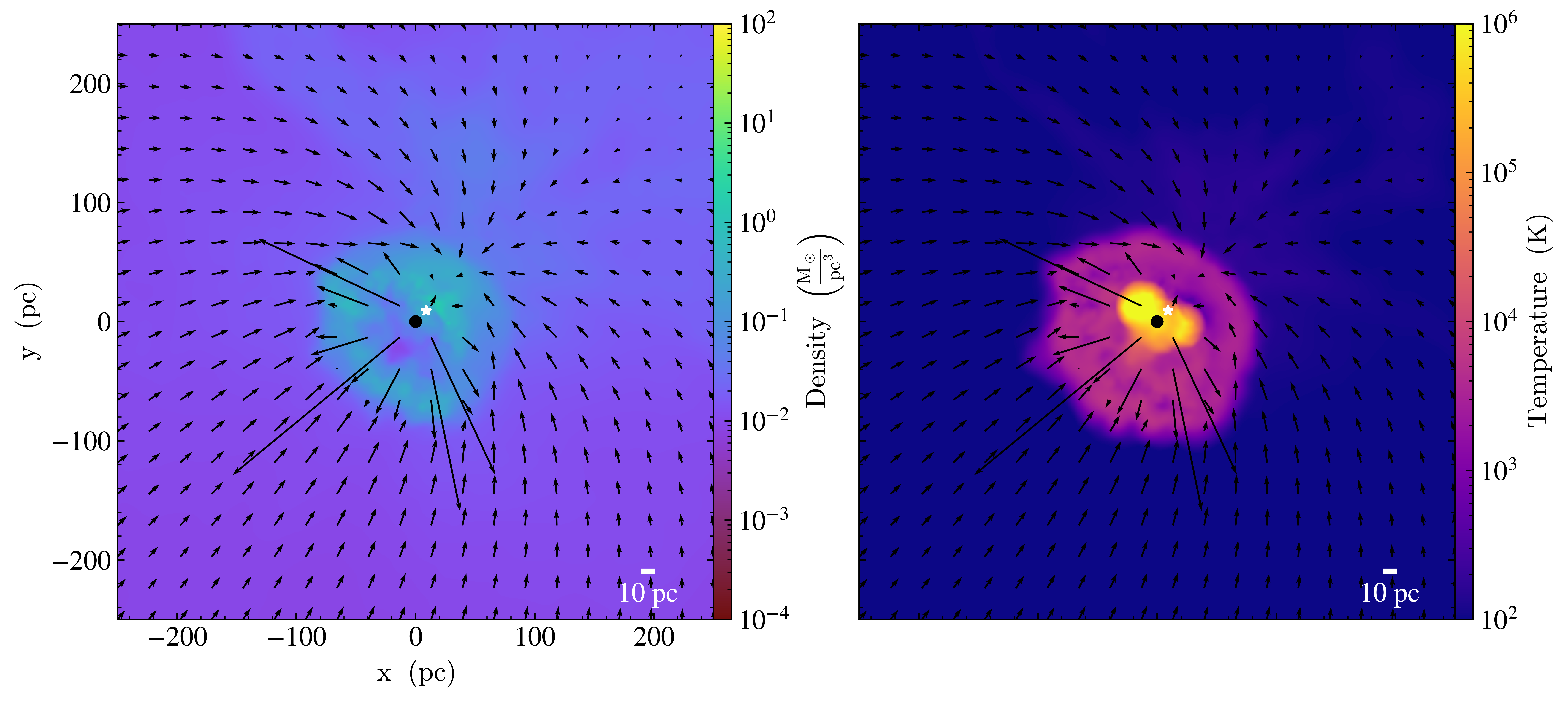}
    \caption{Effect of relative DM-baryon streaming. Output is from the \texttt{PBH\_fd005\_sstr} simulation, which assumes a BH thermal feedback efficiency of $\epsilon_r = 0.5\%$, and a relative streaming velocity of $v_{\rm b\chi} = 1.6~\sigma_{\rm b\chi}$. Similar to Fig.~\ref{fig:SlicePlot}, we show projections of density (\textbf{left panel})
    and temperature (\textbf{right panel}), 
    for the gas surrounding the central PBH within a 500~pc region. The snapshot is taken at $z \simeq 110$, corresponding to the moment when the first collapsing gas particle in the vicinity of the PBH is identified. This ultra-early collapse arises from rare conditions when strong streaming is present (see main text).
 }
    \label{fig:SlicePlotStr}
\end{figure*}
\begin{figure*}[ht!]
\centering
    \includegraphics[width= \linewidth]{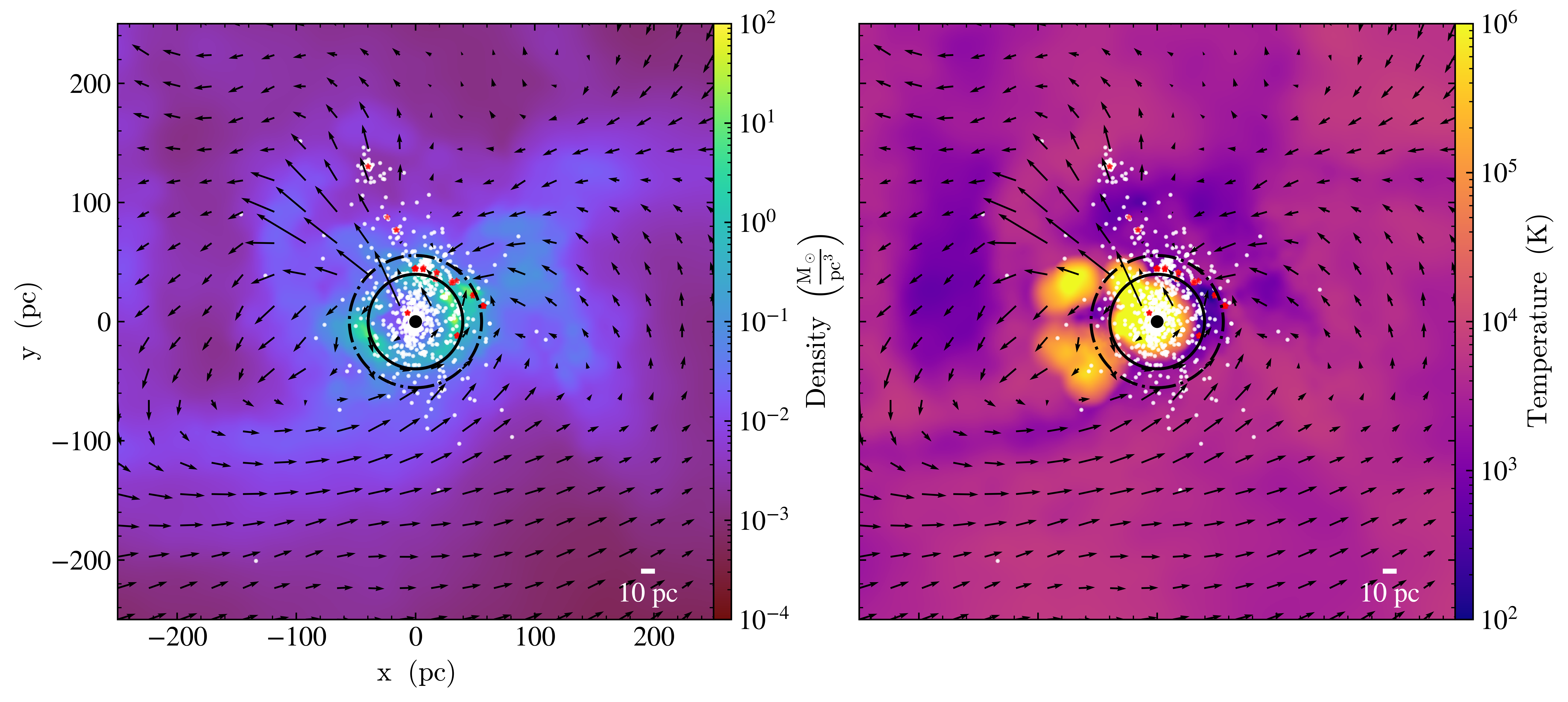}
    \caption{Effect of reduced feedback strength. Similar to Fig.~\ref{fig:SlicePlotFinal}, we present projected density (\textbf{left panel}) and temperature (\textbf{right panel}) distributions, for the gas surrounding the central PBH within a 500~pc region. The snapshot is taken at $z \simeq 12$, corresponding to the formation of star clusters around the PBH. Here, the output is from the \texttt{PBH\_fd0005} simulation, which assumes a reduced BH thermal feedback efficiency of $\epsilon_r = 0.05\%$. For improved visualization, we adjusted the sub-sampling ratio to display one out of every 100 star particles, as compared to Fig.~\ref{fig:SlicePlotFinal}. 
    As can be seen, in this case the PBH-seeded galaxy is able to establish a disk-like configuration for its gaseous and stellar components.
 }
    \label{fig:SlicePlotwf0005}
\end{figure*}

\clearpage
\appendix
Here, we first demonstrate the effect of relative DM-baryon streaming on the onset of first star formation (Fig.~\ref{fig:SlicePlotStr}), where an ultra-early collapse occurs in the vicinity of the PBH in the \texttt{PBH\_fd005\_sstr} simulation with $v_{\rm b\chi} = 1.6~\sigma_{\rm b\chi}$. On the other hand, Fig.~\ref{fig:SlicePlotwf0005} illustrates the formation of star clusters around the PBH under conditions of reduced feedback strength in the \texttt{PBH\_fd0005} simulation with $\epsilon_r=0.05\%$. In this case, the PBH-seeded galaxy develops a disk-like configuration in both its gaseous and stellar components.

\vspace{-5ex}


\clearpage
\bibliography{Main}{}

\begin{thebibliography}{}
\expandafter\ifx\csname natexlab\endcsname\relax\def\natexlab#1{#1}\fi
\providecommand{\url}[1]{\href{#1}{#1}}
\providecommand{\dodoi}[1]{doi:~\href{http://doi.org/#1}{\nolinkurl{#1}}}
\providecommand{\doeprint}[1]{\href{http://ascl.net/#1}{\nolinkurl{http://ascl.net/#1}}}
\providecommand{\doarXiv}[1]{\href{https://arxiv.org/abs/#1}{\nolinkurl{https://arxiv.org/abs/#1}}}

\bibitem[{{Abel} {et~al.}(1997){Abel}, {Anninos}, {Zhang}, \& {Norman}}]{Abel1997NewA....2..181A}
{Abel}, T., {Anninos}, P., {Zhang}, Y., \& {Norman}, M.~L. 1997, \na, 2, 181, \dodoi{10.1016/S1384-1076(97)00010-9}

\bibitem[{{Abel} {et~al.}(2002){Abel}, {Bryan}, \& {Norman}}]{Abel2002}
{Abel}, T., {Bryan}, G.~L., \& {Norman}, M.~L. 2002, Science, 295, 93, \dodoi{10.1126/science.295.5552.93}

\bibitem[{{Adamo} {et~al.}(2024){Adamo}, {Bradley}, {Vanzella}, {Claeyssens}, {Welch}, {Diego}, {Mahler}, {Oguri}, {Sharon}, {Abdurro'uf}, {Hsiao}, {Xu}, {Messa}, {Lassen}, {Zackrisson}, {Brammer}, {Coe}, {Kokorev}, {Ricotti}, {Zitrin}, {Fujimoto}, {Inoue}, {Resseguier}, {Rigby}, {Jim{\'e}nez-Teja}, {Windhorst}, {Hashimoto}, \& {Tamura}}]{Adamo2024}
{Adamo}, A., {Bradley}, L.~D., {Vanzella}, E., {et~al.} 2024, \nat, 632, 513, \dodoi{10.1038/s41586-024-07703-7}

\bibitem[{{Adams} {et~al.}(2024){Adams}, {Conselice}, {Austin}, {Harvey}, {Ferreira}, {Trussler}, {Juod{\v{z}}balis}, {Li}, {Windhorst}, {Cohen}, {Jansen}, {Summers}, {Tompkins}, {Driver}, {Robotham}, {D'Silva}, {Yan}, {Coe}, {Frye}, {Grogin}, {Koekemoer}, {Marshall}, {Pirzkal}, {Ryan}, {Maksym}, {Rutkowski}, {Willmer}, {Hammel}, {Nonino}, {Bhatawdekar}, {Wilkins}, {Bradley}, {Broadhurst}, {Cheng}, {Dole}, {Hathi}, \& {Zitrin}}]{Adams2024}
{Adams}, N.~J., {Conselice}, C.~J., {Austin}, D., {et~al.} 2024, \apj, 965, 169, \dodoi{10.3847/1538-4357/ad2a7b}

\bibitem[{{Afshordi} {et~al.}(2003){Afshordi}, {McDonald}, \& {Spergel}}]{Afshordi2003ApJ...594L..71A}
{Afshordi}, N., {McDonald}, P., \& {Spergel}, D.~N. 2003, \apjl, 594, L71, \dodoi{10.1086/378763}

\bibitem[{{Agarwal} {et~al.}(2014){Agarwal}, {Dalla Vecchia}, {Johnson}, {Khochfar}, \& {Paardekooper}}]{AgarwalDCBH2014MNRAS.443..648A}
{Agarwal}, B., {Dalla Vecchia}, C., {Johnson}, J.~L., {Khochfar}, S., \& {Paardekooper}, J.-P. 2014, \mnras, 443, 648, \dodoi{10.1093/mnras/stu1112}

\bibitem[{{Agarwal} \& {Khochfar}(2015)}]{Agarwal2015MNRAS.446..160A}
{Agarwal}, B., \& {Khochfar}, S. 2015, \mnras, 446, 160, \dodoi{10.1093/mnras/stu1973}

\bibitem[{{Alexander} \& {Natarajan}(2014)}]{Alexander2014Sci...345.1330A}
{Alexander}, T., \& {Natarajan}, P. 2014, Science, 345, 1330, \dodoi{10.1126/science.1251053}

\bibitem[{{Ali-Ha{\"\i}moud} \& {Kamionkowski}(2017)}]{Ali-Haimoud2017PhRvD}
{Ali-Ha{\"\i}moud}, Y., \& {Kamionkowski}, M. 2017, \prd, 95, 043534, \dodoi{10.1103/PhysRevD.95.043534}

\bibitem[{{Arrabal Haro} {et~al.}(2023){Arrabal Haro}, {Dickinson}, {Finkelstein}, {Fujimoto}, {Fern{\'a}ndez}, {Kartaltepe}, {Jung}, {Cole}, {Burgarella}, {Chworowsky}, {Hutchison}, {Morales}, {Papovich}, {Simons}, {Amor{\'\i}n}, {Backhaus}, {Bagley}, {Bisigello}, {Calabr{\`o}}, {Castellano}, {Cleri}, {Dav{\'e}}, {Dekel}, {Ferguson}, {Fontana}, {Gawiser}, {Giavalisco}, {Harish}, {Hathi}, {Hirschmann}, {Holwerda}, {Huertas-Company}, {Koekemoer}, {Larson}, {Lucas}, {Mobasher}, {P{\'e}rez-Gonz{\'a}lez}, {Pirzkal}, {Rose}, {Santini}, {Trump}, {de la Vega}, {Wang}, {Weiner}, {Wilkins}, {Yang}, {Yung}, \& {Zavala}}]{Arrabal2023ApJ...951L..22A}
{Arrabal Haro}, P., {Dickinson}, M., {Finkelstein}, S.~L., {et~al.} 2023, \apjl, 951, L22, \dodoi{10.3847/2041-8213/acdd54}

\bibitem[{{Astropy Collaboration} {et~al.}(2013){Astropy Collaboration}, {Robitaille}, {Tollerud}, {Greenfield}, {Droettboom}, {Bray}, {Aldcroft}, {Davis}, {Ginsburg}, {Price-Whelan}, {Kerzendorf}, {Conley}, {Crighton}, {Barbary}, {Muna}, {Ferguson}, {Grollier}, {Parikh}, {Nair}, {Unther}, {Deil}, {Woillez}, {Conseil}, {Kramer}, {Turner}, {Singer}, {Fox}, {Weaver}, {Zabalza}, {Edwards}, {Azalee Bostroem}, {Burke}, {Casey}, {Crawford}, {Dencheva}, {Ely}, {Jenness}, {Labrie}, {Lim}, {Pierfederici}, {Pontzen}, {Ptak}, {Refsdal}, {Servillat}, \& {Streicher}}]{2013A&A...558A..33A}
{Astropy Collaboration}, {Robitaille}, T.~P., {Tollerud}, E.~J., {et~al.} 2013, \aap, 558, A33, \dodoi{10.1051/0004-6361/201322068}

\bibitem[{{Astropy Collaboration} {et~al.}(2018){Astropy Collaboration}, {Price-Whelan}, {Sip{\H{o}}cz}, {G{\"u}nther}, {Lim}, {Crawford}, {Conseil}, {Shupe}, {Craig}, {Dencheva}, {Ginsburg}, {VanderPlas}, {Bradley}, {P{\'e}rez-Su{\'a}rez}, {de Val-Borro}, {Aldcroft}, {Cruz}, {Robitaille}, {Tollerud}, {Ardelean}, {Babej}, {Bach}, {Bachetti}, {Bakanov}, {Bamford}, {Barentsen}, {Barmby}, {Baumbach}, {Berry}, {Biscani}, {Boquien}, {Bostroem}, {Bouma}, {Brammer}, {Bray}, {Breytenbach}, {Buddelmeijer}, {Burke}, {Calderone}, {Cano Rodr{\'\i}guez}, {Cara}, {Cardoso}, {Cheedella}, {Copin}, {Corrales}, {Crichton}, {D'Avella}, {Deil}, {Depagne}, {Dietrich}, {Donath}, {Droettboom}, {Earl}, {Erben}, {Fabbro}, {Ferreira}, {Finethy}, {Fox}, {Garrison}, {Gibbons}, {Goldstein}, {Gommers}, {Greco}, {Greenfield}, {Groener}, {Grollier}, {Hagen}, {Hirst}, {Homeier}, {Horton}, {Hosseinzadeh}, {Hu}, {Hunkeler}, {Ivezi{\'c}}, {Jain}, {Jenness}, {Kanarek}, {Kendrew}, {Kern}, {Kerzendorf}, {Khvalko}, {King}, {Kirkby}, {Kulkarni},
  {Kumar}, {Lee}, {Lenz}, {Littlefair}, {Ma}, {Macleod}, {Mastropietro}, {McCully}, {Montagnac}, {Morris}, {Mueller}, {Mumford}, {Muna}, {Murphy}, {Nelson}, {Nguyen}, {Ninan}, {N{\"o}the}, {Ogaz}, {Oh}, {Parejko}, {Parley}, {Pascual}, {Patil}, {Patil}, {Plunkett}, {Prochaska}, {Rastogi}, {Reddy Janga}, {Sabater}, {Sakurikar}, {Seifert}, {Sherbert}, {Sherwood-Taylor}, {Shih}, {Sick}, {Silbiger}, {Singanamalla}, {Singer}, {Sladen}, {Sooley}, {Sornarajah}, {Streicher}, {Teuben}, {Thomas}, {Tremblay}, {Turner}, {Terr{\'o}n}, {van Kerkwijk}, {de la Vega}, {Watkins}, {Weaver}, {Whitmore}, {Woillez}, {Zabalza}, \& {Astropy Contributors}}]{2018AJ....156..123A}
{Astropy Collaboration}, {Price-Whelan}, A.~M., {Sip{\H{o}}cz}, B.~M., {et~al.} 2018, \aj, 156, 123, \dodoi{10.3847/1538-3881/aabc4f}

\bibitem[{{Atrio-Barandela}(2022)}]{Atrio2022}
{Atrio-Barandela}, F. 2022, \apj, 939, 69, \dodoi{10.3847/1538-4357/ac983e}

\bibitem[{{Barkana} \& {Loeb}(2001)}]{Barkana2001:FirstStar}
{Barkana}, R., \& {Loeb}, A. 2001, \physrep, 349, 125, \dodoi{10.1016/S0370-1573(01)00019-9}

\bibitem[{{Begelman} {et~al.}(2006){Begelman}, {Volonteri}, \& {Rees}}]{Begelman2006:DCBH}
{Begelman}, M.~C., {Volonteri}, M., \& {Rees}, M.~J. 2006, \mnras, 370, 289, \dodoi{10.1111/j.1365-2966.2006.10467.x}

\bibitem[{{Belotsky} {et~al.}(2019){Belotsky}, {Dokuchaev}, {Eroshenko}, {Esipova}, {Khlopov}, {Khromykh}, {Kirillov}, {Nikulin}, {Rubin}, \& {Svadkovsky}}]{Belotsky2019}
{Belotsky}, K.~M., {Dokuchaev}, V.~I., {Eroshenko}, Y.~N., {et~al.} 2019, Eur. Phys. J. C, 79, 246, \dodoi{10.1140/epjc/s10052-019-6741-4}

\bibitem[{Bernal {et~al.}(2018)Bernal, Raccanelli, Verde, \& Silk}]{Bernal:2017nec}
Bernal, J.~L., Raccanelli, A., Verde, L., \& Silk, J. 2018, JCAP, 05, 017, \dodoi{10.1088/1475-7516/2018/05/017}

\bibitem[{{Bogd{\'a}n} {et~al.}(2024){Bogd{\'a}n}, {Goulding}, {Natarajan}, {Kov{\'a}cs}, {Tremblay}, {Chadayammuri}, {Volonteri}, {Kraft}, {Forman}, {Jones}, {Churazov}, \& {Zhuravleva}}]{Bogdan:2023UHZ1}
{Bogd{\'a}n}, {\'A}., {Goulding}, A.~D., {Natarajan}, P., {et~al.} 2024, Nature Astronomy, 8, 126, \dodoi{10.1038/s41550-023-02111-9}

\bibitem[{Boucenna {et~al.}(2018)Boucenna, Kuhnel, Ohlsson, \& Visinelli}]{Boucenna:2017ghj}
Boucenna, S.~M., Kuhnel, F., Ohlsson, T., \& Visinelli, L. 2018, JCAP, 07, 003, \dodoi{10.1088/1475-7516/2018/07/003}

\bibitem[{Boudaud {et~al.}(2021)Boudaud, Lacroix, Stref, Lavalle, \& Salati}]{Boudaud:2021irr}
Boudaud, M., Lacroix, T., Stref, M., Lavalle, J., \& Salati, P. 2021, JCAP, 08, 053, \dodoi{10.1088/1475-7516/2021/08/053}

\bibitem[{{Boylan-Kolchin}(2023)}]{Boylan2023}
{Boylan-Kolchin}, M. 2023, Nature Astronomy, 7, 731, \dodoi{10.1038/s41550-023-01937-7}

\bibitem[{{Boylan-Kolchin}(2025)}]{boylan2024}
---. 2025, \mnras, 538, 3210, \dodoi{10.1093/mnras/staf471}

\bibitem[{{Bromm}(2013)}]{Bromm2013}
{Bromm}, V. 2013, Reports on Progress in Physics, 76, 112901, \dodoi{10.1088/0034-4885/76/11/112901}

\bibitem[{{Bromm} {et~al.}(2002){Bromm}, {Coppi}, \& {Larson}}]{Bromm2002ApJ...564...23B}
{Bromm}, V., {Coppi}, P.~S., \& {Larson}, R.~B. 2002, \apj, 564, 23, \dodoi{10.1086/323947}

\bibitem[{{Bromm} \& {Loeb}(2003)}]{BrommDCBH2003ApJ...596...34B}
{Bromm}, V., \& {Loeb}, A. 2003, \apj, 596, 34, \dodoi{10.1086/377529}

\bibitem[{{Burns} {et~al.}(2021){Burns}, {Bale}, {Bradley}, {Ahmed}, {Allen}, {Bowman}, {Furlanetto}, {MacDowall}, {Mirocha}, {Nhan}, {Pivovaroff}, {Pulupa}, {Rapetti}, {Slosar}, \& {Tauscher}}]{Burns2021arXiv210305085B}
{Burns}, J., {Bale}, S., {Bradley}, R., {et~al.} 2021, arXiv e-prints, arXiv:2103.05085, \dodoi{10.48550/arXiv.2103.05085}

\bibitem[{{Cammelli} {et~al.}(2025){Cammelli}, {Monaco}, {Tan}, {Singh}, {Fontanot}, {De Lucia}, {Hirschmann}, \& {Xie}}]{Cammelli2025pop3SMBH}
{Cammelli}, V., {Monaco}, P., {Tan}, J.~C., {et~al.} 2025, \mnras, 536, 851, \dodoi{10.1093/mnras/stae2663}

\bibitem[{{Cappelluti} {et~al.}(2022){Cappelluti}, {Hasinger}, \& {Natarajan}}]{Cappelluti2022ApJ}
{Cappelluti}, N., {Hasinger}, G., \& {Natarajan}, P. 2022, \apj, 926, 205, \dodoi{10.3847/1538-4357/ac332d}

\bibitem[{{Carniani} {et~al.}(2024){Carniani}, {Hainline}, {D'Eugenio}, {Eisenstein}, {Jakobsen}, {Witstok}, {Johnson}, {Chevallard}, {Maiolino}, {Helton}, {Willott}, {Robertson}, {Alberts}, {Arribas}, {Baker}, {Bhatawdekar}, {Boyett}, {Bunker}, {Cameron}, {Cargile}, {Charlot}, {Curti}, {Curtis-Lake}, {Egami}, {Giardino}, {Isaak}, {Ji}, {Jones}, {Kumari}, {Maseda}, {Parlanti}, {P{\'e}rez-Gonz{\'a}lez}, {Rawle}, {Rieke}, {Rieke}, {Del Pino}, {Saxena}, {Scholtz}, {Smit}, {Sun}, {Tacchella}, {{\"U}bler}, {Venturi}, {Williams}, \& {Willmer}}]{Carniani2024}
{Carniani}, S., {Hainline}, K., {D'Eugenio}, F., {et~al.} 2024, \nat, 633, 318, \dodoi{10.1038/s41586-024-07860-9}

\bibitem[{{Carr} {et~al.}(2021{\natexlab{a}}){Carr}, {Clesse}, {Garc{\'\i}a-Bellido}, \& {K{\"u}hnel}}]{Carr2021PDU....3100755C}
{Carr}, B., {Clesse}, S., {Garc{\'\i}a-Bellido}, J., \& {K{\"u}hnel}, F. 2021{\natexlab{a}}, Physics of the Dark Universe, 31, 100755, \dodoi{10.1016/j.dark.2020.100755}

\bibitem[{{Carr} {et~al.}(2021{\natexlab{b}}){Carr}, {Kohri}, {Sendouda}, \& {Yokoyama}}]{Carr2021}
{Carr}, B., {Kohri}, K., {Sendouda}, Y., \& {Yokoyama}, J. 2021{\natexlab{b}}, Reports on Progress in Physics, 84, 116902, \dodoi{10.1088/1361-6633/ac1e31}

\bibitem[{{Carr} \& {K{\"u}hnel}(2020)}]{Carr2020ARNPS..70..355C}
{Carr}, B., \& {K{\"u}hnel}, F. 2020, Annual Review of Nuclear and Particle Science, 70, 355, \dodoi{10.1146/annurev-nucl-050520-125911}

\bibitem[{Carr {et~al.}(2021)Carr, Kuhnel, \& Visinelli}]{Carr:2020mqm}
Carr, B., Kuhnel, F., \& Visinelli, L. 2021, Mon. Not. Roy. Astron. Soc., 506, 3648, \dodoi{10.1093/mnras/stab1930}

\bibitem[{{Carr} \& {Silk}(2018)}]{Carr2018MNRAS.478.3756C}
{Carr}, B., \& {Silk}, J. 2018, \mnras, 478, 3756, \dodoi{10.1093/mnras/sty1204}

\bibitem[{{Carr}(1975)}]{Carr1975ApJ...201....1C}
{Carr}, B.~J. 1975, \apj, 201, 1, \dodoi{10.1086/153853}

\bibitem[{{Carr} {et~al.}(2024){Carr}, {Clesse}, {Garc{\'\i}a-Bellido}, {Hawkins}, \& {K{\"u}hnel}}]{2024PhR..1054....1C}
{Carr}, B.~J., {Clesse}, S., {Garc{\'\i}a-Bellido}, J., {Hawkins}, M.~R.~S., \& {K{\"u}hnel}, F. 2024, \physrep, 1054, 1, \dodoi{10.1016/j.physrep.2023.11.005}

\bibitem[{{Carr} \& {Sakellariadou}(1999)}]{Carr1999ApJ...516..195C}
{Carr}, B.~J., \& {Sakellariadou}, M. 1999, \apj, 516, 195, \dodoi{10.1086/307071}

\bibitem[{{Casanueva-Villarreal} {et~al.}(2025){Casanueva-Villarreal}, {Padilla}, {Tissera}, {Liu}, \& {Bromm}}]{Casanueva2025}
{Casanueva-Villarreal}, C., {Padilla}, N., {Tissera}, P.~B., {Liu}, B., \& {Bromm}, V. 2025, arXiv e-prints, arXiv:2505.10706, \dodoi{10.48550/arXiv.2505.10706}

\bibitem[{{Casanueva-Villarreal} {et~al.}(2024){Casanueva-Villarreal}, {Tissera}, {Padilla}, {Liu}, {Bromm}, {Pedrosa}, {Bignone}, \& {Dominguez-Tenreiro}}]{Casanueva-Villarreal2024}
{Casanueva-Villarreal}, C., {Tissera}, P.~B., {Padilla}, N., {et~al.} 2024, \aap, 688, A183, \dodoi{10.1051/0004-6361/202449650}

\bibitem[{{Castellano} {et~al.}(2022){Castellano}, {Fontana}, {Treu}, {Santini}, {Merlin}, {Leethochawalit}, {Trenti}, {Vanzella}, {Mestric}, {Bonchi}, {Belfiori}, {Nonino}, {Paris}, {Polenta}, {Roberts-Borsani}, {Boyett}, {Brada{\v{c}}}, {Calabr{\`o}}, {Glazebrook}, {Grillo}, {Mascia}, {Mason}, {Mercurio}, {Morishita}, {Nanayakkara}, {Pentericci}, {Rosati}, {Vulcani}, {Wang}, \& {Yang}}]{GHz2}
{Castellano}, M., {Fontana}, A., {Treu}, T., {et~al.} 2022, \apjl, 938, L15, \dodoi{10.3847/2041-8213/ac94d0}

\bibitem[{{Castellano} {et~al.}(2023){Castellano}, {Fontana}, {Treu}, {Merlin}, {Santini}, {Bergamini}, {Grillo}, {Rosati}, {Acebron}, {Leethochawalit}, {Paris}, {Bonchi}, {Belfiori}, {Calabr{\`o}}, {Correnti}, {Nonino}, {Polenta}, {Trenti}, {Boyett}, {Brammer}, {Broadhurst}, {Caminha}, {Chen}, {Filippenko}, {Fortuni}, {Glazebrook}, {Mascia}, {Mason}, {Menci}, {Meneghetti}, {Mercurio}, {Metha}, {Morishita}, {Nanayakkara}, {Pentericci}, {Roberts-Borsani}, {Roy}, {Vanzella}, {Vulcani}, {Yang}, \& {Wang}}]{Castellano2023ApJ...948L..14C}
---. 2023, \apjl, 948, L14, \dodoi{10.3847/2041-8213/accea5}

\bibitem[{{Ceverino} \& {Klypin}(2009)}]{Ceverino2009ApJ...695..292C}
{Ceverino}, D., \& {Klypin}, A. 2009, \apj, 695, 292, \dodoi{10.1088/0004-637X/695/1/292}

\bibitem[{{Clesse} \& {Garc{\'\i}a-Bellido}(2018)}]{2018PDU....22..137C}
{Clesse}, S., \& {Garc{\'\i}a-Bellido}, J. 2018, Physics of the Dark Universe, 22, 137, \dodoi{10.1016/j.dark.2018.08.004}

\bibitem[{{Colazo} {et~al.}(2024){Colazo}, {Stasyszyn}, \& {Padilla}}]{Colazo2024}
{Colazo}, P.~E., {Stasyszyn}, F., \& {Padilla}, N. 2024, \aap, 685, L8, \dodoi{10.1051/0004-6361/202449565}

\bibitem[{{Coppola} {et~al.}(2011){Coppola}, {Longo}, {Capitelli}, {Palla}, \& {Galli}}]{CC2011ApJS..193....7C}
{Coppola}, C.~M., {Longo}, S., {Capitelli}, M., {Palla}, F., \& {Galli}, D. 2011, \apjs, 193, 7, \dodoi{10.1088/0067-0049/193/1/7}

\bibitem[{{Dayal}(2024)}]{Dayal2024A&A...690A.182D}
{Dayal}, P. 2024, \aap, 690, A182, \dodoi{10.1051/0004-6361/202451481}

\bibitem[{{Dayal} {et~al.}(2017){Dayal}, {Choudhury}, {Pacucci}, \& {Bromm}}]{Dayal2017}
{Dayal}, P., {Choudhury}, T.~R., {Pacucci}, F., \& {Bromm}, V. 2017, \mnras, 472, 4414, \dodoi{10.1093/mnras/stx2282}

\bibitem[{{De Luca} {et~al.}(2020){De Luca}, {Franciolini}, {Pani}, \& {Riotto}}]{Deluca2020JCAP...06..044D}
{De Luca}, V., {Franciolini}, G., {Pani}, P., \& {Riotto}, A. 2020, \jcap, 2020, 044, \dodoi{10.1088/1475-7516/2020/06/044}

\bibitem[{{De Luca} {et~al.}(2023){De Luca}, {Franciolini}, \& {Riotto}}]{DeLuca2023PhRvL.130q1401D}
{De Luca}, V., {Franciolini}, G., \& {Riotto}, A. 2023, \prl, 130, 171401, \dodoi{10.1103/PhysRevLett.130.171401}

\bibitem[{{De Rossi} \& {Bromm}(2023)}]{DeRossi2023}
{De Rossi}, M.~E., \& {Bromm}, V. 2023, \apjl, 946, L20, \dodoi{10.3847/2041-8213/acc32e}

\bibitem[{{Dekel} {et~al.}(2023){Dekel}, {Sarkar}, {Birnboim}, {Mandelker}, \& {Li}}]{Dekel:2023FFB}
{Dekel}, A., {Sarkar}, K.~C., {Birnboim}, Y., {Mandelker}, N., \& {Li}, Z. 2023, \mnras, 523, 3201, \dodoi{10.1093/mnras/stad1557}

\bibitem[{{D'Eugenio} {et~al.}(2025){D'Eugenio}, {Maiolino}, {Perna}, {Uebler}, {Ji}, {McClymont}, {Koudmani}, {Sijacki}, {Juod{\v{z}}balis}, {Scholtz}, {Bennett}, {Bunker}, {Carniani}, {Charlot}, {Cresci}, {Curtis-Lake}, {Dalla Bont{\`a}}, {Jones}, {Lyu}, {Marconi}, {Mazzolari}, {Nelson}, {Parlanti}, {Robertson}, {Schneider}, {Simmonds}, {Tacchella}, {Venturi}, {Willott}, {Witstok}, \& {Witten}}]{DEugenio2025arXivBlackTHUNDER}
{D'Eugenio}, F., {Maiolino}, R., {Perna}, M., {et~al.} 2025, arXiv e-prints, arXiv:2503.11752, \dodoi{10.48550/arXiv.2503.11752}

\bibitem[{{Diemer}(2018)}]{Diemer2018ApJCOLOSSUS}
{Diemer}, B. 2018, \apjs, 239, 35, \dodoi{10.3847/1538-4365/aaee8c}

\bibitem[{{Donnan} {et~al.}(2024){Donnan}, {McLure}, {Dunlop}, {McLeod}, {Magee}, {Arellano-C{\'o}rdova}, {Barrufet}, {Begley}, {Bowler}, {Carnall}, {Cullen}, {Ellis}, {Fontana}, {Illingworth}, {Grogin}, {Hamadouche}, {Koekemoer}, {Liu}, {Mason}, {Santini}, \& {Stanton}}]{Donnan2024}
{Donnan}, C.~T., {McLure}, R.~J., {Dunlop}, J.~S., {et~al.} 2024, \mnras, 533, 3222, \dodoi{10.1093/mnras/stae2037}

\bibitem[{{Draine} \& {Bertoldi}(1996)}]{DB1996ApJ...468..269D}
{Draine}, B.~T., \& {Bertoldi}, F. 1996, \apj, 468, 269, \dodoi{10.1086/177689}

\bibitem[{{D{\"u}chting}(2004)}]{DuchtingPBH2004PhRvD..70f4015D}
{D{\"u}chting}, N. 2004, \prd, 70, 064015, \dodoi{10.1103/PhysRevD.70.064015}

\bibitem[{Eroshenko(2016)}]{Eroshenko:2016yve}
Eroshenko, Y.~N. 2016, Astron. Lett., 42, 347, \dodoi{10.1134/S1063773716060013}

\bibitem[{{Escriv{\`a}}(2022)}]{Escriva2022}
{Escriv{\`a}}, A. 2022, Universe, 8, 66, \dodoi{10.3390/universe8020066}

\bibitem[{Escriv{\`a} {et~al.}(2024)Escriv{\`a}, Kuhnel, \& Tada}]{ESCRIVA2024261}
Escriv{\`a}, A., Kuhnel, F., \& Tada, Y. 2024, in Black Holes in the Era of Gravitational-Wave Astronomy, ed. M.~A. Sedda, E.~Bortolas, \& M.~Spera (Elsevier), 261--377, \dodoi{https://doi.org/10.1016/B978-0-32-395636-9.00012-8}

\bibitem[{{Finkelstein} {et~al.}(2022){Finkelstein}, {Bagley}, {Haro}, {Dickinson}, {Ferguson}, {Kartaltepe}, {Papovich}, {Burgarella}, {Kocevski}, {Huertas-Company}, {Iyer}, {Koekemoer}, {Larson}, {P{\'e}rez-Gonz{\'a}lez}, {Rose}, {Tacchella}, {Wilkins}, {Chworowsky}, {Medrano}, {Morales}, {Somerville}, {Yung}, {Fontana}, {Giavalisco}, {Grazian}, {Grogin}, {Kewley}, {Kirkpatrick}, {Kurczynski}, {Lotz}, {Pentericci}, {Pirzkal}, {Ravindranath}, {Ryan}, {Trump}, {Yang}, {Almaini}, {Amor{\'\i}n}, {Annunziatella}, {Backhaus}, {Barro}, {Behroozi}, {Bell}, {Bhatawdekar}, {Bisigello}, {Bromm}, {Buat}, {Buitrago}, {Calabr{\`o}}, {Casey}, {Castellano}, {Ch{\'a}vez Ortiz}, {Ciesla}, {Cleri}, {Cohen}, {Cole}, {Cooke}, {Cooper}, {Cooray}, {Costantin}, {Cox}, {Croton}, {Daddi}, {Dav{\'e}}, {de La Vega}, {Dekel}, {Elbaz}, {Estrada-Carpenter}, {Faber}, {Fern{\'a}ndez}, {Finkelstein}, {Freundlich}, {Fujimoto}, {Garc{\'\i}a-Argum{\'a}nez}, {Gardner}, {Gawiser}, {G{\'o}mez-Guijarro}, {Guo}, {Hamblin}, {Hamilton}, {Hathi},
  {Holwerda}, {Hirschmann}, {Hutchison}, {Jaskot}, {Jha}, {Jogee}, {Juneau}, {Jung}, {Kassin}, {Bail}, {Leung}, {Lucas}, {Magnelli}, {Mantha}, {Matharu}, {McGrath}, {McIntosh}, {Merlin}, {Mobasher}, {Newman}, {Nicholls}, {Pandya}, {Rafelski}, {Ronayne}, {Santini}, {Seill{\'e}}, {Shah}, {Shen}, {Simons}, {Snyder}, {Stanway}, {Straughn}, {Teplitz}, {Vanderhoof}, {Vega-Ferrero}, {Wang}, {Weiner}, {Willmer}, {Wuyts}, {Zavala}, \& {Ceers Team}}]{Maisies:2022}
{Finkelstein}, S.~L., {Bagley}, M.~B., {Haro}, P.~A., {et~al.} 2022, \apjl, 940, L55, \dodoi{10.3847/2041-8213/ac966e}

\bibitem[{{Galli} \& {Palla}(1998)}]{Galli1998A&A...335..403G}
{Galli}, D., \& {Palla}, F. 1998, \aap, 335, 403, \dodoi{10.48550/arXiv.astro-ph/9803315}

\bibitem[{{Galli} \& {Palla}(2013)}]{Galli2013ARA&A..51..163G}
---. 2013, \araa, 51, 163, \dodoi{10.1146/annurev-astro-082812-141029}

\bibitem[{{Gardner} {et~al.}(2006){Gardner}, {Mather}, {Clampin}, {Doyon}, {Greenhouse}, {Hammel}, {Hutchings}, {Jakobsen}, {Lilly}, {Long}, {Lunine}, {McCaughrean}, {Mountain}, {Nella}, {Rieke}, {Rieke}, {Rix}, {Smith}, {Sonneborn}, {Stiavelli}, {Stockman}, {Windhorst}, \& {Wright}}]{Gardner:2006JWST}
{Gardner}, J.~P., {Mather}, J.~C., {Clampin}, M., {et~al.} 2006, \ssr, 123, 485, \dodoi{10.1007/s11214-006-8315-7}

\bibitem[{{Goulding} {et~al.}(2023){Goulding}, {Greene}, {Setton}, {Labbe}, {Bezanson}, {Miller}, {Atek}, {Bogd{\'a}n}, {Brammer}, {Chemerynska}, {Cutler}, {Dayal}, {Fudamoto}, {Fujimoto}, {Furtak}, {Kokorev}, {Khullar}, {Leja}, {Marchesini}, {Natarajan}, {Nelson}, {Oesch}, {Pan}, {Papovich}, {Price}, {van Dokkum}, {Wang}, {Weaver}, {Whitaker}, \& {Zitrin}}]{Goulding2023ApJ...955L..24G}
{Goulding}, A.~D., {Greene}, J.~E., {Setton}, D.~J., {et~al.} 2023, \apjl, 955, L24, \dodoi{10.3847/2041-8213/acf7c5}

\bibitem[{{Greene} {et~al.}(2024){Greene}, {Labbe}, {Goulding}, {Furtak}, {Chemerynska}, {Kokorev}, {Dayal}, {Volonteri}, {Williams}, {Wang}, {Setton}, {Burgasser}, {Bezanson}, {Atek}, {Brammer}, {Cutler}, {Feldmann}, {Fujimoto}, {Glazebrook}, {de Graaff}, {Khullar}, {Leja}, {Marchesini}, {Maseda}, {Matthee}, {Miller}, {Naidu}, {Nanayakkara}, {Oesch}, {Pan}, {Papovich}, {Price}, {van Dokkum}, {Weaver}, {Whitaker}, \& {Zitrin}}]{Greene2024}
{Greene}, J.~E., {Labbe}, I., {Goulding}, A.~D., {et~al.} 2024, \apj, 964, 39, \dodoi{10.3847/1538-4357/ad1e5f}

\bibitem[{Guia {et~al.}(2024)Guia, Pacucci, \& Kocevski}]{Guia:2024toq}
Guia, C.~A., Pacucci, F., \& Kocevski, D.~D. 2024, Res. Notes AAS, 8, 207, \dodoi{10.3847/2515-5172/ad7262}

\bibitem[{{Habouzit} {et~al.}(2016){Habouzit}, {Volonteri}, {Latif}, {Dubois}, \& {Peirani}}]{Habouzit2016}
{Habouzit}, M., {Volonteri}, M., {Latif}, M., {Dubois}, Y., \& {Peirani}, S. 2016, \mnras, 463, 529, \dodoi{10.1093/mnras/stw1924}

\bibitem[{{Hahn} \& {Abel}(2011)}]{hahn2011multi}
{Hahn}, O., \& {Abel}, T. 2011, \mnras, 415, 2101, \dodoi{10.1111/j.1365-2966.2011.18820.x}

\bibitem[{Hasinger(2020)}]{Hasinger:2020ptw}
Hasinger, G. 2020, JCAP, 07, 022, \dodoi{10.1088/1475-7516/2020/07/022}

\bibitem[{Hawking(1971)}]{hawking1971gravitationally}
Hawking, S. 1971, Monthly Notices of the Royal Astronomical Society, 152, 75

\bibitem[{{Hawkins}(2022)}]{2022MNRAS.512.5706H}
{Hawkins}, M.~R.~S. 2022, \mnras, 512, 5706, \dodoi{10.1093/mnras/stac863}

\bibitem[{{Hawkins}(2024)}]{2024MNRAS.527.2393H}
---. 2024, \mnras, 527, 2393, \dodoi{10.1093/mnras/stad3346}

\bibitem[{{Hirano} {et~al.}(2018){Hirano}, {Sullivan}, \& {Bromm}}]{Hirano2018}
{Hirano}, S., {Sullivan}, J.~M., \& {Bromm}, V. 2018, \mnras, 473, L6, \dodoi{10.1093/mnrasl/slx146}

\bibitem[{{Hirano} \& {Yoshida}(2024)}]{Hirano2024ApJ...963....2H}
{Hirano}, S., \& {Yoshida}, N. 2024, \apj, 963, 2, \dodoi{10.3847/1538-4357/ad22e0}

\bibitem[{{Hirano} {et~al.}(2015){Hirano}, {Zhu}, {Yoshida}, {Spergel}, \& {Yorke}}]{Hirano2015ApJ...814...18H}
{Hirano}, S., {Zhu}, N., {Yoshida}, N., {Spergel}, D., \& {Yorke}, H.~W. 2015, \apj, 814, 18, \dodoi{10.1088/0004-637X/814/1/18}

\bibitem[{{Hopkins}(2015)}]{Hopkins2015MNRAS.450...53H}
{Hopkins}, P.~F. 2015, \mnras, 450, 53, \dodoi{10.1093/mnras/stv195}

\bibitem[{{Huang} {et~al.}(2024){Huang}, {Wang}, \& {Piao}}]{Huang2024arXiv241005891H}
{Huang}, H.-L., {Wang}, Y.-T., \& {Piao}, Y.-S. 2024, arXiv e-prints, arXiv:2410.05891, \dodoi{10.48550/arXiv.2410.05891}

\bibitem[{{Ibar} {et~al.}(2008){Ibar}, {Cirasuolo}, {Ivison}, {Best}, {Smail}, {Biggs}, {Simpson}, {Dunlop}, {Almaini}, {McLure}, {Foucaud}, \& {Rawlings}}]{Ibar2008MNRAS.386..953I}
{Ibar}, E., {Cirasuolo}, M., {Ivison}, R., {et~al.} 2008, \mnras, 386, 953, \dodoi{10.1111/j.1365-2966.2008.13077.x}

\bibitem[{{Inayoshi} {et~al.}(2022){Inayoshi}, {Harikane}, {Inoue}, {Li}, \& {Ho}}]{Inayoshi2022ApJ...938L..10I}
{Inayoshi}, K., {Harikane}, Y., {Inoue}, A.~K., {Li}, W., \& {Ho}, L.~C. 2022, \apjl, 938, L10, \dodoi{10.3847/2041-8213/ac9310}

\bibitem[{{Inayoshi} \& {Ichikawa}(2024)}]{InayoshiLRD2024ApJ...973L..49I}
{Inayoshi}, K., \& {Ichikawa}, K. 2024, \apjl, 973, L49, \dodoi{10.3847/2041-8213/ad74e2}

\bibitem[{{Inayoshi} {et~al.}(2024){Inayoshi}, {Kashiyama}, {Li}, {Harikane}, {Ichikawa}, \& {Onoue}}]{Inayoshi2024ApJ...966..164I}
{Inayoshi}, K., {Kashiyama}, K., {Li}, W., {et~al.} 2024, \apj, 966, 164, \dodoi{10.3847/1538-4357/ad344c}

\bibitem[{{Inayoshi} {et~al.}(2020){Inayoshi}, {Visbal}, \& {Haiman}}]{Inayoshi:2020}
{Inayoshi}, K., {Visbal}, E., \& {Haiman}, Z. 2020, \araa, 58, 27, \dodoi{10.1146/annurev-astro-120419-014455}

\bibitem[{{Inman} \& {Ali-Ha{\"\i}moud}(2019{\natexlab{a}})}]{Inman2019PhRvD.100h3528I}
{Inman}, D., \& {Ali-Ha{\"\i}moud}, Y. 2019{\natexlab{a}}, \prd, 100, 083528, \dodoi{10.1103/PhysRevD.100.083528}

\bibitem[{{Inman} \& {Ali-Ha{\"\i}moud}(2019{\natexlab{b}})}]{Inman2019}
---. 2019{\natexlab{b}}, \prd, 100, 083528, \dodoi{10.1103/PhysRevD.100.083528}

\bibitem[{{Ito} \& {Omukai}(2024)}]{Ito2024PASJ...76..850I}
{Ito}, M., \& {Omukai}, K. 2024, \pasj, 76, 850, \dodoi{10.1093/pasj/psae054}

\bibitem[{{Jeon} {et~al.}(2025){Jeon}, {Bromm}, {Liu}, \& {Finkelstein}}]{Jeon2025ApJ...979..127J}
{Jeon}, J., {Bromm}, V., {Liu}, B., \& {Finkelstein}, S.~L. 2025, \apj, 979, 127, \dodoi{10.3847/1538-4357/ad9f3a}

\bibitem[{{Jeon} {et~al.}(2023){Jeon}, {Liu}, {Bromm}, \& {Finkelstein}}]{Jeon2023}
{Jeon}, J., {Liu}, B., {Bromm}, V., \& {Finkelstein}, S.~L. 2023, \mnras, 524, 176, \dodoi{10.1093/mnras/stad1877}

\bibitem[{{Jeon} {et~al.}(2012){Jeon}, {Pawlik}, {Greif}, {Glover}, {Bromm}, {Milosavljevi{\'c}}, \& {Klessen}}]{Jeon2012ApJ...754...34J}
{Jeon}, M., {Pawlik}, A.~H., {Greif}, T.~H., {et~al.} 2012, \apj, 754, 34, \dodoi{10.1088/0004-637X/754/1/34}

\bibitem[{{Ji} {et~al.}(2025){Ji}, {Maiolino}, {{\"U}bler}, {Scholtz}, {D'Eugenio}, {Sun}, {Perna}, {Turner}, {Arribas}, {Bennett}, {Bunker}, {Carniani}, {Charlot}, {Cresci}, {Curti}, {Egami}, {Fabian}, {Inayoshi}, {Isobe}, {Jones}, {Juod{\v{z}}balis}, {Kumari}, {Lyu}, {Mazzolari}, {Parlanti}, {Robertson}, {Rodr{\'\i}guez Del Pino}, {Schneider}, {Sijacki}, {Tacchella}, {Trinca}, {Valiante}, {Venturi}, {Volonteri}, {Willott}, {Witten}, \& {Witstok}}]{BlackTHUNDER2025arXiv}
{Ji}, X., {Maiolino}, R., {{\"U}bler}, H., {et~al.} 2025, arXiv e-prints, arXiv:2501.13082, \dodoi{10.48550/arXiv.2501.13082}

\bibitem[{{Johnson} \& {Bromm}(2006)}]{Johnson2006MNRAS.366..247J}
{Johnson}, J.~L., \& {Bromm}, V. 2006, \mnras, 366, 247, \dodoi{10.1111/j.1365-2966.2005.09846.x}

\bibitem[{{Kashlinsky}(2016)}]{Kashlinsky2016ApJ...823L..25K}
{Kashlinsky}, A. 2016, \apjl, 823, L25, \dodoi{10.3847/2041-8205/823/2/L25}

\bibitem[{{Kashlinsky}(2021)}]{Kashlinsky2021PhRvL.126a1101K}
---. 2021, \prl, 126, 011101, \dodoi{10.1103/PhysRevLett.126.011101}

\bibitem[{{Kashlinsky} {et~al.}(2005){Kashlinsky}, {Arendt}, {Mather}, \& {Moseley}}]{2005Natur.438...45K}
{Kashlinsky}, A., {Arendt}, R.~G., {Mather}, J., \& {Moseley}, S.~H. 2005, \nat, 438, 45, \dodoi{10.1038/nature04143}

\bibitem[{{Klessen} \& {Glover}(2023)}]{Klessen:2023FirstStars}
{Klessen}, R.~S., \& {Glover}, S. C.~O. 2023, \araa, 61, 65, \dodoi{10.1146/annurev-astro-071221-053453}

\bibitem[{{Kocevski} {et~al.}(2024){Kocevski}, {Finkelstein}, {Barro}, {Taylor}, {Calabr{\`o}}, {Laloux}, {Buchner}, {Trump}, {Leung}, {Yang}, {Dickinson}, {P{\'e}rez-Gonz{\'a}lez}, {Pacucci}, {Inayoshi}, {Somerville}, {McGrath}, {Akins}, {Bagley}, {Bisigello}, {Bowler}, {Carnall}, {Casey}, {Cheng}, {Cleri}, {Costantin}, {Cullen}, {Davis}, {Donnan}, {Dunlop}, {Ellis}, {Ferguson}, {Fujimoto}, {Fontana}, {Giavalisco}, {Grazian}, {Grogin}, {Hathi}, {Hirschmann}, {Huertas-Company}, {Holwerda}, {Illingworth}, {Juneau}, {Kartaltepe}, {Koekemoer}, {Li}, {Lucas}, {Magee}, {Mason}, {McLeod}, {McLure}, {Napolitano}, {Papovich}, {Pirzkal}, {Rodighiero}, {Santini}, {Wilkins}, \& {Yung}}]{KocevskiLRD2024arXiv240403576K}
{Kocevski}, D.~D., {Finkelstein}, S.~L., {Barro}, G., {et~al.} 2024, arXiv e-prints, arXiv:2404.03576, \dodoi{10.48550/arXiv.2404.03576}

\bibitem[{{Kokorev} {et~al.}(2024){Kokorev}, {Caputi}, {Greene}, {Dayal}, {Trebitsch}, {Cutler}, {Fujimoto}, {Labb{\'e}}, {Miller}, {Iani}, {Navarro-Carrera}, \& {Rinaldi}}]{KokorvLRD2024ApJ...968...38K}
{Kokorev}, V., {Caputi}, K.~I., {Greene}, J.~E., {et~al.} 2024, \apj, 968, 38, \dodoi{10.3847/1538-4357/ad4265}

\bibitem[{{Koopmans} {et~al.}(2021){Koopmans}, {Barkana}, {Bentum}, {Bernardi}, {Boonstra}, {Bowman}, {Burns}, {Chen}, {Datta}, {Falcke}, {Fialkov}, {Gehlot}, {Gurvits}, {Jeli{\'c}}, {Klein-Wolt}, {Lazio}, {Meerburg}, {Mellema}, {Mertens}, {Mesinger}, {Offringa}, {Pritchard}, {Semelin}, {Subrahmanyan}, {Silk}, {Trott}, {Vedantham}, {Verde}, {Zaroubi}, \& {Zarka}}]{Koopmans2021ExA....51.1641K}
{Koopmans}, L. V.~E., {Barkana}, R., {Bentum}, M., {et~al.} 2021, Experimental Astronomy, 51, 1641, \dodoi{10.1007/s10686-021-09743-7}

\bibitem[{{Kormendy} \& {Ho}(2013)}]{Kormendy2013GalaxySMBH}
{Kormendy}, J., \& {Ho}, L.~C. 2013, \araa, 51, 511, \dodoi{10.1146/annurev-astro-082708-101811}

\bibitem[{{Kov{\'a}cs} {et~al.}(2024){Kov{\'a}cs}, {Bogd{\'a}n}, {Natarajan}, {Werner}, {Azadi}, {Volonteri}, {Tremblay}, {Chadayammuri}, {Forman}, {Jones}, \& {Kraft}}]{Kovacs2024ApJ...965L..21K}
{Kov{\'a}cs}, O.~E., {Bogd{\'a}n}, {\'A}., {Natarajan}, P., {et~al.} 2024, \apjl, 965, L21, \dodoi{10.3847/2041-8213/ad391f}

\bibitem[{{Krumholz} {et~al.}(2019){Krumholz}, {McKee}, \& {Bland-Hawthorn}}]{Krumholz2019ARA&A..57..227K}
{Krumholz}, M.~R., {McKee}, C.~F., \& {Bland-Hawthorn}, J. 2019, \araa, 57, 227, \dodoi{10.1146/annurev-astro-091918-104430}

\bibitem[{{Kulkarni} {et~al.}(2022){Kulkarni}, {Visbal}, {Bryan}, \& {Li}}]{Kulkarni2022}
{Kulkarni}, M., {Visbal}, E., {Bryan}, G.~L., \& {Li}, X. 2022, \apjl, 941, L18, \dodoi{10.3847/2041-8213/aca47c}

\bibitem[{{Labb{\'e}} {et~al.}(2023){Labb{\'e}}, {van Dokkum}, {Nelson}, {Bezanson}, {Suess}, {Leja}, {Brammer}, {Whitaker}, {Mathews}, {Stefanon}, \& {Wang}}]{Labbe2023Natur.616..266L}
{Labb{\'e}}, I., {van Dokkum}, P., {Nelson}, E., {et~al.} 2023, \nat, 616, 266, \dodoi{10.1038/s41586-023-05786-2}

\bibitem[{{Labbe} {et~al.}(2025){Labbe}, {Greene}, {Bezanson}, {Fujimoto}, {Furtak}, {Goulding}, {Matthee}, {Naidu}, {Oesch}, {Atek}, {Brammer}, {Chemerynska}, {Coe}, {Cutler}, {Dayal}, {Feldmann}, {Franx}, {Glazebrook}, {Leja}, {Maseda}, {Marchesini}, {Nanayakkara}, {Nelson}, {Pan}, {Papovich}, {Price}, {Suess}, {Wang}, {Weaver}, {Whitaker}, {Williams}, \& {Zitrin}}]{LabbeLRD2025ApJ...978...92L}
{Labbe}, I., {Greene}, J.~E., {Bezanson}, R., {et~al.} 2025, \apj, 978, 92, \dodoi{10.3847/1538-4357/ad3551}

\bibitem[{{Larson} {et~al.}(2023){Larson}, {Finkelstein}, {Kocevski}, {Hutchison}, {Trump}, {Arrabal Haro}, {Bromm}, {Cleri}, {Dickinson}, {Fujimoto}, {Kartaltepe}, {Koekemoer}, {Papovich}, {Pirzkal}, {Tacchella}, {Zavala}, {Bagley}, {Behroozi}, {Champagne}, {Cole}, {Jung}, {Morales}, {Yang}, {Zhang}, {Zitrin}, {Amor{\'\i}n}, {Burgarella}, {Casey}, {Ch{\'a}vez Ortiz}, {Cox}, {Chworowsky}, {Fontana}, {Gawiser}, {Grazian}, {Grogin}, {Harish}, {Hathi}, {Hirschmann}, {Holwerda}, {Juneau}, {Leung}, {Lucas}, {McGrath}, {P{\'e}rez-Gonz{\'a}lez}, {Rigby}, {Seill{\'e}}, {Simons}, {de La Vega}, {Weiner}, {Wilkins}, {Yung}, \& {Ceers Team}}]{Larson_2023_BH}
{Larson}, R.~L., {Finkelstein}, S.~L., {Kocevski}, D.~D., {et~al.} 2023, \apjl, 953, L29, \dodoi{10.3847/2041-8213/ace619}

\bibitem[{{Leung} {et~al.}(2024){Leung}, {Finkelstein}, {P{\'e}rez-Gonz{\'a}lez}, {Morales}, {Taylor}, {Barro}, {Kocevski}, {Akins}, {Carnall}, {Ch{\'a}vez Ortiz}, {Cleri}, {Cullen}, {Donnan}, {Dunlop}, {Ellis}, {Grogin}, {Hirschmann}, {Koekemoer}, {Kokorev}, {Lucas}, {McLeod}, {Papovich}, \& {Yung}}]{Leung2024:LRDarXiv}
{Leung}, G. C.~K., {Finkelstein}, S.~L., {P{\'e}rez-Gonz{\'a}lez}, P.~G., {et~al.} 2024, arXiv e-prints, arXiv:2411.12005.
\newblock \doarXiv{2411.12005}

\bibitem[{{Liu} \& {Bromm}(2018)}]{LiuBromm2018}
{Liu}, B., \& {Bromm}, V. 2018, \mnras, 476, 1826, \dodoi{10.1093/mnras/sty350}

\bibitem[{{Liu} \& {Bromm}(2022)}]{Boyuan2022ApJ}
---. 2022, \apjl, 937, L30, \dodoi{10.3847/2041-8213/ac927f}

\bibitem[{{Liu} \& {Bromm}(2023)}]{Boyuan2023arXiv231204085L}
---. 2023, arXiv e-prints, arXiv:2312.04085, \dodoi{10.48550/arXiv.2312.04085}

\bibitem[{{Liu} {et~al.}(2024{\natexlab{a}}){Liu}, {Gurian}, {Inayoshi}, {Hirano}, {Hosokawa}, {Bromm}, \& {Yoshida}}]{Liu2024MNRAS.534..290L}
{Liu}, B., {Gurian}, J., {Inayoshi}, K., {et~al.} 2024{\natexlab{a}}, \mnras, 534, 290, \dodoi{10.1093/mnras/stae2066}

\bibitem[{{Liu} {et~al.}(2019{\natexlab{a}}){Liu}, {Jaacks}, {Finkelstein}, \& {Bromm}}]{Liu2019}
{Liu}, B., {Jaacks}, J., {Finkelstein}, S.~L., \& {Bromm}, V. 2019{\natexlab{a}}, \mnras, 486, 3617, \dodoi{10.1093/mnras/stz910}

\bibitem[{{Liu} {et~al.}(2019{\natexlab{b}}){Liu}, {Schauer}, \& {Bromm}}]{Liu2019bdms}
{Liu}, B., {Schauer}, A. T.~P., \& {Bromm}, V. 2019{\natexlab{b}}, \mnras, 487, 4711, \dodoi{10.1093/mnras/stz1583}

\bibitem[{{Liu} {et~al.}(2020){Liu}, {Schauer}, \& {Bromm}}]{Liu2020}
---. 2020, \mnras, 495, 1700, \dodoi{10.1093/mnras/staa1307}

\bibitem[{{Liu} {et~al.}(2024{\natexlab{b}}){Liu}, {Sibony}, {Meynet}, \& {Bromm}}]{Liu2024che}
{Liu}, B., {Sibony}, Y., {Meynet}, G., \& {Bromm}, V. 2024{\natexlab{b}}, arXiv e-prints, arXiv:2412.02002, \dodoi{10.48550/arXiv.2412.02002}

\bibitem[{{Liu} {et~al.}(2022){Liu}, {Zhang}, \& {Bromm}}]{Boyuan2022MNRAS.514.2376L}
{Liu}, B., {Zhang}, S., \& {Bromm}, V. 2022, \mnras, 514, 2376, \dodoi{10.1093/mnras/stac1472}

\bibitem[{{Lodato} \& {Natarajan}(2006)}]{Lodato:2006DCBH}
{Lodato}, G., \& {Natarajan}, P. 2006, \mnras, 371, 1813, \dodoi{10.1111/j.1365-2966.2006.10801.x}

\bibitem[{{Loeb} \& {Rasio}(1994)}]{Loeb1994ApJ...432...52L}
{Loeb}, A., \& {Rasio}, F.~A. 1994, \apj, 432, 52, \dodoi{10.1086/174548}

\bibitem[{{Lu} {et~al.}(2021){Lu}, {Takhistov}, {Gelmini}, {Hayashi}, {Inoue}, \& {Kusenko}}]{Lu2021}
{Lu}, P., {Takhistov}, V., {Gelmini}, G.~B., {et~al.} 2021, \apjl, 908, L23, \dodoi{10.3847/2041-8213/abdcb6}

\bibitem[{{LVK Collaboration}(2023)}]{2023MNRAS.526.6234L}
{LVK Collaboration}. 2023, \mnras, 526, 6234, \dodoi{10.1093/mnras/stad3120}

\bibitem[{{Mack} {et~al.}(2007){Mack}, {Ostriker}, \& {Ricotti}}]{Mack2007ApJ}
{Mack}, K.~J., {Ostriker}, J.~P., \& {Ricotti}, M. 2007, \apj, 665, 1277, \dodoi{10.1086/518998}

\bibitem[{{Maiolino} {et~al.}(2024{\natexlab{a}}){Maiolino}, {Scholtz}, {Curtis-Lake}, {Carniani}, {Baker}, {de Graaff}, {Tacchella}, {{\"U}bler}, {D'Eugenio}, {Witstok}, {Curti}, {Arribas}, {Bunker}, {Charlot}, {Chevallard}, {Eisenstein}, {Egami}, {Ji}, {Jones}, {Lyu}, {Rawle}, {Robertson}, {Rujopakarn}, {Perna}, {Sun}, {Venturi}, {Williams}, \& {Willott}}]{Maiolino2024A&A}
{Maiolino}, R., {Scholtz}, J., {Curtis-Lake}, E., {et~al.} 2024{\natexlab{a}}, \aap, 691, A145, \dodoi{10.1051/0004-6361/202347640}

\bibitem[{{Maiolino} {et~al.}(2024{\natexlab{b}}){Maiolino}, {Scholtz}, {Witstok}, {Carniani}, {D'Eugenio}, {de Graaff}, {{\"U}bler}, {Tacchella}, {Curtis-Lake}, {Arribas}, {Bunker}, {Charlot}, {Chevallard}, {Curti}, {Looser}, {Maseda}, {Rawle}, {Rodr{\'\i}guez del Pino}, {Willott}, {Egami}, {Eisenstein}, {Hainline}, {Robertson}, {Williams}, {Willmer}, {Baker}, {Boyett}, {DeCoursey}, {Fabian}, {Helton}, {Ji}, {Jones}, {Kumari}, {Laporte}, {Nelson}, {Perna}, {Sandles}, {Shivaei}, \& {Sun}}]{Maiolino2024}
{Maiolino}, R., {Scholtz}, J., {Witstok}, J., {et~al.} 2024{\natexlab{b}}, \nat, 627, 59, \dodoi{10.1038/s41586-024-07052-5}

\bibitem[{{Maiolino} {et~al.}(2024{\natexlab{c}}){Maiolino}, {Scholtz}, {Witstok}, {Carniani}, {D'Eugenio}, {de Graaff}, {{\"U}bler}, {Tacchella}, {Curtis-Lake}, {Arribas}, {Bunker}, {Charlot}, {Chevallard}, {Curti}, {Looser}, {Maseda}, {Rawle}, {Rodr{\'\i}guez del Pino}, {Willott}, {Egami}, {Eisenstein}, {Hainline}, {Robertson}, {Williams}, {Willmer}, {Baker}, {Boyett}, {DeCoursey}, {Fabian}, {Helton}, {Ji}, {Jones}, {Kumari}, {Laporte}, {Nelson}, {Perna}, {Sandles}, {Shivaei}, \& {Sun}}]{Maiolino2024Natur.627...59M}
---. 2024{\natexlab{c}}, \nat, 627, 59, \dodoi{10.1038/s41586-024-07052-5}

\bibitem[{{Matteri} {et~al.}(2025){Matteri}, {Pallottini}, \& {Ferrara}}]{Matteri:2025klg}
{Matteri}, A., {Pallottini}, A., \& {Ferrara}, A. 2025, \aap, 697, A65, \dodoi{10.1051/0004-6361/202553701}

\bibitem[{{Mediavilla} {et~al.}(2017){Mediavilla}, {Jim{\'e}nez-Vicente}, {Mu{\~n}oz}, {Vives-Arias}, \& {Calder{\'o}n-Infante}}]{2017ApJ...836L..18M}
{Mediavilla}, E., {Jim{\'e}nez-Vicente}, J., {Mu{\~n}oz}, J.~A., {Vives-Arias}, H., \& {Calder{\'o}n-Infante}, J. 2017, \apjl, 836, L18, \dodoi{10.3847/2041-8213/aa5dab}

\bibitem[{{Meszaros}(1975)}]{Meszaros1975A&A....38....5M}
{Meszaros}, P. 1975, \aap, 38, 5

\bibitem[{{Milosavljevi{\'c}} {et~al.}(2009){Milosavljevi{\'c}}, {Bromm}, {Couch}, \& {Oh}}]{Milosavljevic2009}
{Milosavljevi{\'c}}, M., {Bromm}, V., {Couch}, S.~M., \& {Oh}, S.~P. 2009, \apj, 698, 766, \dodoi{10.1088/0004-637X/698/1/766}

\bibitem[{{Morr{\'a}s} {et~al.}(2023){Morr{\'a}s}, {Nu{\~n}o Siles}, {Garc{\'\i}a-Bellido}, {Ruiz Morales}, {Men{\'e}ndez-V{\'a}zquez}, {Karathanasis}, {Martinovic}, {Phukon}, {Clesse}, {Mart{\'\i}nez}, \& {Sakellariadou}}]{2023PDU....4201285M}
{Morr{\'a}s}, G., {Nu{\~n}o Siles}, J.~F., {Garc{\'\i}a-Bellido}, J., {et~al.} 2023, Physics of the Dark Universe, 42, 101285, \dodoi{10.1016/j.dark.2023.101285}

\bibitem[{{Naidu} {et~al.}(2022){Naidu}, {Oesch}, {van Dokkum}, {Nelson}, {Suess}, {Brammer}, {Whitaker}, {Illingworth}, {Bouwens}, {Tacchella}, {Matthee}, {Allen}, {Bezanson}, {Conroy}, {Labbe}, {Leja}, {Leonova}, {Magee}, {Price}, {Setton}, {Strait}, {Stefanon}, {Toft}, {Weaver}, \& {Weibel}}]{GLASSz13}
{Naidu}, R.~P., {Oesch}, P.~A., {van Dokkum}, P., {et~al.} 2022, \apjl, 940, L14, \dodoi{10.3847/2041-8213/ac9b22}

\bibitem[{{Napolitano} {et~al.}(2024){Napolitano}, {Castellano}, {Pentericci}, {Vignali}, {Gilli}, {Fontana}, {Santini}, {Treu}, {Calabr{\`o}}, {Llerena}, {Piconcelli}, {Zappacosta}, {Mascia}, {Bergamini}, {Bakx}, {Dickinson}, {Glazebrook}, {Henry}, {Leethochawalit}, {Mazzolari}, {Merlin}, {Morishita}, {Nanayakkara}, {Paris}, {Puccetti}, {Roberts-Borsani}, {Rojas Ruiz}, {Vanzella}, {Vito}, {Vulcani}, {Wang}, {Yoon}, \& {Zavala}}]{GHZ9Napolitano2024arXiv241018763N}
{Napolitano}, L., {Castellano}, M., {Pentericci}, L., {et~al.} 2024, arXiv e-prints, arXiv:2410.18763, \dodoi{10.48550/arXiv.2410.18763}

\bibitem[{{Natarajan} {et~al.}(2024){Natarajan}, {Pacucci}, {Ricarte}, {Bogd{\'a}n}, {Goulding}, \& {Cappelluti}}]{Natarajan:2023UHZ1}
{Natarajan}, P., {Pacucci}, F., {Ricarte}, A., {et~al.} 2024, \apjl, 960, L1, \dodoi{10.3847/2041-8213/ad0e76}

\bibitem[{{Negri} \& {Volonteri}(2017)}]{Negri2017MNRAS.467.3475N}
{Negri}, A., \& {Volonteri}, M. 2017, \mnras, 467, 3475, \dodoi{10.1093/mnras/stx362}

\bibitem[{{Neumayer} {et~al.}(2020){Neumayer}, {Seth}, \& {B{\"o}ker}}]{Neumayer2020A&ARv..28....4N}
{Neumayer}, N., {Seth}, A., \& {B{\"o}ker}, T. 2020, \aapr, 28, 4, \dodoi{10.1007/s00159-020-00125-0}

\bibitem[{{Niikura} {et~al.}(2019){Niikura}, {Takada}, {Yokoyama}, {Sumi}, \& {Masaki}}]{2019PhRvD..99h3503N}
{Niikura}, H., {Takada}, M., {Yokoyama}, S., {Sumi}, T., \& {Masaki}, S. 2019, \prd, 99, 083503, \dodoi{10.1103/PhysRevD.99.083503}

\bibitem[{{O'Brennan} {et~al.}(2025){O'Brennan}, {Regan}, {Brennan}, {McCaffrey}, {Wise}, {Visbal}, \& {Norman}}]{OBrennan2025}
{O'Brennan}, H., {Regan}, J.~A., {Brennan}, J., {et~al.} 2025, arXiv e-prints, arXiv:2502.00574, \dodoi{10.48550/arXiv.2502.00574}

\bibitem[{{Omukai} {et~al.}(2005){Omukai}, {Tsuribe}, {Schneider}, \& {Ferrara}}]{Omukai2005ApJ...626..627O}
{Omukai}, K., {Tsuribe}, T., {Schneider}, R., \& {Ferrara}, A. 2005, \apj, 626, 627, \dodoi{10.1086/429955}

\bibitem[{{Pacucci} {et~al.}(2023){Pacucci}, {Nguyen}, {Carniani}, {Maiolino}, \& {Fan}}]{PacucciLRD2023ApJ...957L...3P}
{Pacucci}, F., {Nguyen}, B., {Carniani}, S., {Maiolino}, R., \& {Fan}, X. 2023, \apjl, 957, L3, \dodoi{10.3847/2041-8213/ad0158}

\bibitem[{{Padmanabhan} \& {Loeb}(2023)}]{Padmanbha2023}
{Padmanabhan}, H., \& {Loeb}, A. 2023, \apjl, 953, L4, \dodoi{10.3847/2041-8213/acea7a}

\bibitem[{{Pezzulli} {et~al.}(2016){Pezzulli}, {Valiante}, \& {Schneider}}]{Pezzulli2016MNRAS.458.3047P}
{Pezzulli}, E., {Valiante}, R., \& {Schneider}, R. 2016, \mnras, 458, 3047, \dodoi{10.1093/mnras/stw505}

\bibitem[{{Phukon} {et~al.}(2021){Phukon}, {Baltus}, {Caudill}, {Clesse}, {Depasse}, {Fays}, {Fong}, {Kapadia}, {Magee}, \& {Tanasijczuk}}]{2021arXiv210511449P}
{Phukon}, K.~S., {Baltus}, G., {Caudill}, S., {et~al.} 2021, arXiv e-prints, arXiv:2105.11449, \dodoi{10.48550/arXiv.2105.11449}

\bibitem[{{Planck Collaboration} {et~al.}(2020){Planck Collaboration}, {Aghanim}, {Akrami}, {Ashdown}, {Aumont}, {Baccigalupi}, {Ballardini}, {Banday}, {Barreiro}, {Bartolo}, {Basak}, {Battye}, {Benabed}, {Bernard}, {Bersanelli}, {Bielewicz}, {Bock}, {Bond}, {Borrill}, {Bouchet}, {Boulanger}, {Bucher}, {Burigana}, {Butler}, {Calabrese}, {Cardoso}, {Carron}, {Challinor}, {Chiang}, {Chluba}, {Colombo}, {Combet}, {Contreras}, {Crill}, {Cuttaia}, {de Bernardis}, {de Zotti}, {Delabrouille}, {Delouis}, {Di Valentino}, {Diego}, {Dor{\'e}}, {Douspis}, {Ducout}, {Dupac}, {Dusini}, {Efstathiou}, {Elsner}, {En{\ss}lin}, {Eriksen}, {Fantaye}, {Farhang}, {Fergusson}, {Fernandez-Cobos}, {Finelli}, {Forastieri}, {Frailis}, {Fraisse}, {Franceschi}, {Frolov}, {Galeotta}, {Galli}, {Ganga}, {G{\'e}nova-Santos}, {Gerbino}, {Ghosh}, {Gonz{\'a}lez-Nuevo}, {G{\'o}rski}, {Gratton}, {Gruppuso}, {Gudmundsson}, {Hamann}, {Handley}, {Hansen}, {Herranz}, {Hildebrandt}, {Hivon}, {Huang}, {Jaffe}, {Jones}, {Karakci}, {Keih{\"a}nen},
  {Keskitalo}, {Kiiveri}, {Kim}, {Kisner}, {Knox}, {Krachmalnicoff}, {Kunz}, {Kurki-Suonio}, {Lagache}, {Lamarre}, {Lasenby}, {Lattanzi}, {Lawrence}, {Le Jeune}, {Lemos}, {Lesgourgues}, {Levrier}, {Lewis}, {Liguori}, {Lilje}, {Lilley}, {Lindholm}, {L{\'o}pez-Caniego}, {Lubin}, {Ma}, {Mac{\'\i}as-P{\'e}rez}, {Maggio}, {Maino}, {Mandolesi}, {Mangilli}, {Marcos-Caballero}, {Maris}, {Martin}, {Martinelli}, {Mart{\'\i}nez-Gonz{\'a}lez}, {Matarrese}, {Mauri}, {McEwen}, {Meinhold}, {Melchiorri}, {Mennella}, {Migliaccio}, {Millea}, {Mitra}, {Miville-Desch{\^e}nes}, {Molinari}, {Montier}, {Morgante}, {Moss}, {Natoli}, {N{\o}rgaard-Nielsen}, {Pagano}, {Paoletti}, {Partridge}, {Patanchon}, {Peiris}, {Perrotta}, {Pettorino}, {Piacentini}, {Polastri}, {Polenta}, {Puget}, {Rachen}, {Reinecke}, {Remazeilles}, {Renzi}, {Rocha}, {Rosset}, {Roudier}, {Rubi{\~n}o-Mart{\'\i}n}, {Ruiz-Granados}, {Salvati}, {Sandri}, {Savelainen}, {Scott}, {Shellard}, {Sirignano}, {Sirri}, {Spencer}, {Sunyaev}, {Suur-Uski}, {Tauber}, {Tavagnacco},
  {Tenti}, {Toffolatti}, {Tomasi}, {Trombetti}, {Valenziano}, {Valiviita}, {Van Tent}, {Vibert}, {Vielva}, {Villa}, {Vittorio}, {Wandelt}, {Wehus}, {White}, {White}, {Zacchei}, \& {Zonca}}]{Plank2020A&A...641A...6P}
{Planck Collaboration}, {Aghanim}, N., {Akrami}, Y., {et~al.} 2020, \aap, 641, A6, \dodoi{10.1051/0004-6361/201833910}

\bibitem[{{Qin} {et~al.}(2024){Qin}, {Mu{\~n}oz}, {Liu}, \& {Slatyer}}]{Qin2024}
{Qin}, W., {Mu{\~n}oz}, J.~B., {Liu}, H., \& {Slatyer}, T.~R. 2024, \prd, 109, 103026, \dodoi{10.1103/PhysRevD.109.103026}

\bibitem[{{Regan} \& {Volonteri}(2024)}]{Regan2024}
{Regan}, J., \& {Volonteri}, M. 2024, The Open Journal of Astrophysics, 7, 72, \dodoi{10.33232/001c.123239}

\bibitem[{{Regan} {et~al.}(2019){Regan}, {Downes}, {Volonteri}, {Beckmann}, {Lupi}, {Trebitsch}, \& {Dubois}}]{Regan2019}
{Regan}, J.~A., {Downes}, T.~P., {Volonteri}, M., {et~al.} 2019, \mnras, 486, 3892, \dodoi{10.1093/mnras/stz1045}

\bibitem[{{Reines} \& {Volonteri}(2015)}]{Reines2015ApJ...813...82R}
{Reines}, A.~E., \& {Volonteri}, M. 2015, \apj, 813, 82, \dodoi{10.1088/0004-637X/813/2/82}

\bibitem[{{Ricotti}(2007)}]{Ricotti2007ApJI}
{Ricotti}, M. 2007, \apj, 662, 53, \dodoi{10.1086/516562}

\bibitem[{{Ricotti} {et~al.}(2008){Ricotti}, {Ostriker}, \& {Mack}}]{Ricotti2008ApJII}
{Ricotti}, M., {Ostriker}, J.~P., \& {Mack}, K.~J. 2008, \apj, 680, 829, \dodoi{10.1086/587831}

\bibitem[{{Robertson}(2022)}]{Robertson2022}
{Robertson}, B.~E. 2022, \araa, 60, 121, \dodoi{10.1146/annurev-astro-120221-044656}

\bibitem[{Salati \& Lavalle(2025)}]{Salati:2025hpd}
Salati, P., \& Lavalle, J. 2025, EPJ Web Conf., 319, 03001, \dodoi{10.1051/epjconf/202531903001}

\bibitem[{{Schauer} {et~al.}(2023){Schauer}, {Boylan-Kolchin}, {Colston}, {Sameie}, {Bromm}, {Bullock}, \& {Wetzel}}]{Schauer2023}
{Schauer}, A. T.~P., {Boylan-Kolchin}, M., {Colston}, K., {et~al.} 2023, \apj, 950, 20, \dodoi{10.3847/1538-4357/accc2c}

\bibitem[{{Schauer} {et~al.}(2019){Schauer}, {Glover}, {Klessen}, \& {Ceverino}}]{Schauer2019MNRAS.484.3510S}
{Schauer}, A. T.~P., {Glover}, S. C.~O., {Klessen}, R.~S., \& {Ceverino}, D. 2019, \mnras, 484, 3510, \dodoi{10.1093/mnras/stz013}

\bibitem[{{Scholtz} {et~al.}(2024){Scholtz}, {Witten}, {Laporte}, {{\"U}bler}, {Perna}, {Maiolino}, {Arribas}, {Baker}, {Bennett}, {D'Eugenio}, {Simmonds}, {Tacchella}, {Witstok}, {Bunker}, {Carniani}, {Charlot}, {Cresci}, {Curtis-Lake}, {Eisenstein}, {Kumari}, {Robertson}, {Rodr{\'\i}guez Del Pino}, {Smit}, {Venturi}, {Williams}, \& {Willmer}}]{Scholtz2024A&A...687A.283S}
{Scholtz}, J., {Witten}, C., {Laporte}, N., {et~al.} 2024, \aap, 687, A283, \dodoi{10.1051/0004-6361/202347187}

\bibitem[{{Shen} {et~al.}(2023){Shen}, {Vogelsberger}, {Boylan-Kolchin}, {Tacchella}, \& {Kannan}}]{Shen2023}
{Shen}, X., {Vogelsberger}, M., {Boylan-Kolchin}, M., {Tacchella}, S., \& {Kannan}, R. 2023, \mnras, 525, 3254, \dodoi{10.1093/mnras/stad2508}

\bibitem[{{Shen} {et~al.}(2024){Shen}, {Vogelsberger}, {Boylan-Kolchin}, {Tacchella}, \& {Naidu}}]{Shen2024}
{Shen}, X., {Vogelsberger}, M., {Boylan-Kolchin}, M., {Tacchella}, S., \& {Naidu}, R.~P. 2024, \mnras, 533, 3923, \dodoi{10.1093/mnras/stae1932}

\bibitem[{{Silk} {et~al.}(2024){Silk}, {Begelman}, {Norman}, {Nusser}, \& {Wyse}}]{Silk2024ApJ...961L..39S}
{Silk}, J., {Begelman}, M.~C., {Norman}, C., {Nusser}, A., \& {Wyse}, R. F.~G. 2024, \apjl, 961, L39, \dodoi{10.3847/2041-8213/ad1bf0}

\bibitem[{{Smirnov} {et~al.}(2024){Smirnov}, {Goobar}, {Linden}, \& {M{\"o}rtsell}}]{2024PhRvL.132o1401S}
{Smirnov}, J., {Goobar}, A., {Linden}, T., \& {M{\"o}rtsell}, E. 2024, \prl, 132, 151401, \dodoi{10.1103/PhysRevLett.132.151401}

\bibitem[{{Smith} \& {Bromm}(2019)}]{Smith2019:BHreview}
{Smith}, A., \& {Bromm}, V. 2019, Contemporary Physics, 60, 111, \dodoi{10.1080/00107514.2019.1615715}

\bibitem[{{Springel}(2005)}]{springel2005cosmological}
{Springel}, V. 2005, \mnras, 364, 1105, \dodoi{10.1111/j.1365-2966.2005.09655.x}

\bibitem[{{Stacy} {et~al.}(2011){Stacy}, {Bromm}, \& {Loeb}}]{Stacy2011}
{Stacy}, A., {Bromm}, V., \& {Loeb}, A. 2011, \apjl, 730, L1, \dodoi{10.1088/2041-8205/730/1/L1}

\bibitem[{{Su} {et~al.}(2023){Su}, {Li}, \& {Feng}}]{Su2023}
{Su}, B.-Y., {Li}, N., \& {Feng}, L. 2023, arXiv e-prints, arXiv:2306.05364, \dodoi{10.48550/arXiv.2306.05364}

\bibitem[{{Sugimura} {et~al.}(2014){Sugimura}, {Omukai}, \& {Inoue}}]{Sugimura2014MNRAS.445..544S}
{Sugimura}, K., {Omukai}, K., \& {Inoue}, A.~K. 2014, \mnras, 445, 544, \dodoi{10.1093/mnras/stu1778}

\bibitem[{{Sullivan} {et~al.}(2018){Sullivan}, {Hirano}, \& {Bromm}}]{Sullivan2018}
{Sullivan}, J.~M., {Hirano}, S., \& {Bromm}, V. 2018, \mnras, 481, L69, \dodoi{10.1093/mnrasl/sly164}

\bibitem[{{Tacchella} {et~al.}(2023){Tacchella}, {Eisenstein}, {Hainline}, {Johnson}, {Baker}, {Helton}, {Robertson}, {Suess}, {Chen}, {Nelson}, {Pusk{\'a}s}, {Sun}, {Alberts}, {Egami}, {Hausen}, {Rieke}, {Rieke}, {Shivaei}, {Williams}, {Willmer}, {Bunker}, {Cameron}, {Carniani}, {Charlot}, {Curti}, {Curtis-Lake}, {Looser}, {Maiolino}, {Maseda}, {Rawle}, {Rix}, {Smit}, {{\"U}bler}, {Willott}, {Witstok}, {Baum}, {Bhatawdekar}, {Boyett}, {Danhaive}, {de Graaff}, {Endsley}, {Ji}, {Lyu}, {Sandles}, {Saxena}, {Scholtz}, {Topping}, \& {Whitler}}]{Tacchella_2023}
{Tacchella}, S., {Eisenstein}, D.~J., {Hainline}, K., {et~al.} 2023, \apj, 952, 74, \dodoi{10.3847/1538-4357/acdbc6}

\bibitem[{{Takhistov} {et~al.}(2022){Takhistov}, {Lu}, {Gelmini}, {Hayashi}, {Inoue}, \& {Kusenko}}]{Takhistov2022}
{Takhistov}, V., {Lu}, P., {Gelmini}, G.~B., {et~al.} 2022, \jcap, 2022, 017, \dodoi{10.1088/1475-7516/2022/03/017}

\bibitem[{{Taylor} {et~al.}(2024){Taylor}, {Finkelstein}, {Kocevski}, {Jeon}, {Bromm}, {Amorin}, {Arrabal Haro}, {Backhaus}, {Bagley}, {Ba{\~n}ados}, {Bhatawdekar}, {Brooks}, {Calabro}, {Chavez Ortiz}, {Cheng}, {Cleri}, {Cole}, {Davis}, {Dickinson}, {Donnan}, {Dunlop}, {Ellis}, {Fernandez}, {Fontana}, {Fujimoto}, {Giavalisco}, {Grazian}, {Guo}, {Hathi}, {Holwerda}, {Hirschmann}, {Inayoshi}, {Kartaltepe}, {Khusanova}, {Koekemoer}, {Kokorev}, {Larson}, {Leung}, {Lucas}, {McLeod}, {Napolitano}, {Onoue}, {Pacucci}, {Papovich}, {P{\'e}rez-Gonz{\'a}lez}, {Pirzkal}, {Somerville}, {Trump}, {Wilkins}, {Yung}, \& {Zhang}}]{Taylor2024}
{Taylor}, A.~J., {Finkelstein}, S.~L., {Kocevski}, D.~D., {et~al.} 2024, arXiv e-prints, arXiv:2409.06772, \dodoi{10.48550/arXiv.2409.06772}

\bibitem[{Tremmel {et~al.}(2015)Tremmel, Governato, Volonteri, \& Quinn}]{tremmel2015off}
Tremmel, M., Governato, F., Volonteri, M., \& Quinn, T.~R. 2015, \mnras, 451, 1868

\bibitem[{Tremmel {et~al.}(2017)Tremmel, Karcher, Governato, Volonteri, Quinn, Pontzen, Anderson, \& Bellovary}]{tremmel2017romulus}
Tremmel, M., Karcher, M., Governato, F., {et~al.} 2017, \mnras, 470, 1121

\bibitem[{{Trinca} {et~al.}(2024){Trinca}, {Schneider}, {Valiante}, {Graziani}, {Ferrotti}, {Omukai}, \& {Chon}}]{Trinca2024}
{Trinca}, A., {Schneider}, R., {Valiante}, R., {et~al.} 2024, \mnras, 529, 3563, \dodoi{10.1093/mnras/stae651}

\bibitem[{{Wang} {et~al.}(2024){Wang}, {Leja}, {Atek}, {Labb{\'e}}, {Li}, {Bezanson}, {Brammer}, {Cutler}, {Dayal}, {Furtak}, {Greene}, {Kokorev}, {Pan}, {Price}, {Suess}, {Weaver}, {Whitaker}, \& {Williams}}]{Wang2024ApJ...963...74W}
{Wang}, B., {Leja}, J., {Atek}, H., {et~al.} 2024, \apj, 963, 74, \dodoi{10.3847/1538-4357/ad187c}

\bibitem[{{Wang} {et~al.}(2025){Wang}, {Ma}, {Li}, {Cai}, {Wang}, \& {Wu}}]{Wang2025TDE250418144W}
{Wang}, Z., {Ma}, Y., {Li}, Y., {et~al.} 2025, arXiv e-prints, arXiv:2504.18144, \dodoi{10.48550/arXiv.2504.18144}

\bibitem[{{Wolcott-Green} {et~al.}(2011){Wolcott-Green}, {Haiman}, \& {Bryan}}]{Wolcott2011MNRAS.418..838W}
{Wolcott-Green}, J., {Haiman}, Z., \& {Bryan}, G.~L. 2011, \mnras, 418, 838, \dodoi{10.1111/j.1365-2966.2011.19538.x}

\bibitem[{{Woods} {et~al.}(2019){Woods}, {Agarwal}, {Bromm}, {Bunker}, {Chen}, {Chon}, {Ferrara}, {Glover}, {Haemmerl{\'e}}, {Haiman}, {Hartwig}, {Heger}, {Hirano}, {Hosokawa}, {Inayoshi}, {Klessen}, {Kobayashi}, {Koliopanos}, {Latif}, {Li}, {Mayer}, {Mezcua}, {Natarajan}, {Pacucci}, {Rees}, {Regan}, {Sakurai}, {Salvadori}, {Schneider}, {Surace}, {Tanaka}, {Whalen}, \& {Yoshida}}]{Woods2019}
{Woods}, T.~E., {Agarwal}, B., {Bromm}, V., {et~al.} 2019, \pasa, 36, e027, \dodoi{10.1017/pasa.2019.14}

\bibitem[{{Wyrzykowski} \& {Mandel}(2020)}]{2020A&A...636A..20W}
{Wyrzykowski}, {\L}., \& {Mandel}, I. 2020, \aap, 636, A20, \dodoi{10.1051/0004-6361/201935842}

\bibitem[{Xiang {et~al.}(2018)}]{Xiang:2018azk}
Xiang, M., {et~al.} 2018, Astrophys. J. Suppl., 237, 33, \dodoi{10.3847/1538-4365/aad237}

\bibitem[{{Yoshida} {et~al.}(2008){Yoshida}, {Omukai}, \& {Hernquist}}]{Yoshida2008:ProtoStar}
{Yoshida}, N., {Omukai}, K., \& {Hernquist}, L. 2008, Science, 321, 669, \dodoi{10.1126/science.1160259}

\bibitem[{{Yuan} {et~al.}(2024){Yuan}, {Lei}, {Wang}, {Wang}, {Wang}, {Chen}, {Shen}, {Cai}, \& {Fan}}]{Yuan2024SCPMA..6709512Y}
{Yuan}, G.-W., {Lei}, L., {Wang}, Y.-Z., {et~al.} 2024, Science China Physics, Mechanics, and Astronomy, 67, 109512, \dodoi{10.1007/s11433-024-2433-3}

\bibitem[{{Zel'dovich}(1970)}]{Zeldovich1970A&A.....5...84Z}
{Zel'dovich}, Y.~B. 1970, \aap, 5, 84

\bibitem[{{Zel'dovich} \& {Novikov}(1967)}]{Zeldovich1967SvA....10..602Z}
{Zel'dovich}, Y.~B., \& {Novikov}, I.~D. 1967, \sovast, 10, 602

\bibitem[{{Zhang} {et~al.}(2024{\natexlab{a}}){Zhang}, {Bromm}, \& {Liu}}]{Zhang:2024PBH}
{Zhang}, S., {Bromm}, V., \& {Liu}, B. 2024{\natexlab{a}}, \apj, 975, 139, \dodoi{10.3847/1538-4357/ad7b0d}

\bibitem[{{Zhang} {et~al.}(2024{\natexlab{b}}){Zhang}, {Liu}, \& {Bromm}}]{Zhang2024MNRAS.528..180Z}
{Zhang}, S., {Liu}, B., \& {Bromm}, V. 2024{\natexlab{b}}, \mnras, 528, 180, \dodoi{10.1093/mnras/stad3986}

\bibitem[{{Ziparo} {et~al.}(2025){Ziparo}, {Gallerani}, \& {Ferrara}}]{Ziparo2025JCAP...04..040Z}
{Ziparo}, F., {Gallerani}, S., \& {Ferrara}, A. 2025, \jcap, 2025, 040, \dodoi{10.1088/1475-7516/2025/04/040}

\bibitem[{{Ziparo} {et~al.}(2022){Ziparo}, {Gallerani}, {Ferrara}, \& {Vito}}]{Ziparo2022}
{Ziparo}, F., {Gallerani}, S., {Ferrara}, A., \& {Vito}, F. 2022, \mnras, 517, 1086, \dodoi{10.1093/mnras/stac2705}

\end{thebibliography}
\bibliographystyle{aasjournal}



\end{document}